\documentclass{emulateapj}
\pdfoutput=1

\usepackage{url}
\usepackage{natbib}

\newcommand{\ts}{184} 
\newcommand{\lms}{170} 
\newcommand{\rejsrcs}{14}
\newcommand{\hiitemprange}{30 to 40~K}
\newcommand{\lmles}{28} 
\newcommand{\hiis}{16} 
\newcommand{\hiibs}{11} 
\newcommand{\stats}{100} 

\newcommand{\baderrorelipse}{18}

\newcommand{\beq}{\begin{equation}}
\newcommand{\eeq}{\end{equation}}
\newcommand{\bdi}{\begin{displaymath}}
\newcommand{\edi}{\end{displaymath}}

\newcommand{\degree}{$^{\circ}$}
\newcommand{\kms}{km\,s$^{-1}$}
\newcommand{\Cyg}{Cyg~X}
\newcommand{\IRAS}{\textit{IRAS}}
\newcommand{\MAMBO}{MAMBO}
\newcommand{\MSX}{\emph{MSX}}
\newcommand{\MIPS}{MIPS}
\newcommand{\SCUBA}{\textit{SCUBA}}
\newcommand{\IRAC}{\textit{IRAC}}
\newcommand{\lm}{$L$-$M$}

\newcommand{\lsol}{${\rm L}_{\odot}$}
\newcommand{\lom}{ L$_{\odot}$$/$M$_{\odot}$}

\newcommand{\HII}{\ion{H}{2}}
\newcommand{\msol}{${\rm M}_{\odot}$}

\slugcomment{Accepted to the Astrophysical Journal}
\shorttitle{BLAST Survey of Cygnus X}
\shortauthors{Roy et al.}
\begin{document}


\title{The Balloon-borne Large Aperture Submillimeter Telescope (BLAST) 2005:
  A 10~deg$^2$ Survey of Star Formation in Cygnus X}
\author{Arabindo~Roy,\altaffilmark{1,\dag}
        Peter~A.~R.~Ade,\altaffilmark{2}
        James~J.~Bock,\altaffilmark{3,4}
        Edward~L.~Chapin,\altaffilmark{5}
        Mark~J.~Devlin,\altaffilmark{6}
        Simon~R.~Dicker,\altaffilmark{6}
        Kevin~France,\altaffilmark{7}
        Andrew~G.~Gibb,\altaffilmark{5}
        Matthew~Griffin,\altaffilmark{2}
        Joshua~O.~Gundersen,\altaffilmark{8}
        Mark~Halpern,\altaffilmark{5}
        Peter~C.~Hargrave,\altaffilmark{2}
        David~H.~Hughes,\altaffilmark{9}
        Jeff~Klein,\altaffilmark{6}
        Gaelen~Marsden,\altaffilmark{5}
        Peter~G.~Martin,\altaffilmark{10}
        Philip~Mauskopf,\altaffilmark{2}
        Jorge~L.~Morales Ortiz,\altaffilmark{11}
        Calvin~B.~Netterfield,\altaffilmark{1,12}
        Alberto Noriega-Crespo,\altaffilmark{13}
        Luca~Olmi,\altaffilmark{11,14}
        Guillaume~Patanchon,\altaffilmark{15}
        Marie~Rex,\altaffilmark{16}
        Douglas~Scott,\altaffilmark{5}
        Christopher~Semisch,\altaffilmark{6}
        Matthew~D.~P.~Truch,\altaffilmark{6}
        Carole~Tucker,\altaffilmark{2}
        Gregory~S.~Tucker,\altaffilmark{17}
        Marco~P.~Viero,\altaffilmark{1,4}
        Donald~V.~Wiebe\altaffilmark{5}}

\altaffiltext{1}{Department of Astronomy \& Astrophysics, University of Toronto, 50 St. George Street, Toronto, ON  M5S~3H4, Canada}

\altaffiltext{2}{Department of Physics \& Astronomy, Cardiff University, 5 The Parade, Cardiff, CF24~3AA, UK}

\altaffiltext{3}{Jet Propulsion Laboratory, Pasadena, CA 91109-8099}

\altaffiltext{4}{Observational Cosmology, MS 59-33, California Institute of Technology, Pasadena, CA 91125}

\altaffiltext{5}{Department of Physics \& Astronomy, University of British Columbia, 6224 Agricultural Road, Vancouver, BC V6T~1Z1,Canada}

\altaffiltext{6}{Department of Physics and Astronomy, University of Pennsylvania, 209 South 33rd Street, Philadelphia, PA 19104}

\altaffiltext{7}{Center for Astrophysics and Space Astronomy, University of Colorado, Boulder CO, 8030}

\altaffiltext{8}{Department of Physics, University of Miami, 1320 Campo Sano Drive, Carol Gables, FL 33146}

\altaffiltext{9}{Instituto Nacional de Astrof{\'i}sica {\'O}ptica y Electr{\'o}nica (INAOE), Aptdo. Postal 51 y 72000 Puebla, Mexico}

\altaffiltext{10}{Canadian Institute for Theoretical Astrophysics, University of Toronto, 60 St. George Street, Toronto, ON M5S~3H8, Canada}

\altaffiltext{11} {University of Puerto Rico, Rio Piedras Campus, Physics Dept., Box 23343, UPR station, San Juan, Puerto Rico}

\altaffiltext{12}{Department of Physics, University of Toronto, 60 St. George Street, Toronto, ON M5S~1A7, Canada}

\altaffiltext{13}{\emph{Spitzer} Science Center, California Institute of Technology, Mail Code 314-6, Pasadena, CA 91125 }

\altaffiltext{14}{Istituto di Radioastronomia, Largo E. Fermi 5, I-50125, Firenze, Italy}

\altaffiltext{15}{Laboratoire APC, 10, rue Alice Domon et L{\'e}onie Duquet 75205 Paris, France}

\altaffiltext{16}{Steward Observatory, University of Arizona, 933 N. Cherry Ave, Tucson, AZ 85721, USA}

\altaffiltext{17}{Department of Physics, Brown University, 182 Hope Street, Providence, RI 02912}

\altaffiltext{\dag}{\url{aroy@cita.utoronto.ca}}



\begin{abstract}
We present Cygnus X in a new multi-wavelength perspective based on an
unbiased BLAST survey at 250, 350, and 500~\micron, combined with rich
datasets for this well-studied region.  Our primary goal is to
investigate the early stages of high mass star formation.  We have
detected \ts\ compact sources in various stages of evolution across all
three BLAST bands.  From their well-constrained spectral energy
distributions, we obtain the physical properties mass, surface density,
bolometric luminosity, and dust temperature.  Some of the bright sources
reaching 40~K contain well-known compact \HII\ regions.  We relate these
to other sources at earlier stages of evolution via the energetics as
deduced from their position in the luminosity-mass (\lm) diagram.  The
BLAST spectral coverage, near the peak of the spectral energy
distribution of the dust, reveals fainter sources too cool ($\sim10$~K)
to be seen by earlier shorter-wavelength surveys like \IRAS.  We detect
thermal emission from infrared dark clouds and investigate the
phenomenon of cold ``starless cores'' more generally.  \textit{Spitzer}
images of these cold sources often show stellar nurseries, but these
potential sites for massive star formation are ``starless'' in the sense
that to date there is no massive protostar in a vigorous accretion
phase.  We discuss evolution in the context of the \lm\ diagram.  Theory
raises some interesting possibilities: some cold massive compact sources
might never form a cluster containing massive stars; and clusters with
massive stars might not have an identifiable compact cold massive
precursor.
\end{abstract}



\section{Introduction}

We report on an unbiased survey of the Cygnus X (\Cyg) high-mass star
formation region, conducted in 2005 by the Balloon-borne Large Aperture
Submillimeter Telescope (BLAST), a 2-m stratospheric telescope that maps
simultaneously at 250, 350, and 500~\micron\ \citep{pascale2008}. A
primary ambition for BLAST was to study the earliest stages of massive
protostellar evolution. Massive stars play an important role in Galactic
ecology, initially through protostellar outflows and later through
radiation pressure, ionization, stellar winds, and supernova explosions,
and yet, in spite of their pivotal role, very little is known about
their formation \citep{Zinn2007}.

Massive molecular clouds are favorable sites for massive star formation.
If they are sufficiently dense to be self-gravitating, then they also
have high extinction.  Even in the near to mid infrared they have
significant optical depth, affecting observations of any embedded
protostars, and leading to the ``infrared dark cloud'' (IRDC) phenomenon
seen by ISO \citep{per96}, \MSX\ \citep{ega98,simon2006}, and
\textit{Spitzer} \IRAC\ and \MIPS\ \citep{car2005,kraemer2010}. Some
IRDCs are nurseries of young protostellar objects and protoclusters
\citep{simon2006}.  However, IRDCs, by definition found in silhouette,
require a luminous diffuse background. Massive clouds can be discovered
directly, without this detection bias, by their far-infrared to mm-wave
optically-thin thermal dust emission. BLAST exploits this fact.

We introduce the BLAST imaging of \Cyg\ in \S~\ref{imaging}.
BLAST detects thermal emission from dust in largely neutral regions and
so provides a complementary view of the interstellar medium (ISM)
compared to, for example, radio emission from ionized gas
(\S~\ref{free-free}; though both are ultimately dependent on the local
stellar radiation field and so are related spatially in some predictable
ways).
An important feature of BLAST imaging is its capability of observing and
characterizing extended structures in the star-forming environment at
different spatial scales (\S\S~\ref{structures} and \ref{sec:diffuse}).
In \S\S~\ref{sources} and \ref{blastphot} we describe the
identification and quantification of BLAST compact sources.
Appendix~\ref{multiphot} discusses measurement of their flux densities
at different wavelengths for use in the multi-wavelength spectral energy
distribution (SED).

Blind submillimeter surveys with BLAST
\citep{chapin2008,netterfield2009}, precursors to those with
\textit{Herschel} (e.g., \citealp{molinari2010}), are ideally suited for
finding and characterizing cold sources, because of the designed
coverage near the peak of the cold dust spectral energy distribution
(SED).  This allows us to determine where the dust temperature $T$ is
low, a key requirement before determining the cold dust column densities
and masses.  BLAST of course also sees the later evolutionary stages
when the dust is warmed up by the forming protostars, and again the
multi-wavelength coverage tightly constrains both dust temperature and
bolometric luminosity ($L_{\rm bol}$).

In order to interpret the BLAST emission more fully, and place the
compact sources in context, we make use of many different surveys of
\Cyg\ in other tracers, including both continuum emission and
molecular lines.  These rich multi-wavelength, multi-species surveys
of \Cyg\ are the basis of a brief overview of the physical environment
in \S~\ref{overview}.

The spectral energy distributions (SEDs) and deduced properties $T$,
$L$, and mass $M$ of the compact sources are presented in
\S~\ref{compact}.
As we discuss in \S~\ref{sec:lum}, the mass of the compact source, along
with its luminosity (which maps into $T$), determine its position on an
evolutionary track in the \lm\ diagram \citep{mol2008}.  Investigating
the stages of high mass protostellar and protocluster evolution is a
prime aspect of this paper.

In the earliest stages, before there is any (significant) internal
source of energy from accretion or nuclear burning, the dust is heated
only by the external radiation field. Because of the high extinction the
dust is cold, with $T$ typically $\sim 15$~K or lower, but in any case
colder than the ``ambient'' temperature of diffuse dust in the same
external radiative environment (which near these star forming molecular
clouds might be much more intense than the local interstellar radiation
field).
Such cold clouds, and the initial gravitationally-bound condensations
within them are therefore rather invisible to \IRAS, but are in principle
detectable by BLAST and ground-based surveys targeting regions of high
extinction (e.g., \citealp{motte2007} using \MAMBO\ at 1.2~mm).
According to the time sequence and nomenclature used for low mass star
formation, sources detected in this earlier evolutionary stage could be
called ``class $-1$.''  However, as a reminder of the energetics, we
prefer the physically-motivated shorthand ``stage E'' (from Externally
heated, but also usefully Earliest -- see \S~\ref{sec:early}). Likewise,
we call sources in the subsequent evolutionary stage, when there is
sufficiently vigorous accretion power internally to raise the dust to
the ambient or higher temperature, ``stage A'' (see
\S~\ref{sec:accretion}; the low mass analog would be ``class 0'').

An illustrated view of the evolutionary stages of massive star
formation, illuminated from the BLAST submillimeter perspective, is
presented in \S~\ref{evol}.

A key factor that is important to the interpretation of our particular
observations is that massive stars form in stellar groups and star
clusters \citep{lada2003}.
The approximately 1\arcmin\ angular resolution (FWHM) of the BLAST
images corresponds to 0.5~pc at 1.7~kpc.  This is the size scale
(sometimes called a ``clump'' -- \citealp{bergin2007}) of the observed
embedded and newly-emerging clusters in \Cyg\ (\citealp{leduigou2002};
\S~\ref{clusters}).  Thus it is relatively straightforward to search for
the precursors to these protoclusters, within which massive protostars
will be the dominant source of luminosity and ionizing radiation.
However, resolving protoclusters into unconfused condensations (a
smaller linear scale sometimes called ``cores'') that are destined to
form single (proto)stars requires more angular resolution than has been
available in surveys (even with the 11\arcsec\ beam of MAMBO2 which is
smaller than achieved here with BLAST).  Submillimeter observations with
the larger telescope of \textit{Herschel} will be better (18\arcsec\ at
250~\micron\ with the SPIRE camera), but not immune to these
considerations.

Empirically, \cite{bontemps2009} have made an extensive study of
fragmentation and sub-fragmentation inside targeted molecular dense
cores in the \Cyg\ region with interferometric observations at Plateau
de Bure.
Theoretically, it is still unclear how clusters containing massive stars
form
\citep{MckeeTan2003,krumholz2008,bonnell1997,bonnell2001,bonnell2007,smith2009,wang2010}.
Some insight provided by the BLAST observations and ancillary data,
coupled with current theory is the focus of the discussion in
\S~\ref{sec:discussion}.


\section{BLAST imaging of \Cyg} \label{imaging}

\Cyg, positioned in the Galactic Plane at about $l = \ 80$\degree\ and
about $1.7$~kpc away from the Sun \citep{sch2006}, has long been known
for its massive star formation.  With BLAST
\citep{pascale2008,truch2008} we surveyed 10 deg$^2$ in \Cyg\ for 10.6
hr during the June 2005 flight (BLAST05), mapping the area on three
visits to provide cross-linked scanning.  In addition, a significant
amount of calibration-related time was spent observing a circular cap of
radius 1\degree\ centred on W75N, resulting in even higher signal to
noise and dynamic range there.

Final maps on 15\arcsec\ pixels were produced using the map-maker
SANEPIC \citep{pat08}.  The combination of high scan speed and low $1/f$
knee, together with the cross-linking and common-mode removal in
SANEPIC, produces a map retaining diffuse low spatial frequency
emission, in contrast to current ground-based mm-wave mapping which
contends with the atmosphere and therefore is spatially filtered and
emphasizes the compact structures.  However, preprocessing of the
time-ordered data to remove very low frequency drifts makes the SANEPIC
map average near zero, the DC level having been effectively removed.

BLAST05 was designed to produce diffraction-limited beams with FWHM
30\arcsec, 42\arcsec, and 60\arcsec\ at 250, 350, and 500~\micron,
respectively. However, BLAST05 suffered from an uncharacterized failure
relating to the primary mirror figure and focus. Nevertheless, even with
the corrupted beam the resolution is about 3\arcmin\ full width half
power \citep{truch2008,roy2010}, comparable to \IRAS, though
disappointing compared to the diffraction-limited images obtained in the
2006 flight (\citealp{truch2009,netterfield2009}).

\begin{figure}
\centering
\includegraphics[width=\linewidth]{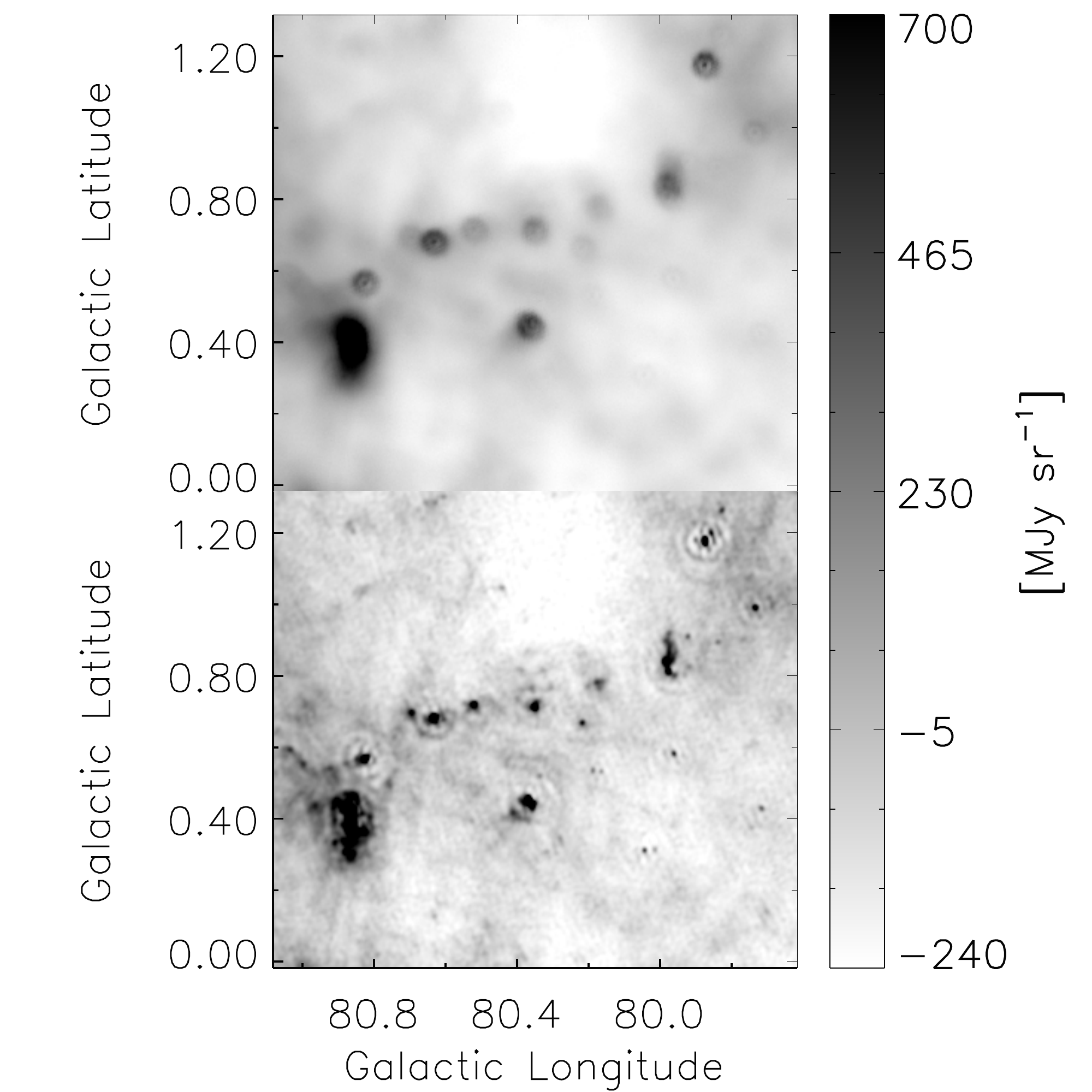}
\caption{Upper: enlarged section of 250~\micron\ BLAST05 image in
  Fig.~\ref{fig:omap}. Lower: deconvolved image using a Lucy-Richardson
  algorithm. Full width half maxima of the deconvolved sources are about
  1\arcmin.  
}
\label{fig:dmap}
\end{figure}

The characteristic beam shape can be seen clearly at many locations
throughout the field, indicating the presence of point sources with
angular size comparable to the diffraction limit.
Obviously we want to work with images with as good resolution as
possible.  We removed the worst effects of the corrupted beam (see Roy
et al. 2010, in preparation) by a Lucy-Richardson deconvolution
algorithm \citep{richardson1972,lucy1974}.  The significant improvement
is illustrated in Figure~\ref{fig:dmap}.
Even so the effective resolution is about 1\arcmin, less dependent on
wavelength than for diffraction (which has the silver-lining benefit
that beam matching is not an issue). Moreover, there are artefacts
introduced near very strong sources.
Mathematically, Lucy-Richardson deconvolution should conserve flux, and
we have verified this through simulations and measurements on isolated
compact sources (\S~\ref{blastphot}).
The astrometric accuracy is good, about 5\arcsec\ when the deconvolved
maps (\S~\ref{blastphot}) are aligned with compact sources in common
with MAMBO2 \citep{motte2007} and/or \SCUBA\ \citep{difrancesco}.

\begin{figure*}
\includegraphics[scale=0.6,angle=0]{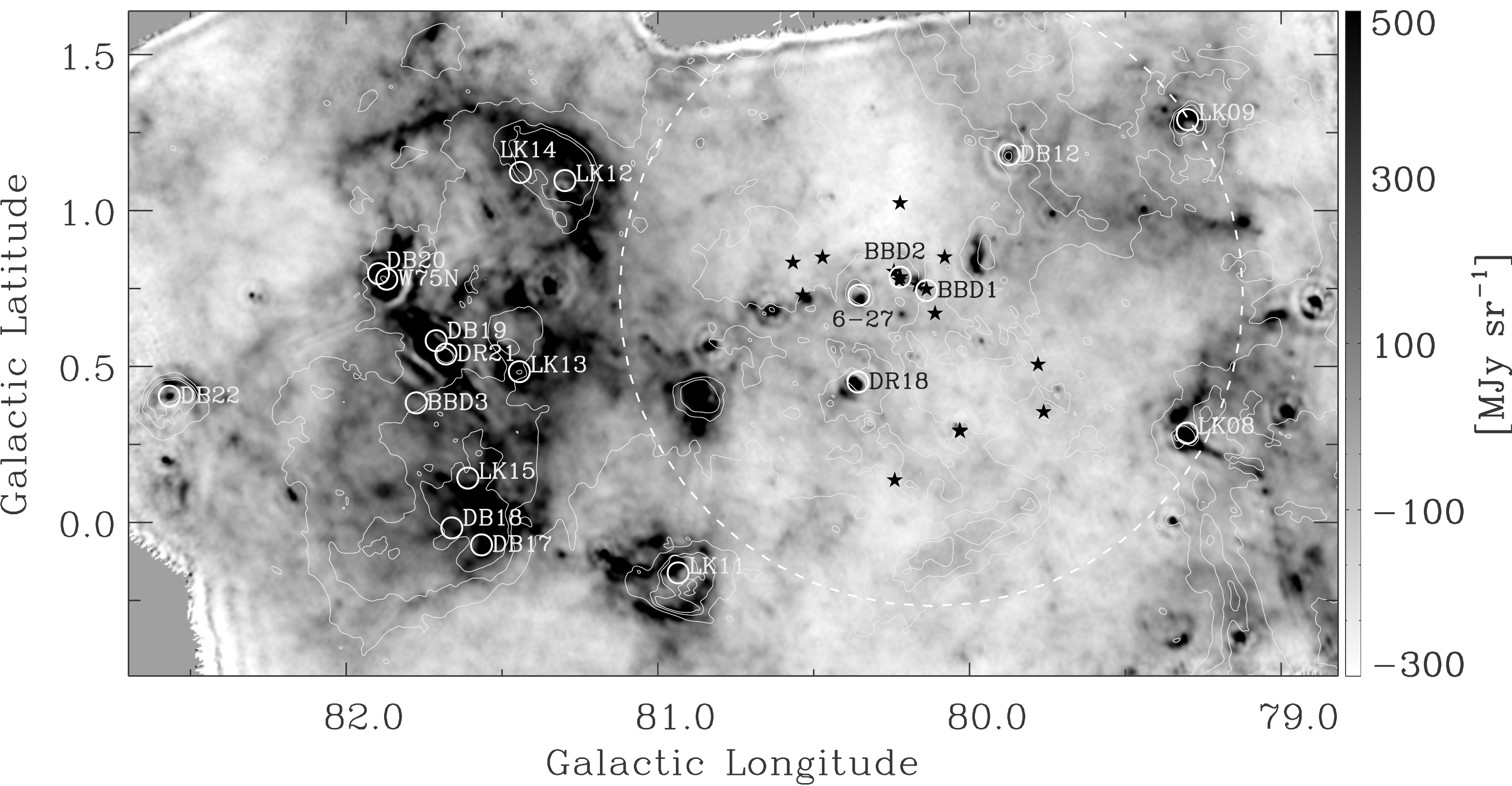}
\includegraphics[scale=0.6,angle=0]{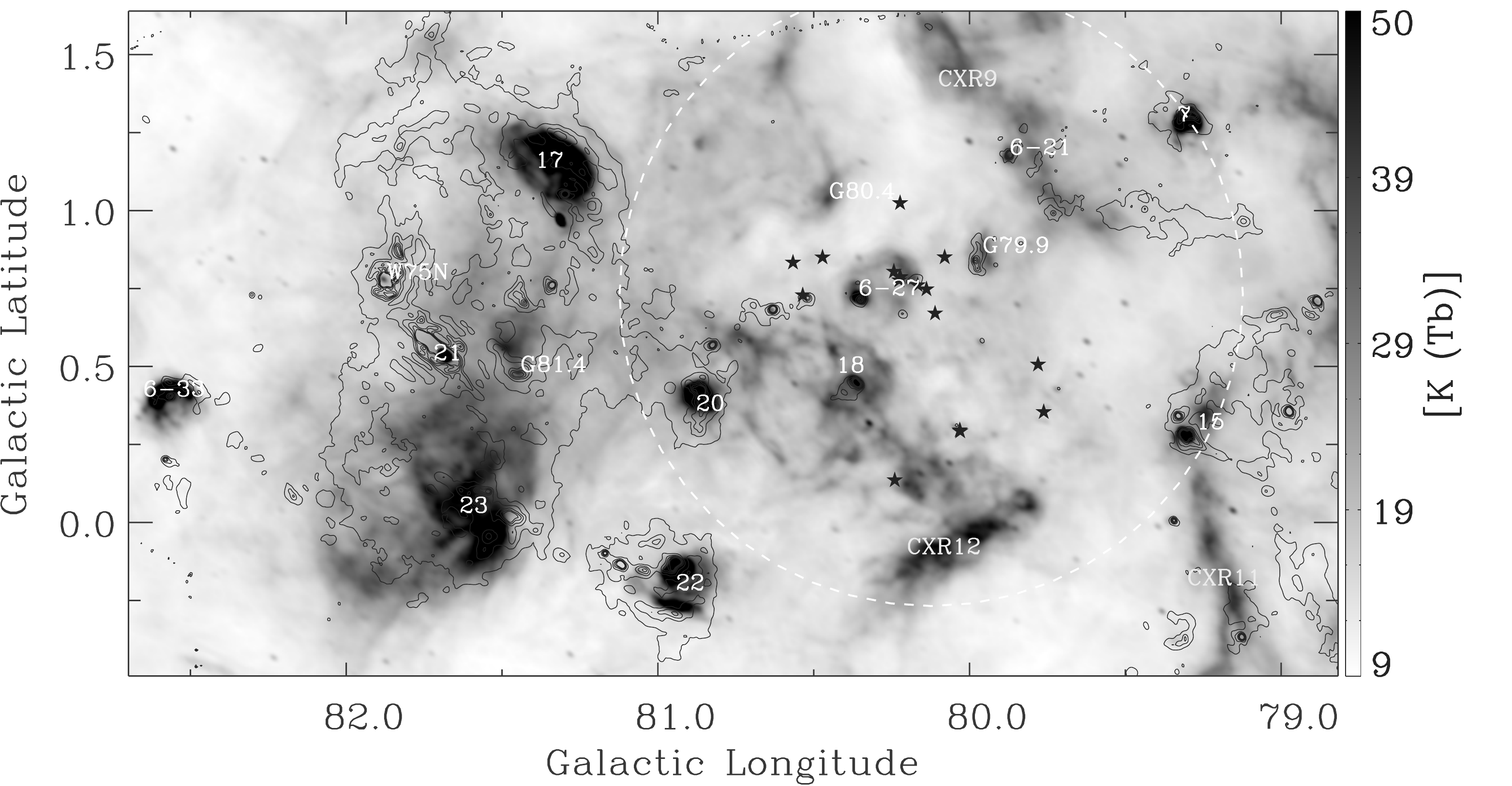}
\caption{Upper panel: BLAST05 deconvolved 250~\micron\ dust emission map
  of \Cyg.  Contours are of 21-cm radio continuum emission from the CGPS
  in the lower panel. The large dashed circle of radius 60\arcmin\ shows
  the extent of the Cyg OB2 association \citep{knodlseder2000} and the
  stars are the most luminous O stars in the list of \citet{neg2008}
  (see text).  Circles denote the positions (not extent) of star
  clusters (see \S~\ref{clusters} for nomenclature).
Lower panel: The radio image in turn has contours from the BLAST05
image. Prominent \HII\ regions and diffuse ridges are marked (see
\S~\ref{free-free} for nomenclature).
}
\label{fig:omap}
\end{figure*}

The full 250-\micron\ map is shown in the upper panel of
Figure~\ref{fig:omap}.
BLAST measures the dust continuum emission along the line of sight and
probes the major star-forming sites in the region by detecting reprocessed 
stellar or protostellar radiation.
The positions of luminous stars in the Cyg OB association
\citep{neg2008} and a number of stellar clusters (\S~\ref{clusters}) are
indicated for reference.
Expanding \HII\ regions, bubbles, and stellar outflows, and the
interplay with structures in the interstellar medium make the morphology
of extended emission in the \Cyg\ region quite complex.
A comprehensive review covering important physical aspects of \Cyg\ is 
presented by \cite{reipurth2008}.
\subsection{A Complementary View using Radio Continuum Emission}\label{free-free}

A complementary way to see the effects of the high-energy radiation from
hot massive stars is via radio emission from the ionized gas. The lower
panel of Figure~\ref{fig:omap} shows the 21-cm continuum map from the
CGPS \citep{tay2003}, which has 1\arcmin\ resolution, about the same 
resolution as the BLAST maps.  A
number of \HII\ regions are labelled for reference: numerals n are DRn
regions from \citet{mic1966}, ECX6-* from \citet{wendker1991}, and
otherwise G$l$ (accurate coordinates for these three regions being
G$79.957+0.866$, G$80.451+1.041$ and G$81.445+0.480$).  Note that the
recombination line velocity for G79.9 is $-$14.8~\kms and for ECX6-27 is
$-$64.4~\kms \citep{lockman1989}, putting these beyond the Cyg~OB2
association.

In diffuse regions the free-free emission is optically thin, but for the
most dense compact \HII\ regions, it can be self-absorbed.
There is rough correspondence with the dust emission, as might be
expected given the impact of OB star radiation on both.  Note, however,
that unlike the dust emission, which depends on the column density, the
\HII\ emission depends on the emission measure, involving the square of
the local density.  Furthermore, ionizing radiation is obviously
essential to produce the plasma.  This radiation, the rest of the
non-ionizing starlight, trapped Lyman-$\alpha$, and cooling lines all
heat any dust surviving in the plasma. Most of the sub-ionizing
radiation escapes the plasma to heat any dust outside the \HII\ region.
In an edge-on geometry, as occurs in extended regions like
DR22, the warm dust immediately outside the arc-shaped ionization front
(I-front) can be seen (Fig.~\ref{fig:omap} and magnified views in
figures below). In a related way, in DR23 the ionization is clearly
stopped on the right by a dense cloud which forms a ``bay'' in the radio
emission, and a corresponding bright feature in the dust map.

There are also large scale ionized structures denoted CXRn
(\Cyg\ Ridges) by \citet{wendker1991}.  Those in the BLAST region are
labelled in Figure~\ref{fig:omap}.  These are seen dimly in the BLAST
image because they have relatively low column density (they also do not
show very strongly in the extinction map; \S~\ref{extinction}).  A much
better dust emission tracer (\S~\ref{structures}) of these ionized
ridges, indeed of all of the diffuse ionized gas, is the
24~\micron\ \MIPS\ image, from the \textit{Spitzer} \Cyg\ Legacy Survey
\citep{hora2009}, which looks remarkably like the radio continuum image.

\subsection{Planck Cold Core Survey}\label{pcc}

In the midst of this turmoil, we search for relatively cold structures
which could be the precursors of the next generation of stars.
This is also the goal of the lower-resolution Planck Cold Core survey
\citep{juv2010}, an unbiased search using the Planck all-sky maps of
submillimeter dust emission.  Cold cores are characterized by a lack of
corresponding \IRAS\ 100~\micron\ emission.  Because the Planck
high-frequency angular resolution is 4\arcmin, clumps -- even if more
extended than the BLAST clumps -- will appear as point sources, and smaller
structures will be beam-diluted.  Even for relatively nearby regions
like \Cyg, the Planck beam would correspond to a spatial extent of about
2~pc.

\begin{figure*}
\centering
\includegraphics[scale=0.6,angle=0]{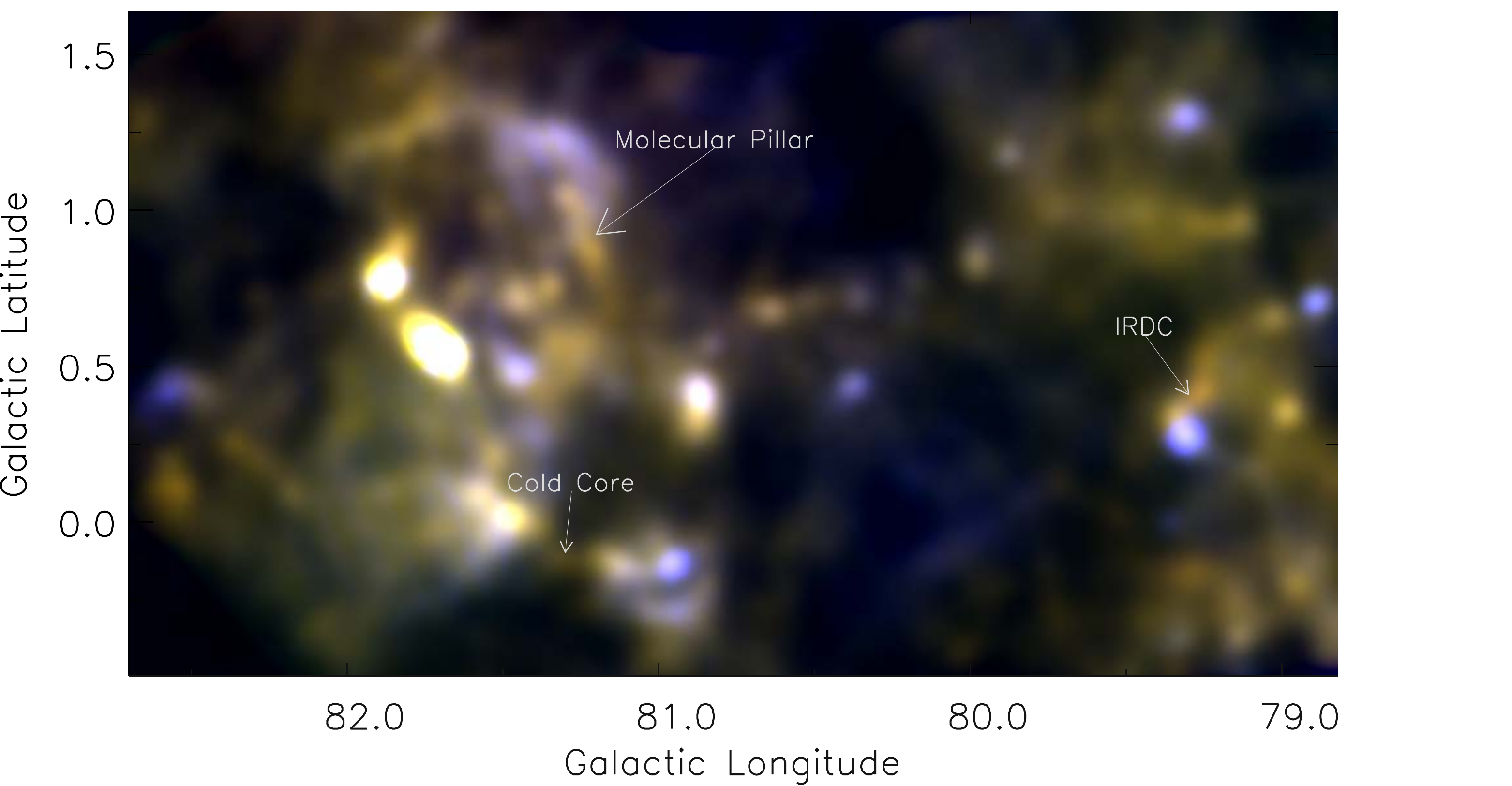}
\caption{Three color BLAST plus \IRAS\ image of \Cyg\ (500, 350, and 100
  \micron\ are represented by red, green, and blue,
  respectively). Images are first convolved to the Planck high frequency
  4\arcmin\ resolution.}
\label{fig:planck}
\end{figure*}

The highest-frequency Planck channels correspond approximately to 350
and 500~\micron.  Prior to the availability of Planck imaging data, we
explored the possibilities by convolving the two longer-wavelength BLAST
images to the Planck resolution and combining them with \IRAS\
100~\micron\ in a three-color image (Fig.~\ref{fig:planck}). At this
resolution, BLAST stage E sources like C116 (\S~\ref{compact} and
\S~\ref{coldcloudsf}) still stand out as being cool.  More extended,
elongated structures like the cool IRDC ridge near DR15 (\S~\ref{irdcs})
and the DR17 molecular pillar (\S~\ref{posters} and \S~\ref{pillars})
are also preserved.  Of course the regions of warmer dust are
highlighted in this image as well.

\subsection{Diffuse Emission}\label{structures}

BLAST maps measure surface brightness (MJy sr$^{-1}$), hence column
density of the dust $M_{\rm {d}}$:
\begin{equation}
I_\nu = M_{\rm {d}} \kappa_\nu B_{\nu}(T) = r \mu m_{\rm{H}}
N_{\rm{H}}\kappa_\nu B_{\nu}(T),
\label{column}
\end{equation}
where $r$ is the dust to gas ratio, $N_{\rm{H}}$ is the H column
density, and $\kappa_\nu$ is the mass absorption coefficient of the
dust.  For later characterization of the SED we note that the spectral
dependence of $\kappa_\nu$ is usually scaled with respect to a fiducial
frequency as $\kappa_{\nu}/\kappa_0= (\nu/\nu_0)^\beta$.  Note also that
for absolute measures of column density we would have to restore the
zero point (DC level) of the maps, as we did for the Cas~A region
\citep{sibthorpe2010} but not here.

In the Galactic Plane, cirrus-like structures in the form of bright
diffuse emission, termed ``interstellar froth'' \citep{waller1994}, are
distributed everywhere.  With the compact structures removed, either
explicitly or by clipping \citep{md07}, the diffuse emission has a
cirrus-like power spectrum, with lower fluctuations at high spatial
frequencies.  In fact, we have found that for this region the power
spectrum is quite like that for \IRAS\ 100~\micron, with amplitude simply
scaled by the SED$^2$ for the appropriate cirrus temperature
\citep{roy2010}.

Somehow, through turbulent motions, thermal instabilities, and phase
transitions, massive localized regions become self-gravitating and so at
high spatial frequencies they stand out from the dwindling cirrus
structure as compact sources. A distribution of masses is
expected, but the cirrus confusion limits our ability to distinguish
lower mass sources (\S~\ref{sources}).  This situation improves with
angular resolution, but even for \textit{Herschel} it is a dominant
limitation \citep{martin2010}.

Actually, the diffuse emission in the submillimeter and mid-infrared
wavelengths comes from three different dust components, distinguished
principally by their size distribution
(\citealp{desert1990,lidraine2001}, Compiegne et al., in preparation):
Big Grains (BGs), Very Small Grains (VSGs), and polycyclic aromatic
hydrocarbons (PAHs).  The BGs, which account for the most of the dust
mass and therefore most of the longer wavelength emission, are in
thermal equilibrium. The VSGs have a relatively lower share of the total
dust mass, even smaller in dense regions.  They are small enough to
experience non-equilibrium heating and so broaden the spectrum toward
shorter wavelengths, beyond the spectral peak of the BG emission.
Compared to what is expected from equilibrium BGs alone, this excess
non-equilibrium emission appears typically at 60~\micron\ and shorter
wavelengths. PAHs are also non-equilibrium emitters, in strong spectral
bands pumped by far-ultraviolet radiation.  The relative amounts of
emission in these three components can vary from region to region.
\citet{sch2006} used the \MSX\ band A images (8.3~\micron;
\citealp{car2005}) effectively as a tracer of PAHs and the
environment. We use \textit{Spitzer} \IRAC\ band 4 images (8.6~\micron;
\citealp{werner2004}) from the \Cyg\ Legacy Survey \citep{hora2009} for
the same purpose (\S~\ref{fuv}), since they have better sensitivity and
resolution.  \textit{Spitzer} MIPS 24~\micron\ images trace VSG emission
generally, and the hotter grains associated with ionized gas, and so, as
commented above, look remarkably like the radio continuum images.

\subsection{Dust Temperature from the Diffuse Emission} \label{sec:diffuse}

From the above equation it is clear that the BG emission that BLAST sees
is modulated by the dust temperature $T$ which is in equilibrium with
the local radiation field, whether in the diffuse medium or within a
compact source.  We concentrate here on the former.  For our exploratory
work here on the BGs, we adopt a single-temperature SED and therefore
fit only data at 100~\micron\ and larger to avoid contamination by VSG
emission.

For a sufficiently large and homogeneous region an estimate of the
characteristic temperature can be obtained by fitting an SED to the
square root of the amplitudes of power spectra. By this method using
data at 100, 250, 350, and 500~\micron, \citet{roy2010} obtained a
cirrus dust temperature of 19.9$\pm$1.3~K for a relatively uncrowded
region (basically the right half of the BLAST map).
However, this more global power spectrum method is not practical when
the region considered is relatively small, inhomogeneous, and/or
focusing primarily on particular sub-structures.

An alternative approach for smaller regions begins with pixel-by-pixel
correlations of images with respect to some reference image (here
250~\micron). Small scale structures are remarkably well correlated
across the three BLAST bands.  The slopes of these correlations describe
the relative SED of the spatially-varying dust emission that is changing
in common across these images.  Note that this approach cannot give the
dust temperature pixel by pixel (that would require the DC offsets
too). On the other hand, the emission in each pixel is from dust of
different temperatures along the line of sight, whereas our approach
isolates and characterizes certain more localized spatial components.

For quantification of the size of the changes in temperature, we
selected regions that appeared to have different colors in a map like
Figure~\ref{fig:planck}.
\begin{figure}
\centering
\includegraphics[width=\linewidth]{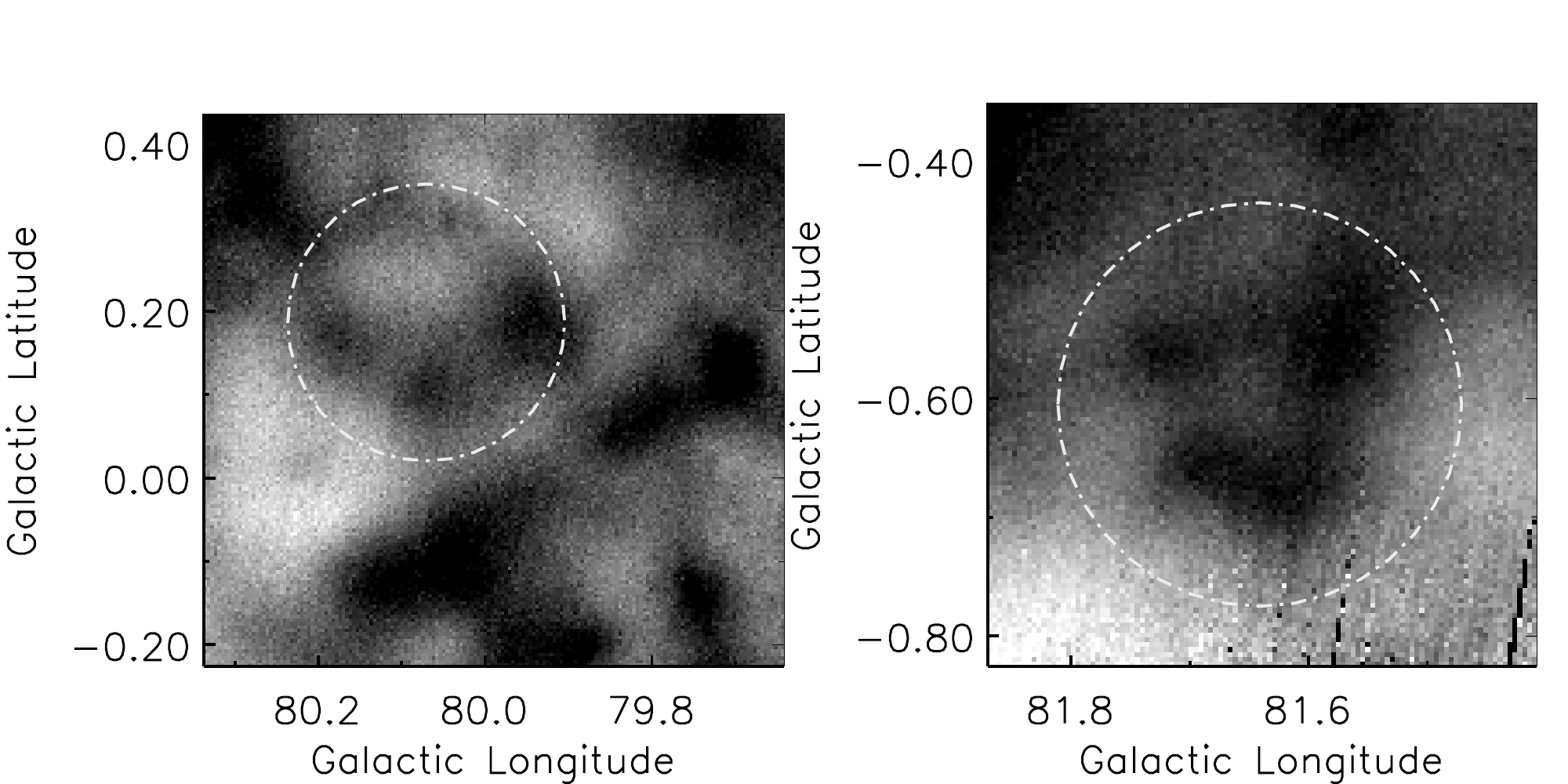}
\caption{Two small regions in \Cyg\ selected on the basis of a
  three-color 500, 250, and 100~\micron\ image to have warm (left) and
  relatively cold dust (images here at 500~\micron).  Note that as in
  other figures, white represents lower intensity.
}
\label{fig:diffuse}
\end{figure}

\begin{figure*}
\centering
\plottwo{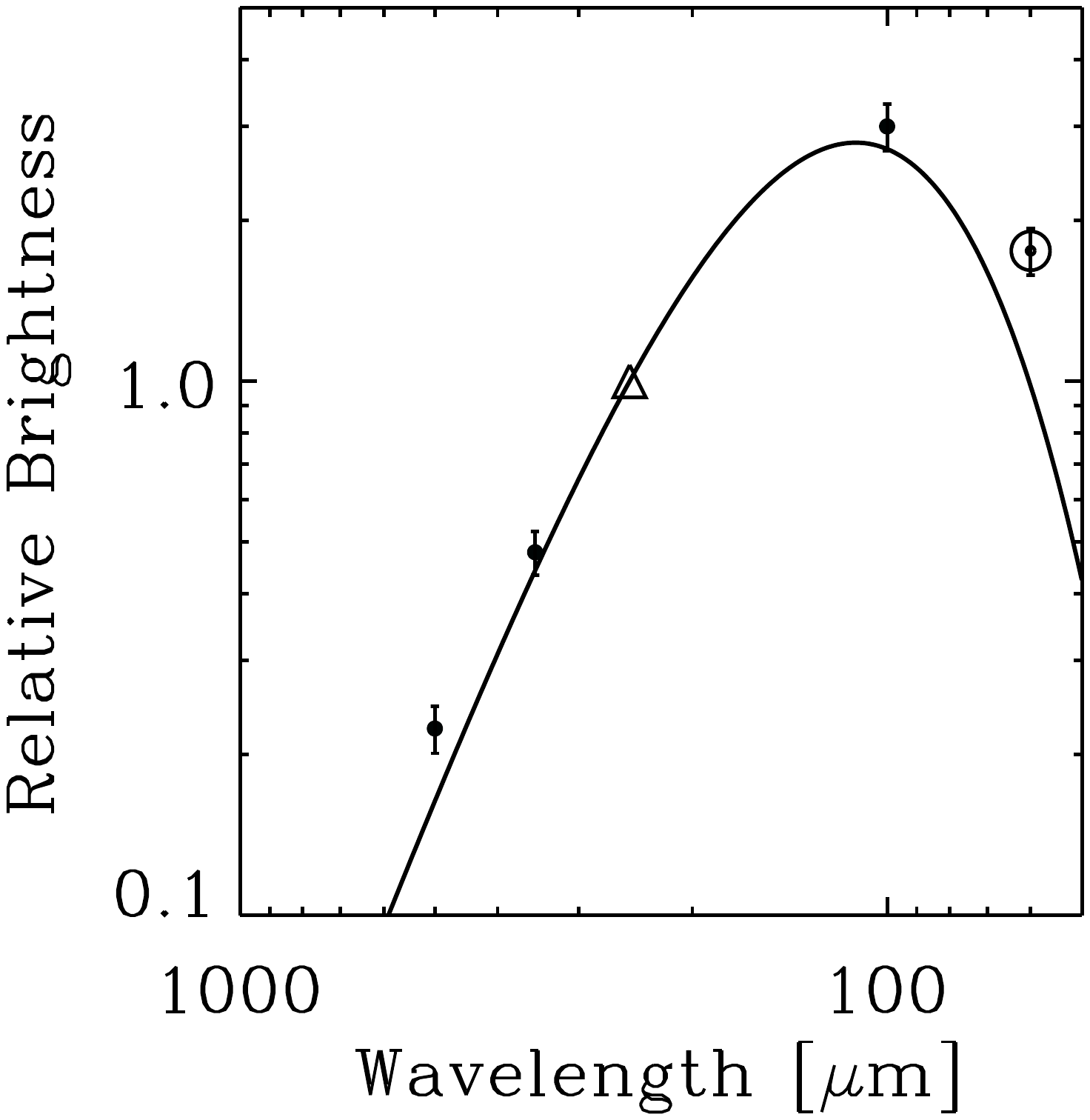}{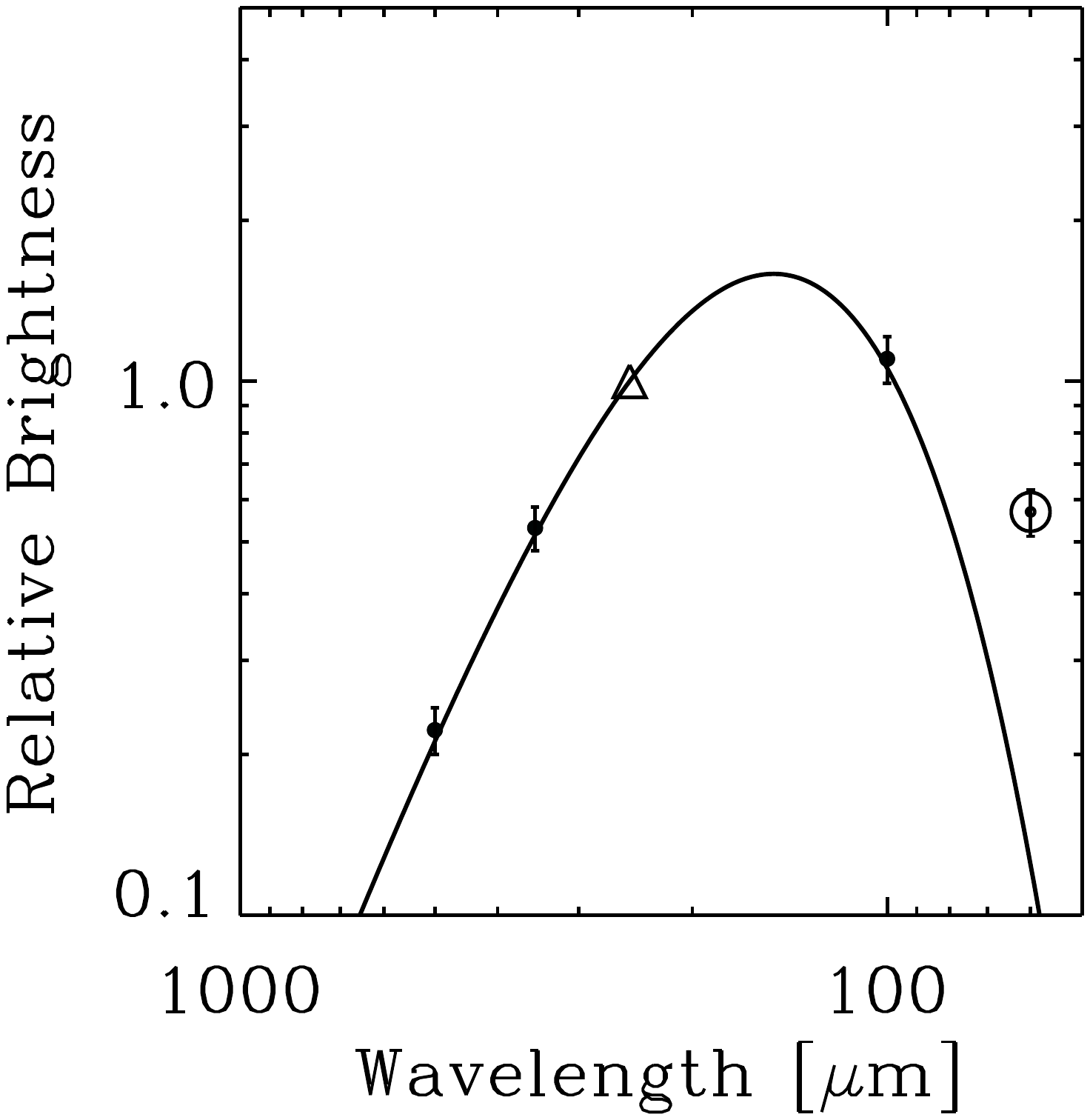}
\caption{Left: Relative SED normalized at 250~\micron\ corresponding to
  the circular region shown in the left-hand image of
  Fig.~\ref{fig:diffuse}. Triangle shows the relative value unity at
  250~\micron.  The SED fitted using $\beta=1.5$ has a temperature of
  29.7 $\pm$ 2.3~K. The 60~\micron\ relative brightness (represented by
  a circle) is not used in the fit.  Right: Relative SED corresponding
  to circular region shown in right-hand image of
  Fig.~\ref{fig:diffuse}. The fit temperature is $21.6\pm 0.5$~K.}
\label{fig:sed_diffuse2}
\end{figure*}

The first region, shown in 500~\micron\ emission in the left panel of
Figure~\ref{fig:diffuse}, is within the relative void created by the Cyg
OB2 stars, near CXR12 (Fig.~\ref{fig:omap}).  Here there is relatively
strong 100 and 60~\micron\ emission.  When the SED (left panel of
Fig.~\ref{fig:sed_diffuse2}) is fit using $\beta=1.5$, which is
consistent with the value adopted for the analyses of compact sources in
this paper, the derived dust equilibrium temperature $T = 29.7 \pm
2.3$~K.  For $\beta=2$, appropriate for local diffuse dust in the atomic
gas at high latitudes \citep{boulanger1996,lidraine2001}, $T$ is
25.4$\pm$2.3~K.  This can be compared to the local high-latitude
equilibrium temperature, 17.5~K.  The higher $T$ in \Cyg\ results from
the much higher effective interstellar radiation field from the OB
association.
The ridge of warm dust emission and this particular selected warm region
above and to the left are also prominent in diffuse \HII\ emission
(Fig.~\ref{fig:omap}) and 24~\micron\ emission, which supports the view
that the local radiation field is high here.  This region is
particularly devoid of CO emission (\S~\ref{molecules}).  However, the
material emitting in the submillimeter must be neutral, because the
associated extinction $\Delta A_V$ is up to 3 mag above the local
background (and is spatially well correlated with the BLAST emission;
\S~\ref{extinction}).
To the lower right of the CXR12 ridge, the dust is cooler, and there is
some CO emission. CXR12 is oriented roughly tangentially to the
direction to the center of the OB association, and might involve
material swept up.  However, the geometry is unclear, not immediately
recalling a classical edge-on photodissociation region (PDR) in a
molecular cloud.

The selected cooler region, judged from relatively low 100 and
60~\micron\ emission, is shown by a circle in the right panel of
Figure~\ref{fig:diffuse}.  Its SED (right panel of
Fig.~\ref{fig:sed_diffuse2}) yields an equilibrium temperature of 21.6
$\pm$ 0.5~K.  The lower $T$ would imply a lower effective interstellar
radiation field, due to the high extinction, with $A_V$ for this
molecular structure (\S~\ref{molecules}) up to 6 mag above the local
background.

\subsection{Compact Sources}\label{sources}

When $I_{\nu}$ in equation~(\ref{column}) is integrated over a suitable
solid angle, accounting for the background, the flux density $S_\nu$
(typically in Jy) of a ``source'' is obtained. Likewise, the source mass
$M$ (gas plus dust) is related to the integrated column density when the
distance $D$ is known, so that
\begin{equation}
S_\nu = L_\nu/(4\pi D^2) = MD^{-2} r \kappa_\nu B_\nu(T).
\label{mass}
\end{equation}
Note that ``source'' refers to the dust emission that is being observed,
not what heats the dust, and so in particular does not imply that there
is an embedded source of energy (a star or protostar) within the volume
of dust being measured.

The definition and characterization of what is a compact source is a
difficult challenge, and different extraction schemes, like Clumpfind
\citep{williams1994}, Gaussclump \citep{stutzki1990,kramer1998}, or the
multi-scale method of \cite{motte2007}, can produce different catalogs
and source characteristics.  Operationally what is called a ``compact
source'' is often basically a structure comparable in size to the beam.
This immediately cautions that what is a compact source to one
instrument could become resolved into multiple components at higher
resolution, as has been illustrated by submm/mm interferometric
observations of \IRAS\ protoclusters and hot cores (e.g., with the SMA and
IRAM Plateau de Bure, \citealp{beuther2007a,beuther2007b,bontemps2009}).

As mentioned above, massive stars form in clusters.  What BLAST can hope
to detect in this survey at 1\arcmin\ resolution is unlikely to be a
single protostar or its progenitor, but rather a protocluster or stellar
nursery possibly containing many unresolved objects.  Of course,
depending on the mass function, a single object among these might
dominate the luminosity and ionizing flux.

We have taken a pragmatic view that since whatever underlying structure
that is smeared out as a compact structure does appear quite like a
Gaussian, this is an adequate model for extracting flux
densities. Furthermore, the deconvolved beam for this BLAST05 survey is
of similar size at all three bands, so that the same volume of dust is
being examined.
A corollary is that when we use other initially higher-resolution data
to expand the spectral coverage of multi-wavelength photometry
(Appendix~\ref{multiphot}), we measured the flux density consistently by
first convolving these maps to the BLAST05 resolution.

\subsection{Compact Source Identification and Photometry}\label{blastphot}

For compact source identification we used the IDL-implemented DAOPHOT
``FIND'' routine, keeping as candidates detected peaks that were above
the local background in both the 250 and 350~\micron\ images by
$3\sigma$ (empirical local rms fluctuation).
Both source identification and photometry on individual sources are
affected by the presence of cirrus noise; for example, at
250~\micron\ the $1-\sigma$ cirrus noise level was evaluated to be $\sim
4$~Jy in the fainter right portion of the map \citep{roy2010}, although
it is expected to vary over the map according to the median brightness
of the region (making catalog completeness a function of position in the
map).  The 500~\micron\ map was not productive for further independent
identifications due to at least as great apparent relative confusion
noise.  In fact, for \rejsrcs\ faint sources among the above it was not
possible to measure a reliable 500~\micron\ flux density.
Given this noise and also the deconvolution artefacts near the brightest
sources, we visually checked our candidates from the automated list.
Since we were not depending on a complete survey -- for example, to find
a mass function or relative lifetimes of different stages -- we
were conservative in rejecting sources and a few were even added to the
list. We were most concerned with having sources with accurately
determined SEDs, so that reliable physical parameters could be
determined (\S~\ref{compact}).

Photometry was carried out using a Gaussian non-linear fit, with
amplitude, FWHM, position, and linear background as free parameters.
Model-independent aperture photometry for isolated sources gave the same
fluxes within 7\%, confirming that this is an acceptable approach.
\Cyg\ has a sufficiently flat background locally that fitting a linear
background is an adequate approximation.
However, in a complex crowded star-forming region like much of \Cyg, it
is often not possible to extract fluxes by aperture photometry or by
fitting a single Gaussian.  In such cases, we adopted a multiple
Gaussian technique, fitting simultaneously to all candidates within a
2.5\arcmin\ radius about the parent source. If a parent source in a
crowded region had a low signal to noise ratio compared to its neighbors
then we fixed their positions and/or FWHMs as required.

Positions and flux densities for 184 sources are given in
Table~\ref{tab:bsource}.  These \Cyg\ sources are referred to below as
`Cn' where `n' is an up-to-three digit number, ranging from 0 to 183.


\section{Context: Global Environment and Structural Detail}\label{overview}

\subsection{Cyg OB2 Association and Star Clusters}\label{clusters}

Cyg~OB2 is among the more impressive assemblies of OB stars in the
Galaxy.  It is more compact than a normal association and rich enough to
have been called a ``young globular cluster'' \citep{knodlseder2000}.
The latest census indicates that it contains some 60-70 O-type stars
\citep{neg2008}. From their semi-observational HR diagram,
\citet{neg2008} favor a distance 1.5~kpc, with 1.8~kpc being noticeably
too large.  Our adopted distance of 1.7~kpc, consistent with
\citet{sch2006} and \citet{motte2007}, is close to this and a 10\%
uncertainty in distance is of no consequence in the analysis that
follows.

The most luminous hot stars
(selected as $M_{\rm bol} \leq -9.2$, $\log T_{\rm eff} \geq 4.56$ and
$M_{\rm bol} \leq -9.9$, $4.46 \leq \log T_{\rm eff} \leq 4.53$)
are plotted in Figure~\ref{fig:omap} with positions from
\citet{massey1991} and \citet{comeron2002}.  Note that star A37 (at
G80.240+0.135) has what appears to be a bow-shock to its right in the
\textit{Spitzer} images and so might be a runaway star from LK8 in DR15.
\cite{kobulnicky2010} studied the bow-shock morphology of A37 (and
others), placing this star slightly behind Cyg~OB2 at a distance of
2.1~kpc.

According to \citet{knodlseder2000}, Cyg~OB2 extends to a radius of
1\degree\ (30~pc), with half-light radius 13\arcmin\ (6.4~pc).
Examination of the sub-structure reveals two open clusters in the core
(BBD1 and BBD2; \citealp{bica2003}), each of radius $\sim 2$\arcmin\ yet
separated by only 6\arcmin\ and appearing to form a physical pair.

A number of highly reddened OB clusters and stellar groups have been
found in the surrounding molecular clouds using 2MASS.
\citet{dutra2001} performed a targeted search toward large angular size
\HII\ regions in \Cyg; eight clusters and four stellar groups are in
this BLAST survey region.
Five of these have been recovered and characterized by
\citet{leduigou2002}, including the half-population radius $R_{50}$ and
central density.  They also describe two new clusters.
\citet{come1999} studied DR18, finding a stellar group, and
\citet{comeron2002} targeted other ECX \citep{wendker1991} compact
\HII\ regions finding four clusters within the BLAST area. The new one,
ECX6-27, has a recombination line velocity of
$-62.4$~\kms\ \citep{lockman1989} and so is much more distant than
Cyg~OB2, more like 10~kpc.  It has a ``double concentration'' of stars
\citep{comeron2002} and also a complex double structure in the
\textit{Spitzer} images.
The catalog of \citet{bica2003} also includes the W75N and DR21 (W75S)
infrared clusters.
In total there are thirteen clusters and five stellar groups (see
Table~\ref{tab:clusters}).
Their positions (not extents) are shown in Figure~\ref{fig:omap}, with
names in precedence LK, DB, and ECX6-* where the lists overlap.  In
subsequent figures, to indicate the cluster angular size we use $R_{50}$
for the OB star population where available \citep{leduigou2002},
otherwise a radius of 1\arcmin.

\subsection{The Radiative Environment: Ionization}\label{ionization}

The signature of radio emission in this region of the Galactic Plane, as
distinct from the radio galaxy Cygnus A, led to the name
\Cyg\ \citep{piddington1952}.  A more contemporary large scale view is
given in Figure~5 of \citet{uy2001}.  Radio surveys conducted at
different frequencies have catalogued the main \HII\ region complexes
(1390 MHz, \citealp{westerhout}; 5 GHz, \citealp{mic1966};
multifrequency, \citealp{wendker1991}).
What we have observed with BLAST is only the central portion, all
projected within the Cygnus superbubble.  The corresponding 1420~MHz
image from the CGPS, including the diffuse emission, is shown in
Figure~\ref{fig:omap}.

The free-free emission traces the compact sources and extended
structures like I-fronts, elephant trunks and pillars, and material that
is being sculpted by winds.  Figure~\ref{fig:omap} shows a striking
ridge of emission CXR11 extending downward from DR15.  Sculpting seen in
the \textit{Spitzer} images indicates illumination from the upper left
(Cyg~OB2), but like CXR12, this does not appear to be a classical
I-front, because the expected neutral tracers in the PDR (\S~\ref{fuv})
are not present immediately to the right.

In a closed geometry, the bolometric luminosity of the dust emission is
a calorimeter for radiation from any embedded (proto)stars.  Likewise,
the spatially-integrated radio flux from the same region is a measure of
the number of ionizing photons.  Thus one can check for consistency,
which is particularly useful in assessing the distribution of
masses within an unresolved cluster.  For example, dividing the
luminosity among several lower mass stars will produce a lower ionizing
flux.  There are complications of course, in addition to the relative
covering factors.  For example, one must know the distance.  If the
distance is overestimated, then the luminosity will imply more massive
stars and so a relatively higher ionizing flux \citep{comeron2001}.
Furthermore, the radio emission is also
affected by self-absorption.  Thus the ionized zone around young OB
stars deeply embedded in dense gas (ultracompact and hypercompact
\HII\ regions) can appear to be underluminous.  Multifrequency
observations (not pursued here) can of course reveal self-absorption by
its spectral signature in the radio SED (see the discussion of DR21 in
\citealp{wendker1991}), or by the high brightness temperature if the
source is resolved.

\subsection{The Radiative Environment: Far Ultraviolet}\label{fuv}

The non-ionizing radiation heats dust beyond any ionization front
(I-front). This is the main source of the extended submillimeter
emission surrounding the \HII\ regions.  As described above, in cases
with ``edge-on'' geometries, like in DR22, there is a clear displacement
of the BLAST emission to the side of the I-front away from the ionizing
stars.

The non-ionizing radiation induces non-equilibrium emission when the
relevant species, VSGs and PAHs, are present.  Diffuse PAH emission is
well traced by \MSX\ band A; the \MSX\ map of \Cyg\ is discussed by
\citet{sch2006}.  PAH emission can now be seen in more exquisite detail
in the \IRAC\ band 4 images.  Where the geometry is favorable, this
reveals impressive PAH-fronts (defined by where the exciting radiation
is attenuated) outside the I-fronts.
There is also a sweeping arc to the lower right of DR15.  This is not
immediately adjacent to the ionized ridge CXR11 mentioned above (thus
not the standard I-front-PDR geometry).  The arc (and ridge) can be seen
as well in \MIPS\ 24~\micron\ emission \citep{hora2009}, which we
interpret there as VSG emission.

\MIPS\ 24~\micron\ dust emission also turns out to be a good tracer of
the plasma, although because of the different physics there is not a
complete morphological or brightness correspondence.  In the figures for
the selected regions below (\S~\ref{posters}), we show the
24~\micron\ emission overplotted with 21-cm radio emission contours to
illustrate their correlation, and also the relationship to clusters and
the BLAST compact sources.

\subsection{Extinction}\label{extinction}

Another tracer of column density is near-infrared extinction (often
expressed as its equivalent in $A_V$), which can be estimated from the
colors of 2MASS sources.  Such a map by S.\ Bontemps (private
communication) is presented in Figure~1 in \citet{motte2007}.  $A_V$ and
BLAST emission are well correlated.  This correlation can be used to
calibrate the dust opacity (Martin et al.\ in preparation).  In their
Figure~1, \citet{sch2006} show an H$\alpha$ image which strikingly shows
the foreground optical extinction called the Great Cygnus Rift.
%

\subsection{The Molecular Reservoir}\label{molecules}

Numerous surveys of the \Cyg\ giant molecular cloud (GMC) have been
carried out in the molecular line transition
$^{12}$CO~J~=~1~$\rightarrow$ 0 \citep{cong1977,dame1987,leung1992}.
\citet{sch2006} report on an extensive multi-transition survey with
KOSMA ($^{13}$CO~J~=~2~$\rightarrow$ 1, 3~$\rightarrow$ 2, and $^{12}$CO
J~=~3~$\rightarrow$ 2) to study in detail the spatial structural
variations and physical conditions.
The relative intensity of $^{13}$CO~J~=~3~$\rightarrow$ 2 to
$^{13}$CO~J~=~2~$\rightarrow$ 1 depends on the local density and kinetic
temperature.
Their $^{13}$CO~J~=~2~$\rightarrow$ 1 channel maps show elongated
filamentary structure at scales of 10\arcmin\ to 20\arcmin, and smaller
sub-structures (CO clumps) embedded in the larger cloud fragments.  The
peaks of the CO emission profiles occur over a wide range of radial
velocities surveyed from $+20$ to $-20~\rm km~s^{-1}$ (there is not much
gas beyond $-10~\rm km~s^{-1}$ and gas at more negative ``Perseus arm
velocities'' is not included).

\citet{sch2006} provide a comprehensive treatment of the region from
this molecular line perspective.  They estimate 4$\times 10^6$
M$_{\odot}$ of molecular gas, at the large end of GMC masses in the
Milky Way. Thus, despite the prior star formation, there remains a
tremendous reservoir of gas pregnant with star formation, offering the
opportunity to study many evolutionary stages.
To this end, \citet{motte2007} have carried out a continuum survey at
1.2~mm with \MAMBO\, further revealing through dust emission the complex
and hierarchical morphology in the region.  They find that \Cyg\ hosts
about 40 massive protostars destined to be OB stars.  As discussed in
\S~\ref{sec:HII}, the more massive protostars are already forming
(ultra)compact \HII\ regions, famous examples being DR21
\citep{mic1966}, W75N \citep{westerhout}, and AFGL2591, the strongest
submillimeter compact sources within the BLAST coverage.
Table~\ref{tab:clump_cores} lists the correspondences between the cores and
clumps of \citet{motte2007} and our BLAST sources.

CO emission integrated over the line profile ($W$) is often taken as a
surrogate of the column density of molecular hydrogen and so should be
correlated with BLAST emission. We note the good correspondence and
explore this in the examples below. Radiation from massive stars can of
course in time destroy CO and the detailed correspondence with dust
emission.

Neither dust continuum emission nor $W$ is sufficient for describing
the complete physical environment and geometry of the star formation
region. Additional insight can be gained by examination of the CO
velocity cubes \citep{sch2006}.
Where there is CO coverage in our BLAST map, most of the identified
BLAST compact sources are correlated with CO emission features (within
the CO clumps mentioned above), thus associating the dust emission with
gas at a certain velocity.  In \S~\ref{posters}, we provide a few
examples using their $ ^{13}$CO J~=~2~$\rightarrow$ 1 cube to emphasize
that objects seen closely together in projection in a dust emission
image can have quite different velocities.  In principle the velocity
associations could be used to sort the sources with respect to distance.
In \Cyg\ \citet{sch2006} have argued that the main velocity systems are
all at roughly the same distance, 1.7~kpc.  Nevertheless, the
association with different CO clumps having differing velocities does
indicate that the BLAST compact sources are in distinguishable
environments, and there certainly are some sources behind the main
Cyg~OB2 complex (\S~\ref{largedistance}).

\subsection{Selected Regions}\label{posters}

\begin{figure*}
\centering
\includegraphics[width=\linewidth]{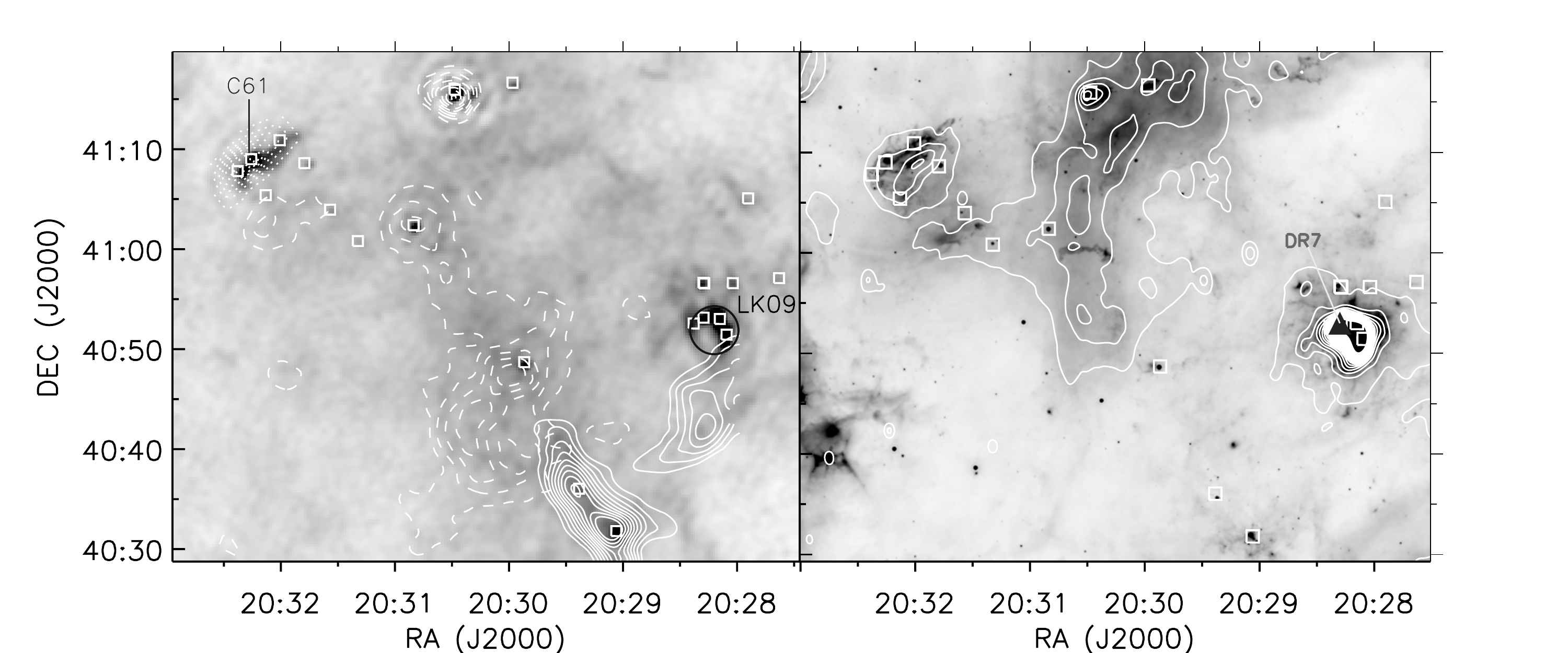} 
\caption{ Left: BLAST 250~\micron\ image including DR7 overlaid with
  contours of $^{13}$CO~J~=~2~$\rightarrow$ 1 emission integrated over
  velocity ranges chosen to highlight the clumps discussed by
  \citet{sch2006}, Fig.~B.4.  Solid, dash, and dot contour lines
  correspond to the ranges 3 to 9, $-7$ to $-1$, and $-13.5$ to
  $-8$~\kms, respectively.  Circle represents position and extent
  $R_{50}$ of star cluster \citep{leduigou2002}.
Right: 24~\micron\ \MIPS\ map of the same region, with contours of the
21-cm radio continuum emission from the CGPS (see Fig.~\ref{fig:omap}).
BLAST compact sources are marked in each image by square boxes of
1.3\arcmin\ width.  Note the \MIPS\ counterparts to many BLAST sources.  }
\label{fig:dr7}
\end{figure*}

In this section we show the BLAST compact sources in the context of the
diffuse submillimeter emission and other tracers discussed above.  They
are marked in the figures with a 1.3~\arcmin\ square, characteristic of
their apparent size in the deconvolved image.  In each example, starting
with Figure~\ref{fig:dr7}, the left panel shows the BLAST
250~\micron\ image. On this are contours of the
$^{13}$CO~J~=~2~$\rightarrow$~1 emission integrated over velocity ranges
which highlight the CO clumps discussed in Appendix~C of
\citet{sch2006}. Good correlations between BLAST emission and CO are
revealed.
The coordinate system chosen for these enlarged views is Equatorial to
facilitate comparison with the cutout regions in the annotated figures
in \citet{sch2006}.  Star clusters, local sources of power, ionization,
and pressure, are also noted (\S~\ref{clusters}).
In the right panel is the \MIPS\ 24~\micron\ image with contours of the
21-cm radio emission from the CGPS map shown in Figure~\ref{fig:omap}.
\citet{motte2007} show pairs consisting of their 1.2~mm \MAMBO\ image and
the corresponding \MSX\ 8.3~\micron\ image, but with no contours
of CO or radio emission.  The \MIPS\ image here is more sensitive to
point-like sources, many coinciding with BLAST sources.

The regions are presented in order of the number of the DR \HII\ region
in the field, whose nominal position is marked with an arrow or filled
triangle in the right hand panel.  To locate the regions in the large
overview image in Galactic coordinates (Fig.~\ref{fig:omap}), see the DR
numbers marked there.
Details of the nature of the BLAST compact sources and their evolution
are deferred to \S\S~\ref{compact} and \ref{evol}.


%

\textit{DR7.} The first region highlighted is a field including DR7, as
shown in Figure~\ref{fig:dr7}.  Immediately outside the DR7 \HII\ region
itself is a rim of BLAST emission containing several compact sources.
The typical rim shaped geometry of the DR7 region has been formed by the
star cluster (LK09) now residing in the cavity.  Interestingly, the DR7
\HII\ region along with its cluster are not associated with the contours
of CO shown.
This \HII\ region has a recombination line velocity of
$-40$~\kms\ \citep{lockman1989} and CO velocity $-50$~\kms\ beyond the
velocity coverage in \citet{sch2006}, placing it well behind the Cyg~OB2
complex, at 3.6 to 7.5~kpc, possibly in the Perseus arm.
Another indicator of the larger distance is the lack of signature in the
extinction map, which is based on colors of detectable 2MASS stars.

There are, however, several CO clumps that do coincide with BLAST
emission and compact sources.  The strong BLAST emission (with compact
sources with \MIPS\ counterparts) to the south coincides with clump 5.
Not all clumps (which have different W) are easily seen in the BLAST
emission (e.g., clumps 7, 8, and 9).  The bright BLAST source at the
top, in clump 1, has a compact \HII\ region and a cometary tail linking
it to Cyg~OB2 (\S~\ref{tails}).

The emission at the upper left, associated with clump 2, is perhaps the
most interesting, having a chaotic morphology in both \MIPS\ and BLAST
images (see also Fig.~\ref{fig:dmap}).  In the ridge we find three BLAST
sources.  Although in projection the center of Cyg~OB2 is close by, about
15\arcmin\ to the east (left), there is no sign of interaction;
furthermore, there is a parallel ridge of \HII\ emission on the right,
suggesting illumination from that side.  It coincides with G79.957+0.866
for which the recombination line velocity is $-14.8$~\kms
\citep{lockman1989}, close to the CO velocity of $-11$~\kms.  In the
entire map, this CO cloud is unique in this velocity range $<
-10$~\kms\ and not obviously connected to the other \Cyg\ molecular
complexes.  This then seems to be an object somewhat more distant,
beyond the influence of Cyg~OB2 (\S~\ref{tails}).

%
\begin{figure*}
\centering
\includegraphics[width=\linewidth]{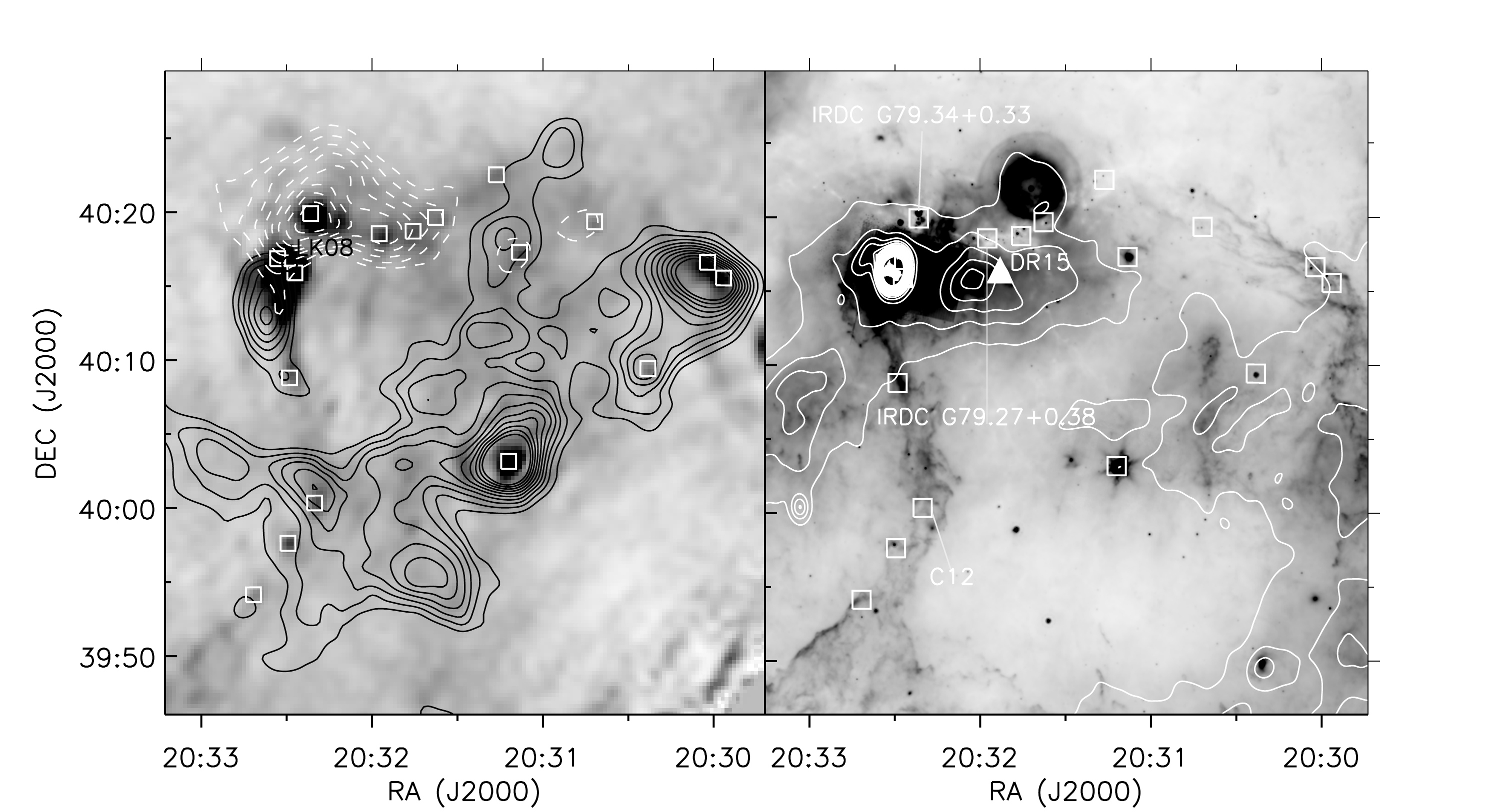} 
\caption{ Same as Fig.~\ref{fig:dr7} but for the DR15 region.  Dashed
  and solid CO contours for ranges $-3$ to 3 and 3 to 20~\kms, as used
  in Fig.~B.6 of \cite{sch2006}.  }
\label{fig:dr15}
\end{figure*}


\textit{DR15.} Figure~\ref{fig:dr15} shows the fascinating region
containing DR15 and the above-mentioned PAH arc.  \cite{sch2006} argue
that the DR15 molecular cloud complex near 0 \kms\ is being influenced
by Cyg OB2 off to the upper left (see also \S~\ref{tails}).  This range
also traces the IRDC ridge containing G79.34+0.33 and G79.27+0.38. As
shown in Figure~\ref{fig:planck} and in more detail in \S~\ref{irdcs},
BLAST sees this cold ridge in emission and resolves several compact
sources; in the \MIPS\ image, some embedded protostars shine faintly
through.
Toward the center of this field is the protostar IRAS~20293+3952,
part of a different CO system at positive velocity.
The pair of BLAST sources to the upper right are associated with
streamers in the \MIPS\ image which appear to be ``blown'' from the
right.

%
\begin{figure*}
\centering
\includegraphics[width=\linewidth]{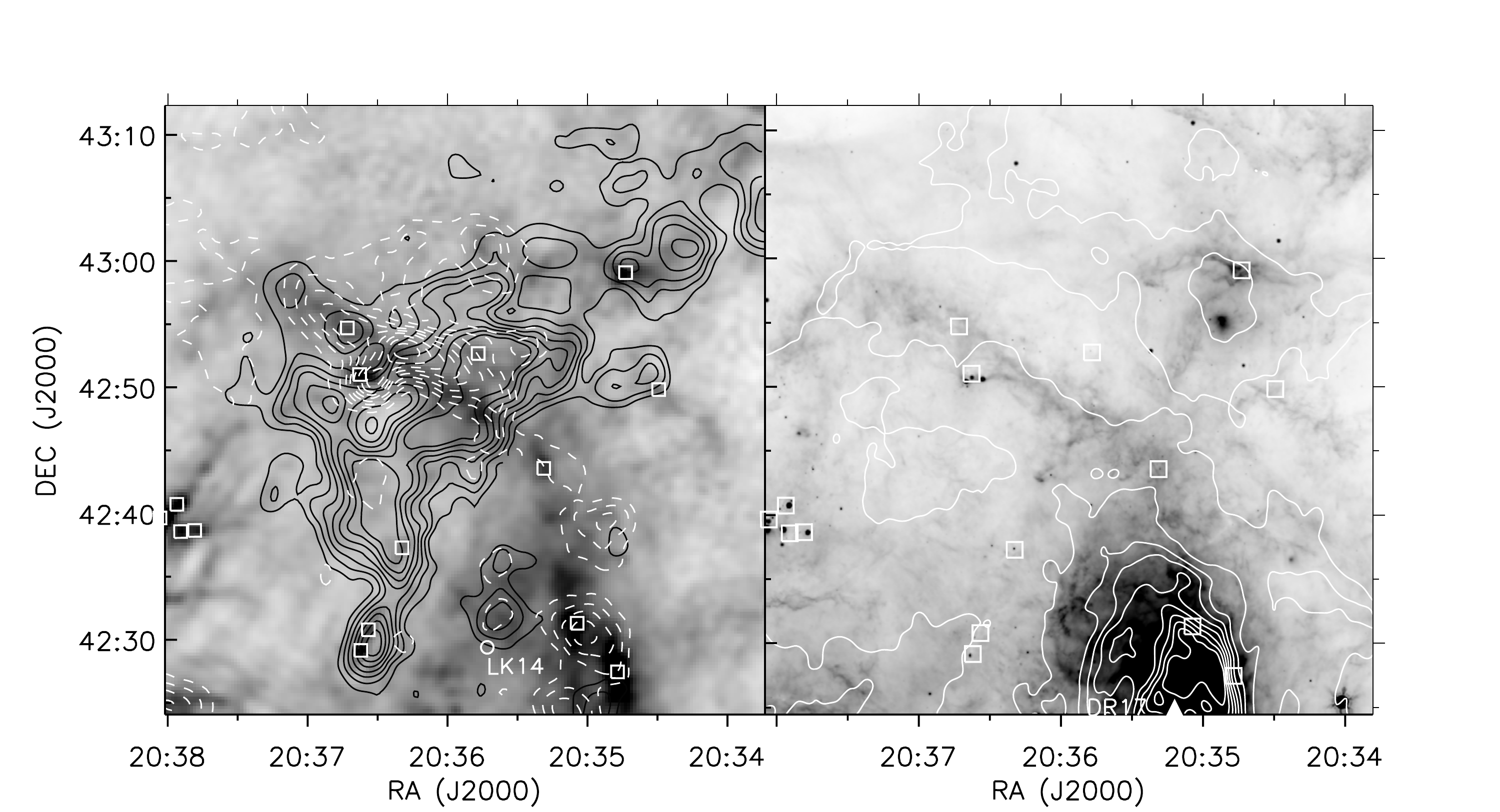} 
\caption{Same as Fig.~\ref{fig:dr7} but for DR17N.  
Solid and dashed CO contours are for ranges 1 to 6 and 6 to 14~\kms, as
used in Fig.~B.1 of \cite{sch2006}.}
\label{fig:dr17N}
\end{figure*}

\textit{DR17N.} The lower portion of Figure~\ref{fig:dr17N} shows the
northern part of the extended DR17 region (see also Fig.~\ref{fig:omap}
and \S~\ref{sec:HII}).  A major arc-shaped extended structure crosses
BLAST and \MIPS\ images, and is seen in 8~\micron\ PAH emission as well,
confining the \HII\ emission influenced by the OB clusters inside DR17.
CO in the higher velocity range 6 to 14~\kms\ best traces the arc.  The
lower velocity system projects across this, and has its own BLAST
emission and compact sources.

%
\begin{figure*}
\centering
\includegraphics[width=\linewidth]{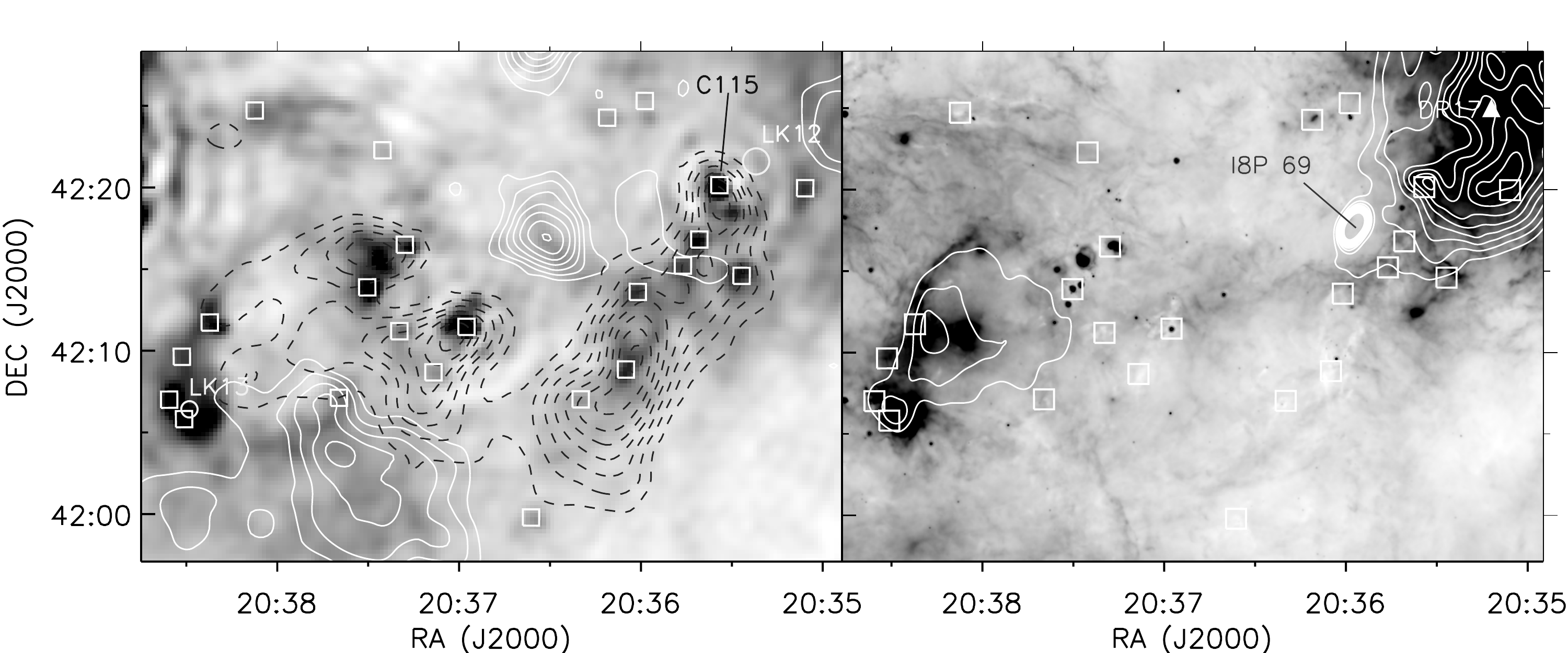} 
\caption{Same as Fig.~\ref{fig:dr17N} but for the DR17-Pillars region to
  its south \citep{sch2006}.  Solid and dashed CO contours are for
  ranges 1 to 6 and 14 to 20~\kms, as used in Fig.~B.1 of
  \cite{sch2006}. The long molecular pillar on the right is that
  labelled in Fig.~\ref{fig:planck}.  A non-thermal extragalactic radio
  source 18P~69 \citep{wendker1991} also prominent in
  Fig.~\ref{fig:omap} is marked.}
\label{fig:dr17}
\end{figure*}    

\textit{DR17-Pillars.} \cite{sch2006} identified ``molecular pillars''
(their Fig.~B.1) in the higher velocity range whose orientation points
to the influence of the OB stars in DR17.  BLAST finds compact sources
associated with these pillars as shown in Figure~\ref{fig:dr17}.
Compared to other dust in DR17, the above-mentioned arc and these
molecular pillars are cool (Fig.~\ref{fig:planck}). There is another
\HII\ region to the left in the image, with BLAST emission and compact
sources along its interface as in DR7.  This is the ``Diamond Ring''
\citep{marston2004} at an intermediate velocity (8~\kms), as described
further in the DR21 discussion below.

%
\begin{figure*}
\centering
\includegraphics[width=\linewidth]{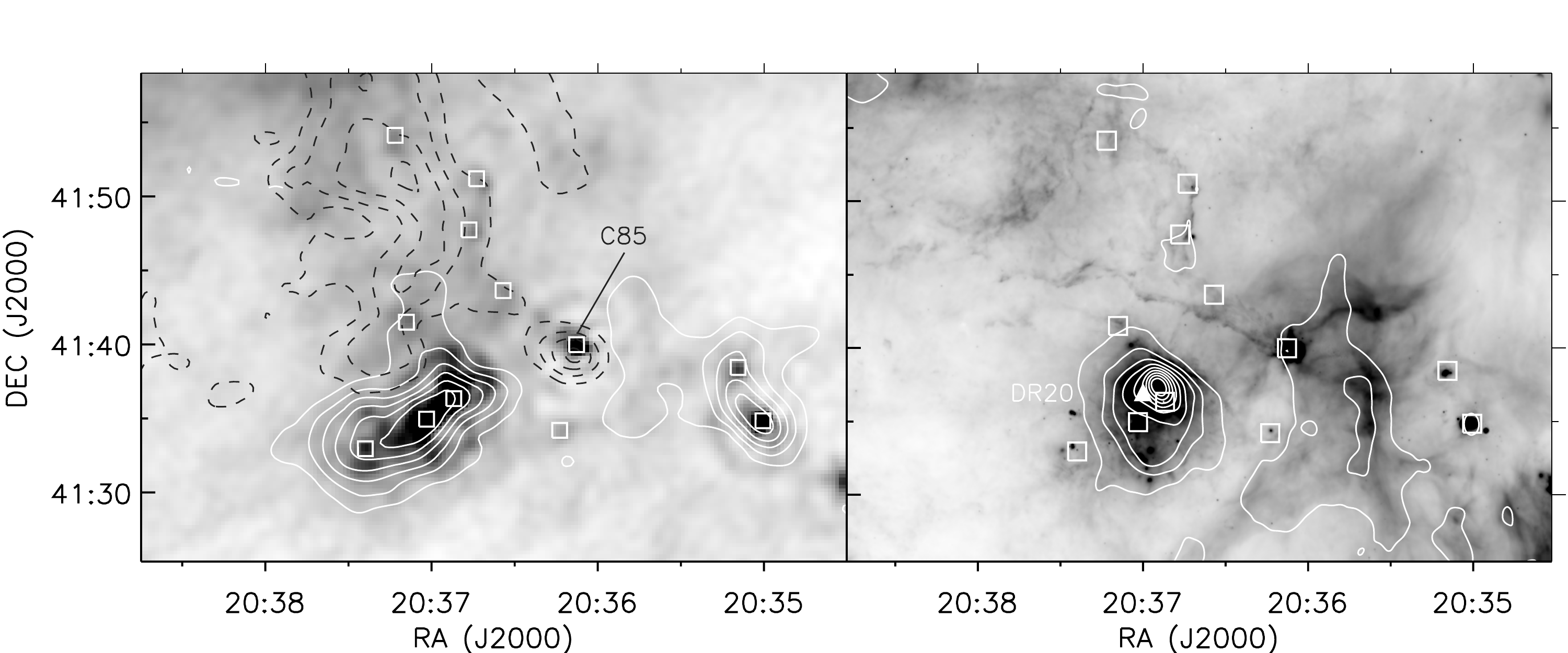} 
\caption{Same as Fig.~\ref{fig:dr7} but for DR20.
Solid and dashed CO contours for ranges $-10$ to 1 and 6 to 14~\kms, as used in
Fig.~B.2 of \cite{sch2006}.
They called the molecular clump  to the
right DR20W, in the same velocity system as DR20, and that containing
C85, in a different velocity system of a prominent elephant trunk
(\S~\ref{pillars}) DR20NW.  }
\label{fig:dr20}
\end{figure*}

\textit{DR20.} Figure~\ref{fig:dr20} shows DR20, which is at the end of
a prominent ridge seen in the low-velocity CO (clump 1) and BLAST
emission (see Fig.~\ref{fig:thumb} in \S~\ref{sec:sed} below).  There
are several BLAST compact sources here and in the complex to the west
(clump 2, DR20W).  The source in the middle of this image is associated
with higher velocity gas (clump 3, DR20NW), as is more diffuse emission
and other compact sources along the pillar toward the upper left
(\S~\ref{pillars}), possibly connected to the velocity system of the
DR17 molecular pillars and arc.

%
\begin{figure*}
\centering
\includegraphics[width=\linewidth]{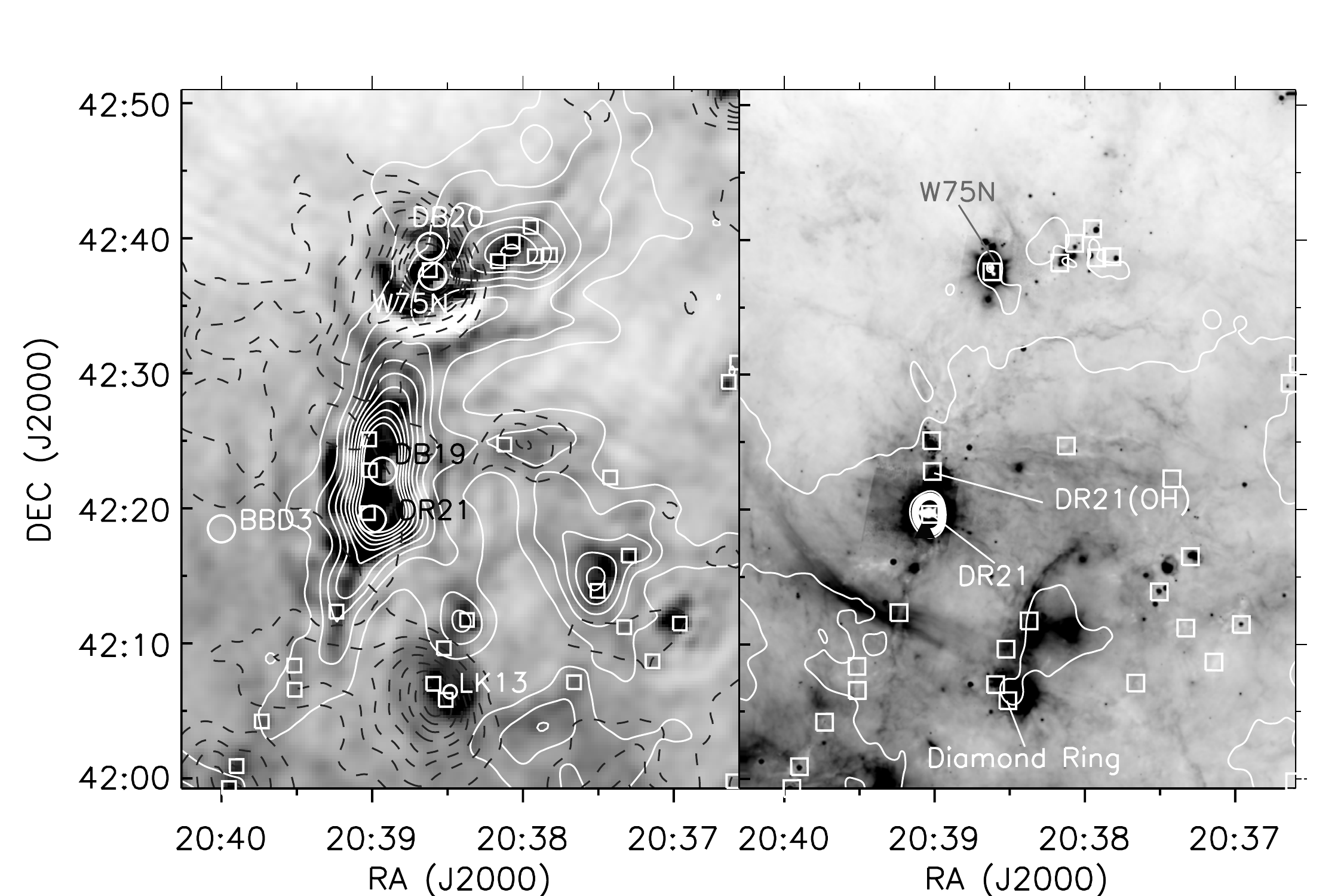}
\caption{Same as Fig.~\ref{fig:dr7} but for DR21 and W75N regions.
Solid and dashed CO contours are for ranges $-10$ to 1 and 6 to
14~\kms\ as used in Fig.~B.3 of \cite{sch2006}.}
\label{fig:dr21}
\end{figure*}

\begin{figure*}
\centering
\includegraphics[width=\linewidth]{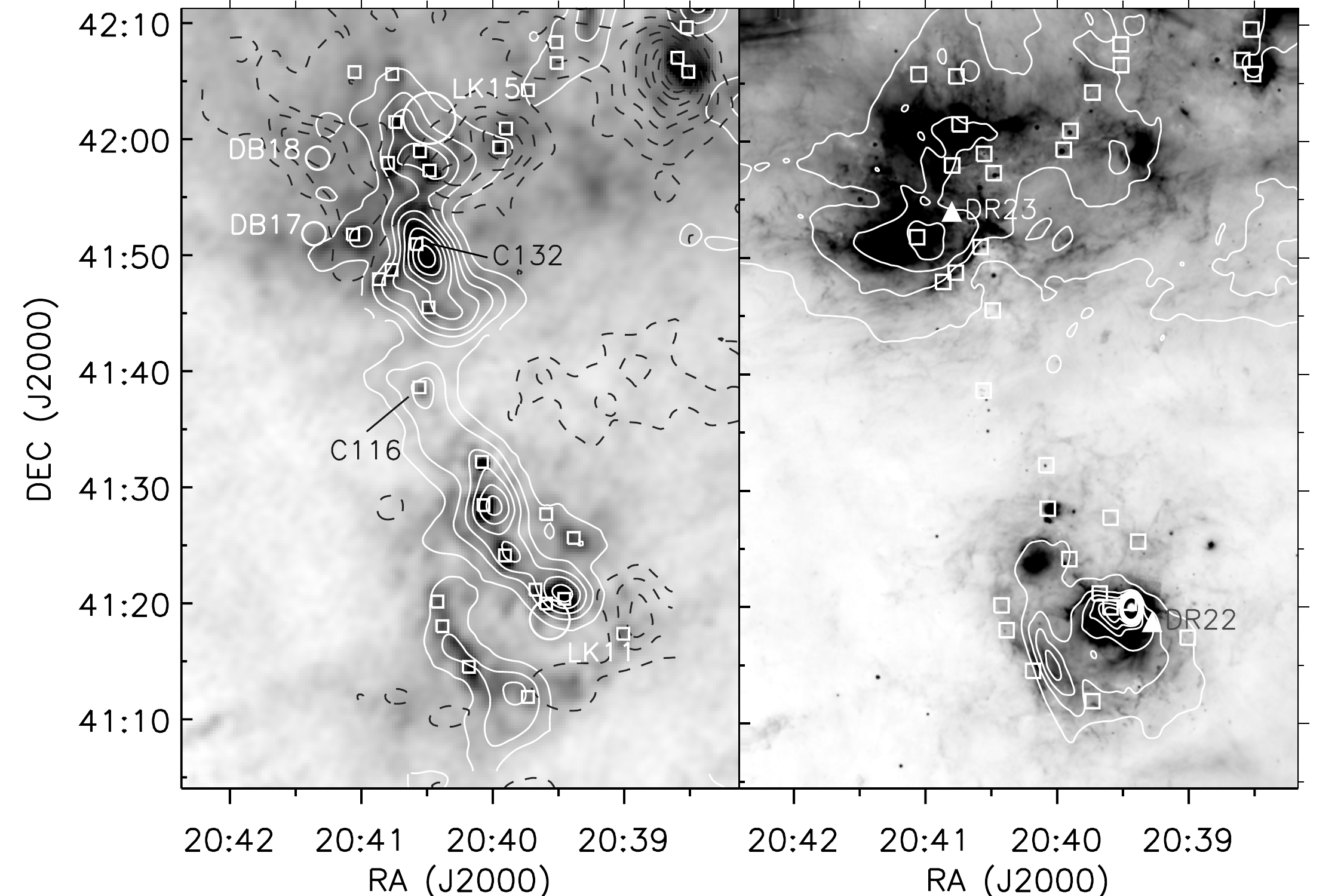} 
\caption{Same as Fig.~\ref{fig:dr21} but for DR22 and DR23 regions to
  its south.  }
\label{fig:dr22}
\end{figure*}

\textit{DR21 and W75N.} The DR21 and W75N regions shown in
Figure~\ref{fig:dr21} is obviously very active.
The ridge of BLAST emission including DR21 and DR21(OH) points
deceptively to W75N. However, the peak CO emission on the ridge (clumps
3 and 4) and clump 8 containing the BLAST sources to the north west of
W75N are at $-3$~\kms\ (the low velocity system), whereas clump 10 of
W75N is in a quite different higher velocity component at $+10$~\kms.
The low velocity system is fairly widespread, including a complex to the
south west (overlapping the lower edge of the eastern-most DR17
molecular pillar) and extending down to DR23 and DR22
(Fig.~\ref{fig:dr22}).  The mass column density traced by BLAST
correlates well with the CO emission.
A fascinating feature is the extension of the DR21 ridge to the south,
which shows up as a very narrow dark lane in the 24~\micron\ image. This
linear feature is at $-4.5$~\kms (see also Fig.~\ref{fig:dark} in
\S~\ref{irdcs} below).

The higher velocity gas also contains prominent sources other than W75N,
including the ``Diamond'' (clump 5) of the ``Diamond Ring''
\citep{marston2004}.  The more extended \HII\ region along the ``Ring''
appears to be defined by both velocity systems.

%

\textit{DR22 and DR23.} The CO cloud structures in the DR22 and DR23
regions shown in Figure~\ref{fig:dr22} correlate well with the BLAST
dust emission.  Immediately to the south west of DR23 is a dim bay in
the \MIPS\ image, where the main CO cloud (clump 1) of the low-velocity
system confines the \HII\ region.  Along this interface are three BLAST
compact sources, including C132, perhaps triggered by the earlier
generation of stars in the DR23 cluster DB17.
The two DR regions are interconnected by a CO filament (several clumps)
in the lower velocity range.  Along this filament there are BLAST
compact sources and dark clouds, one (clump 3) with a protostellar
nursery visible with \MIPS\ and \IRAC\ (\S~\ref{irdcs}).
For DR22, the complementary detail provided by \IRAC\ emission (see
Fig.~\ref{fig:obstar} in \S~\ref{extended}) confirms a classical PDR
geometry, with the lower left being most edge on.  The location of the
BLAST sources, not just the compact \HII\ region but all around the
periphery, is suggestive of triggering. \\ \\ \\
 

\section{Properties of Compact Sources}\label{compact}

\subsection{Submillimeter-MIR SEDs}\label{sec:sed}

\begin{figure}
\centering
\includegraphics[width=\linewidth]{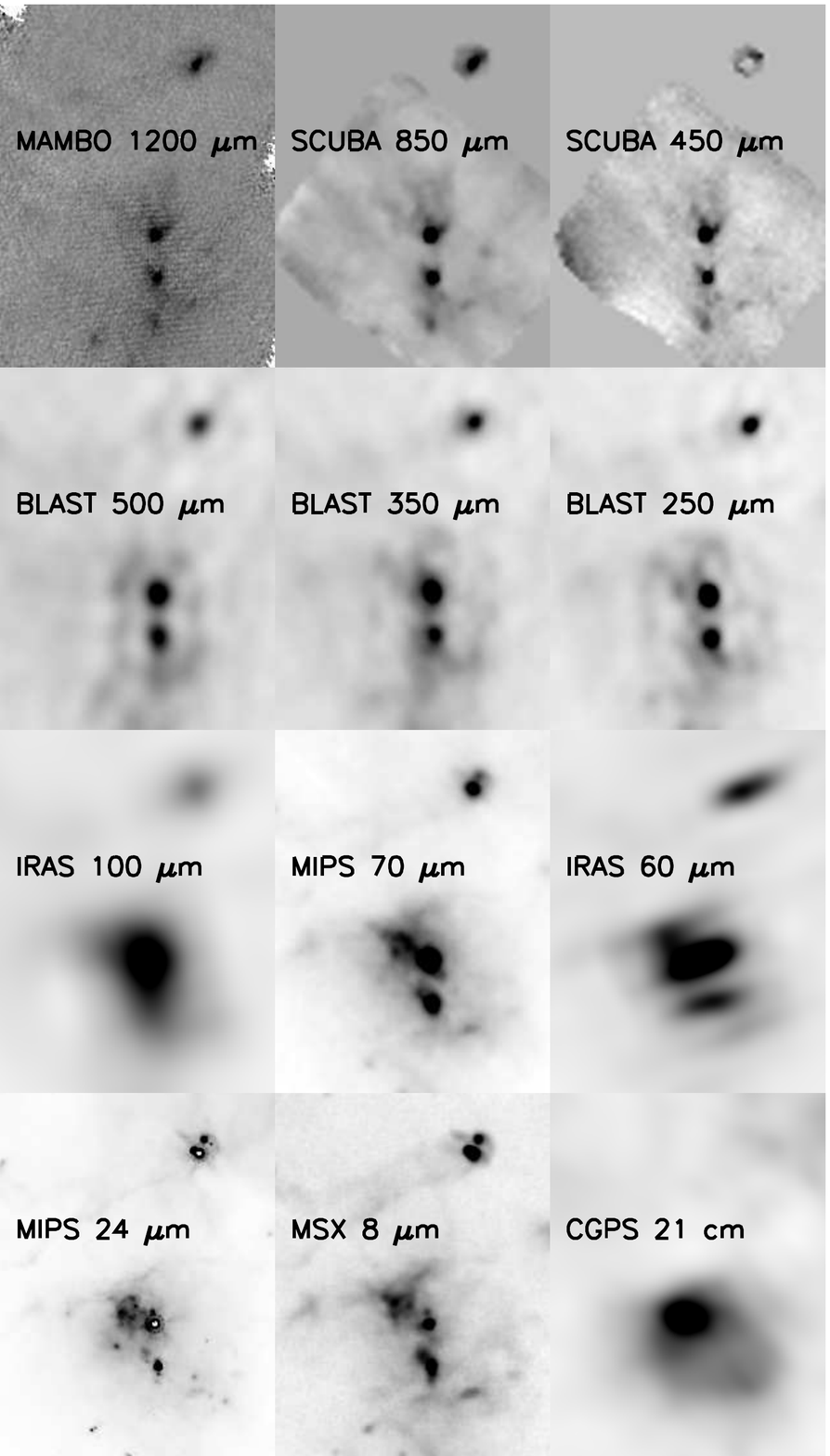}
\caption{Thumbnails zooming in on a 14\arcmin\ by 18\arcmin\ section of
  the BLAST survey area containing DR20 (Galactic coordinates;
  cf.\ Fig.~\ref{fig:dr20}) at multiple wavelengths available for
  photometry. Relative appearance of structures changes because of
  different dust temperatures.  The lowest in the vertical chain of
  three sources, seen only at wavelengths 500~\micron\ and longer and so
  not catalogued as a BLAST source, must be quite cold.
At \IRAS\ 60~\micron\ (IGA, after HIRES processing), sources appear
elliptical across the scan direction.  At \IRAS\ 100~\micron, the
emission of crowded sources is often blended.}
\label{fig:thumb}
\end{figure}

The SED for cold dust emission at temperature $\sim$ 13~K and
emissivity index $\beta = 1.5$ peaks at 250~\micron, and so the
combination of the three BLAST filters alone is well suited for
determining the dust temperature.  Nevertheless, there is a range of
temperature among sources (\S~\ref{temp}) and it is always preferable to
have a broad wavelength coverage spanning both sides of the peak.
Fortunately, \Cyg\ is well covered by both blind and more-targeted
surveys.  An example of the multi-wavelength coverage is given in
Figure~\ref{fig:thumb}.  Clearly, resolution changes from map to map.
Also, for the central source here, the centroid changes at shorter
wavelengths (and in the radio).  As described in \S~\ref{sources}, we
have extracted flux densities consistently within the same size beam, to
characterize the same physical structure.

We fit an idealized single-temperature SED expressed by
equation~(\ref{mass}) to data at all available wavelengths to determine
temperature, mass, and total luminosity of each source (see
\citealp{chapin2008} and \citealp{truch2008} for details).  We adopt the
same parameter values, namely $\kappa_{0} = 10$\,cm$^2$\ g$^{-1}$, $r =
0.01$, and $\beta = 1.5$, and $D = 1.7$~kpc except for a few distant
sources (\S~\ref{largedistance}). Single-temperature SEDs based on this
value of $\beta$ fit the data of typical sources well.  One of the main
systematic uncertainties is the value of $\kappa_0 r$ which is probably
not known within a factor of two.

We treated ancillary photometric data for wavelengths less than
100~\micron\ (see Appendix~\ref{multiphot}) in
the specific context of each individual SED, including them as upper
limits where available/necessary.  Upper limits come with an additional
penalty function in the $\chi^2$ minimization through a survival
equation \citep{chapin2008,truch2008}.

BLAST filters have large spectral widths about the central wavelength.
Color correction to monochromatic flux density is carried out as part of
the SED fitting described by \citet{truch2008}.  The factors are close
to unity and fairly consistent from source to source for the range of
temperatures encountered here: 1.04, 1.07, and 0.99 for 250, 350, and
500~\micron, respectively.

\begin{figure}
\centering
\includegraphics[width=\linewidth]{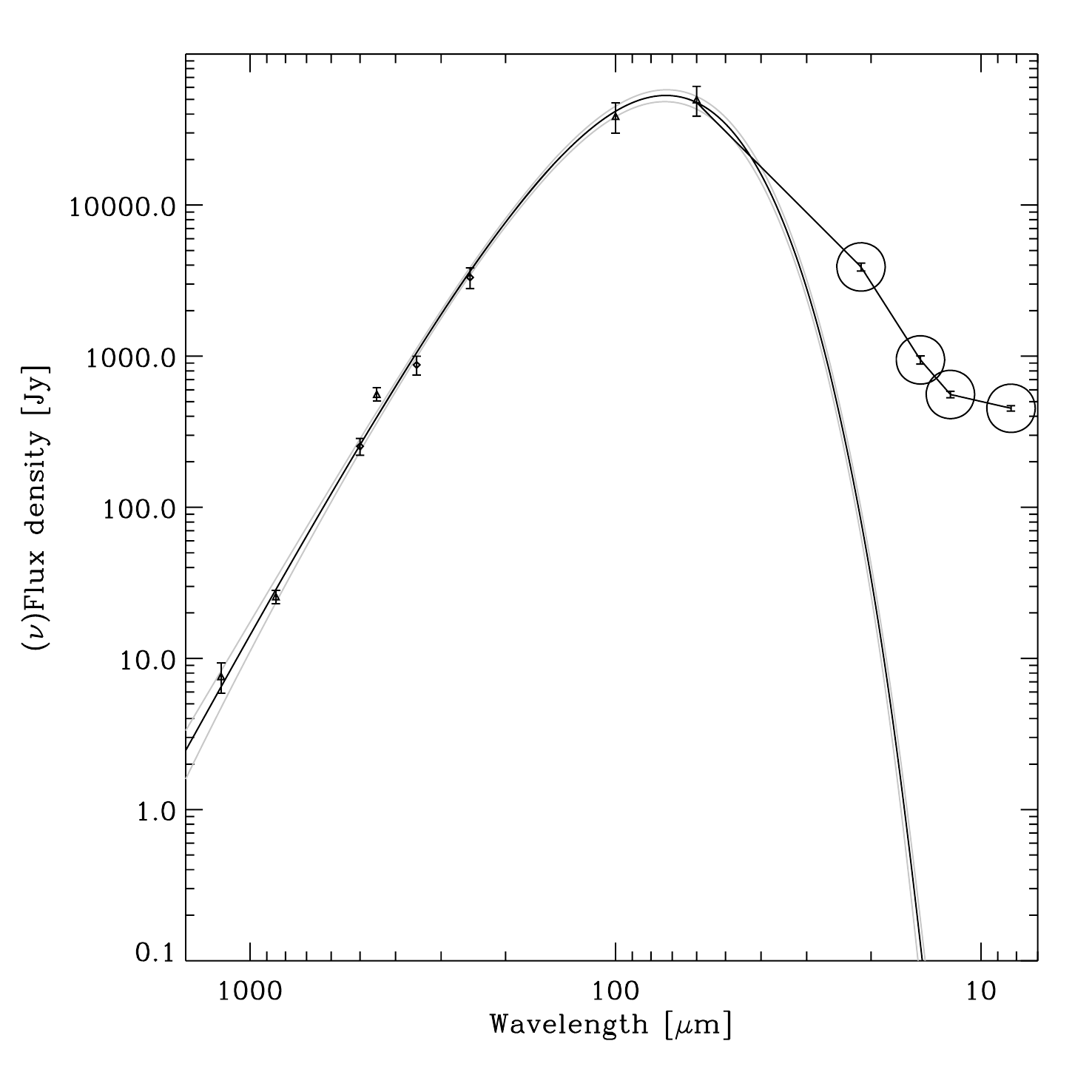}
\caption{SED of C169 (W75N).  To show intuitively where the most
  important contributions to the bolometric luminosity arise, this
  particular log-log plot uses $\nu F_\nu$, obtained by multiplying flux
  densities by $\nu/\nu_{250}$ with $\nu_{250}$ corresponding to
  250~\micron.  The central solid curve shows the best-fit modified
  blackbody using data at $\lambda \ge\ 60$~\micron\ with $\beta = 1.5$.
  The \MSX\ data shown by the circles are not used in the fit but are
  important in constraining $L_{\rm bol}$ (\S~\ref{sec:lumin}).  The
  bracketing curves represent the 68\% confidence envelope of modified
  blackbody models obtained from Monte Carlo simulations.  Best fit
  parameters are $T = 36.2 \pm\ 3.6$~K, $M = (7.6 \pm\ 1.2) \times 10^2$
  M$_\odot$, and $L = (6.2 \pm\ 2.2) \times 10^4$ L$_\odot$.  }
\label{fig:sed1}
\end{figure}

\begin{figure}
\centering
\includegraphics[width=\linewidth]{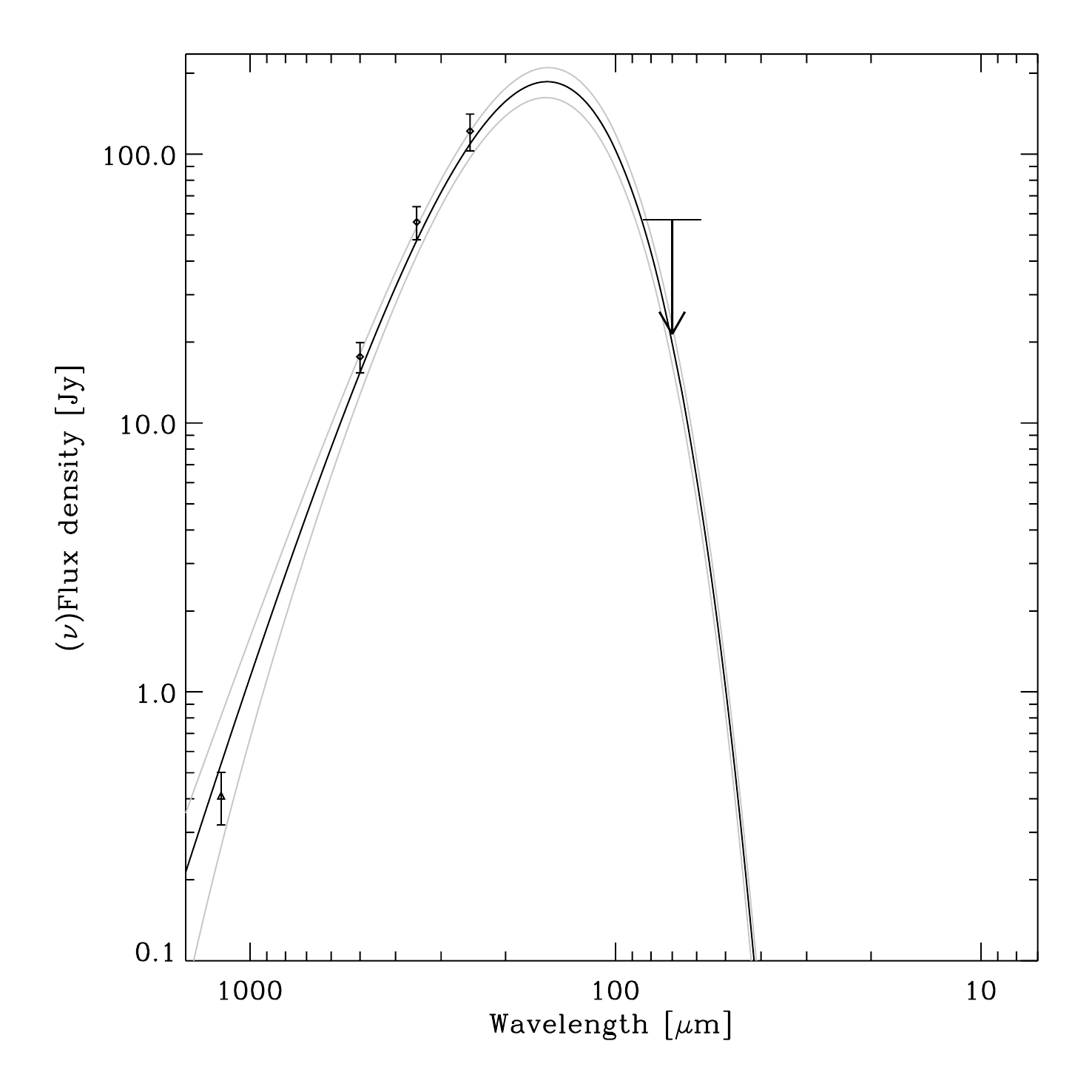}
\caption{Like Fig.~\ref{fig:sed1} but for one of the colder sources in
  \Cyg, C116.  At 70~\micron\ we plot the 1-$\sigma$ (arrow) and
  3-$\sigma$ upper limits (Appendix~\ref{multiphot}); upper limits
  constrain the SED through a penalty function \citep{chapin2008}.  Best
  fit parameters are $T = 17.1 \pm\ 0.9$~K, $M = (1.7 \pm\ 0.2) \times
  10^2$ M$_\odot$, and $L = (2.2 \pm\ 0.4) \times 10^2$ L$_\odot$.  The
  24~\micron\ \MIPS\ image in the upper panel of Fig.~\ref{fig:cold}
  reveals a deeply embedded stellar nursery but it is not yet very
  luminous.  }
 \label{fig:sed2}
\end{figure}

An example of a multi-wavelength SED fit is shown for C169 in
Figure~\ref{fig:sed1}; this is one of the most luminous sources, W75N,
and one of the hottest, with a best fit temperature for mm-FIR data of
36~K.
In contrast to this, the SED of one of the colder sources is shown in
Figure~\ref{fig:sed2}; this is C116 with a temperature of 17~K.

In Table~\ref{tab:sed} we record $T$, $M$, $\Sigma$, and $L$ and their
uncertainties for \lms\ individual sources, there being no entries for
the \rejsrcs\ faint sources either near the map edge or with unreliable
500~\micron\ flux densities.  The uncertainties in $T$, $M$, and $L$,
and the corresponding 68\% confidence envelope of possible modified
blackbodies, were obtained by the Monte Carlo technique described by
\cite{chapin2008}.

\subsection{Sources Behind Cyg~OB2}\label{largedistance}

There are eight sources at distances larger than the main Cyg~OB2
complex, C27, C30, C33, and C35 in DR7 (\S~\ref{posters}), C59, C60, and C61
in G$79.957+0.866$ (\S\S~\ref{posters} and \ref{tails}), and C71 in
ECX6-27 (\S\S~\ref{free-free} and \ref{clusters}).  For calculating $L$
and $M$ we assigned these rough distances of 6.8, 3.4, and 8.5~kpc,
respectively, making $L$ and $M$ 16, 4, and 25 times larger than if the
Cyg~OB2 distance of 1.7~kpc were adopted. This makes C30 the most
luminous source and C71 the most massive.  Both are of course integrated
over much larger spatial scales than for the sources at 1.7~kpc.  Note
that $L/M$ and $T$ are preserved.

\subsection{Temperature} \label{temp}

\begin{figure}
\centering
\includegraphics[width=\linewidth]{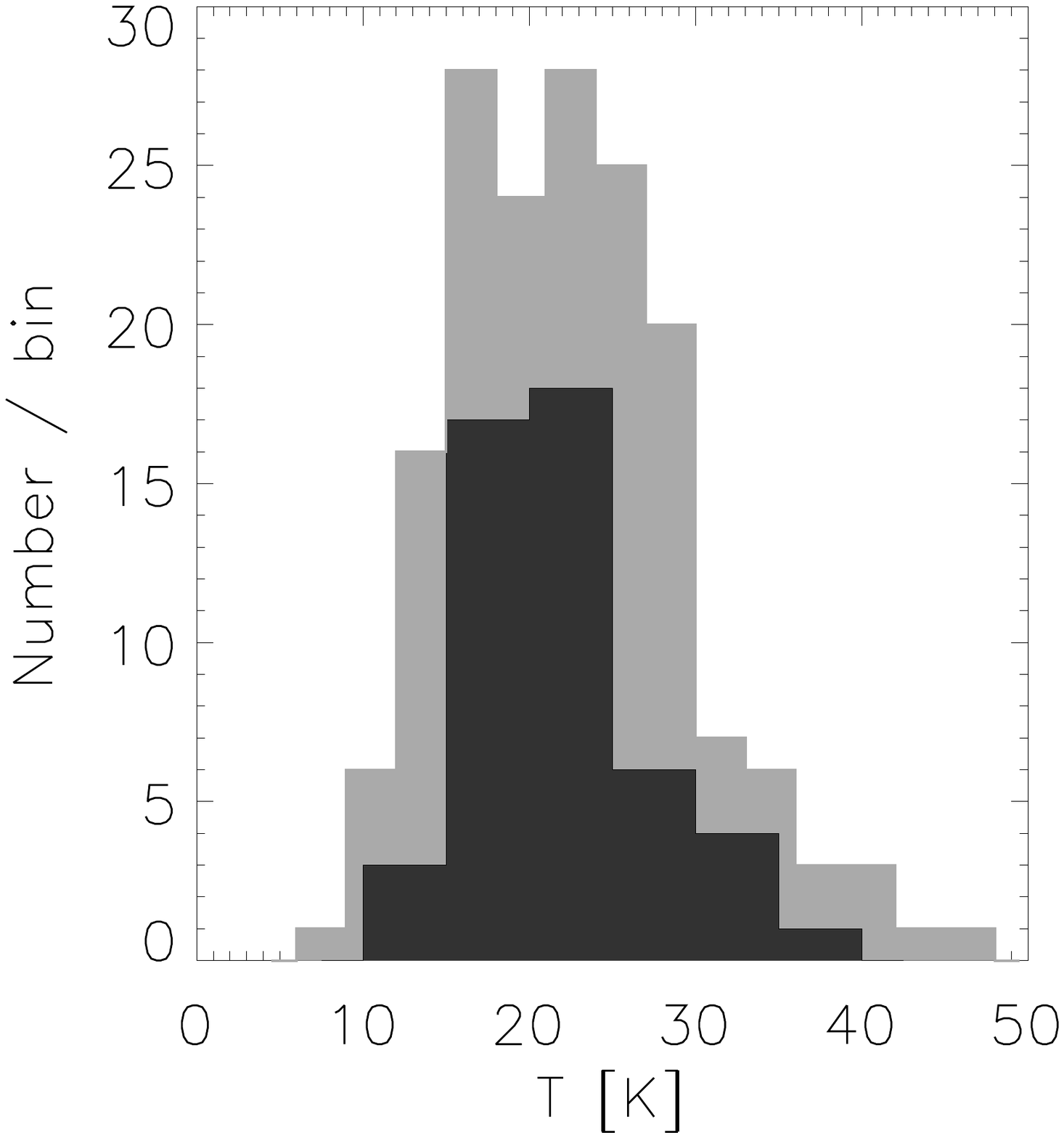}
\caption{Light grey histogram shows the distribution of source
  temperatures for the \Cyg\ field. Dark over-plotted histogram is for
  the BLAST05 Vulpecula field \citep{chapin2008}.
}
\label{fig:temp}
\end{figure}
Temperature is obtained directly as a free parameter of the SED fit.
Figure~\ref{fig:temp} shows the temperature histogram of BLAST sources
detected in the \Cyg\ field.  The superimposed histogram is for the
Vulpecula region \citep{chapin2008}. The median of the distribution in
\Cyg\ is about 22~K which is comparable to the median for Vulpecula at
21~K.
Recall that these are for $\beta = 1.5$.  A somewhat higher $\beta$
combined with a temperature distribution might be a more realistic model
\citep{netterfield2009}, though in the simple model here there is a
single $T$.  For $\beta = 2$, the temperatures would be about 10\%
lower.

\subsection{Mass} \label{sec:mass}

Source mass $M$ is obtained from the best fit amplitude and
temperature of the SED and equation~(\ref{mass}), which shows how any
uncertainties in $\kappa$, $r$, and $D$ are propagated as systematic
errors.  Recall that our sources typically have an apparent FWHM of
1\arcmin, which corresponds to about 0.5~pc.  This is more
characteristic of a ``clump'' mass, as distinguished from the high
resolution measurements of ``cores'' by \citet{motte2007}.  (Note that
the values adopted for $\kappa_0$ and $\beta$ do provide consistency
with the 1.2~mm opacity of 1~cm$^{2}$~g$^{-1}$ adopted by
\citet{motte2007}, following \citet{ossenkopf1994}.  However, without
the multi-wavelength coverage, they were forced to adopt a
temperature.)
From these derived masses, ranging from 10 to 2000~\msol, the BLAST
sources have the potential to form (many) massive stars under
favorable physical conditions.  For example, in the context of the
turbulent core model \cite{krumholz2008} predicted a critical surface
density for the formation of massive stars, as opposed to
fragmentation into much lower masses (see \S~\ref{sec:surfsigma} for
further discussion).

The surface column density for fragments of radius $R$ is given by
\begin{equation}
\Sigma=2.7\times 10^{-2} \left(\frac{M}{100~\rm
  M_{\odot}}\right)\left(\frac{R}{0.5~\rm pc}\right)^{-2}~\rm g~cm^{-2}.
\label{eq:sigma}
\end{equation}
We adopt the deconvolved FWHM of the brightness profile as the radius.
Surface densities of the cores (and clumps) of \citet{motte2007} are
listed in Table~\ref{tab:clump_cores}.  (Note that the assumption of a
temperature of 20~K for the cores means that the surface density
reported might be an upper limit.)
As discussed below, the surface density of BLAST clumps is somewhat
lower, in part because of the low angular resolution which limits the
ability to discern sub-structure at small $R$.

Following equation~(2) of \citet{motte2007}, the volume-averaged
molecular hydrogen density is
\begin{equation}
\langle n_{{\rm H}_2} \rangle = 3.3 \times 10^3 M/(100\ {\rm M}_{\odot})\
        {\rm cm}^{-3}. 
\label{density}
\end{equation}
Because of the BLAST beam size, these clump densities are more than an
order of magnitude lower than for the embedded cores found by
\citet{motte2007}.
The corresponding free-fall time scale \citep{stahler2004} is also
longer,
\begin{equation}
t_{\rm ff} = 5.6 \times 10^5 [M/(100\ {\rm M}_{\odot})]^{-1/2}\ {\rm
yr.}
\label{freefall}
\end{equation}

\subsection{Luminosity} \label{sec:lumin}

The best fit single-temperature SED is a good and integrable
interpolating function from which the far-infrared luminosity $L$ is
derived analytically.  The frequencies at which the most important
contributions arise can be visualized most readily when the SED is
plotted in $\nu F_{\nu}$ form like Figure~\ref{fig:sed1}.
 
Often the short-wavelength mid-infrared data lie in excess of this
simple SED and to allow for that we simply integrate the piecewise curve
connecting that short-wavelength data where available.  This is normally
a small contribution to the total bolometric luminosity $L_{\rm bol}$
(Table~\ref{tab:sed}).  For example, for W75N (C169,
Fig.~\ref{fig:sed1}) $L$ rises from $ 6.2 $  to $ 6.6 \times10^4$ 
L$_\odot$.  In the context of position in the logarithmic
\lm\ plot below, this is not a big correction.

The luminosity range in \Cyg\ is large, extending over three orders of
magnitude from $7 \times 10^4$~\lsol\ down to 40~\lsol, beyond which we
lose many sources in the (cirrus) noise (note that the approximate
250~\micron\ completeness line is about 30~Jy; see
\citealp{chapin2008}).

\subsection{The $L - M$ Diagram}\label{sec:lum}

\begin{figure}
\centering
\includegraphics[width=\linewidth]{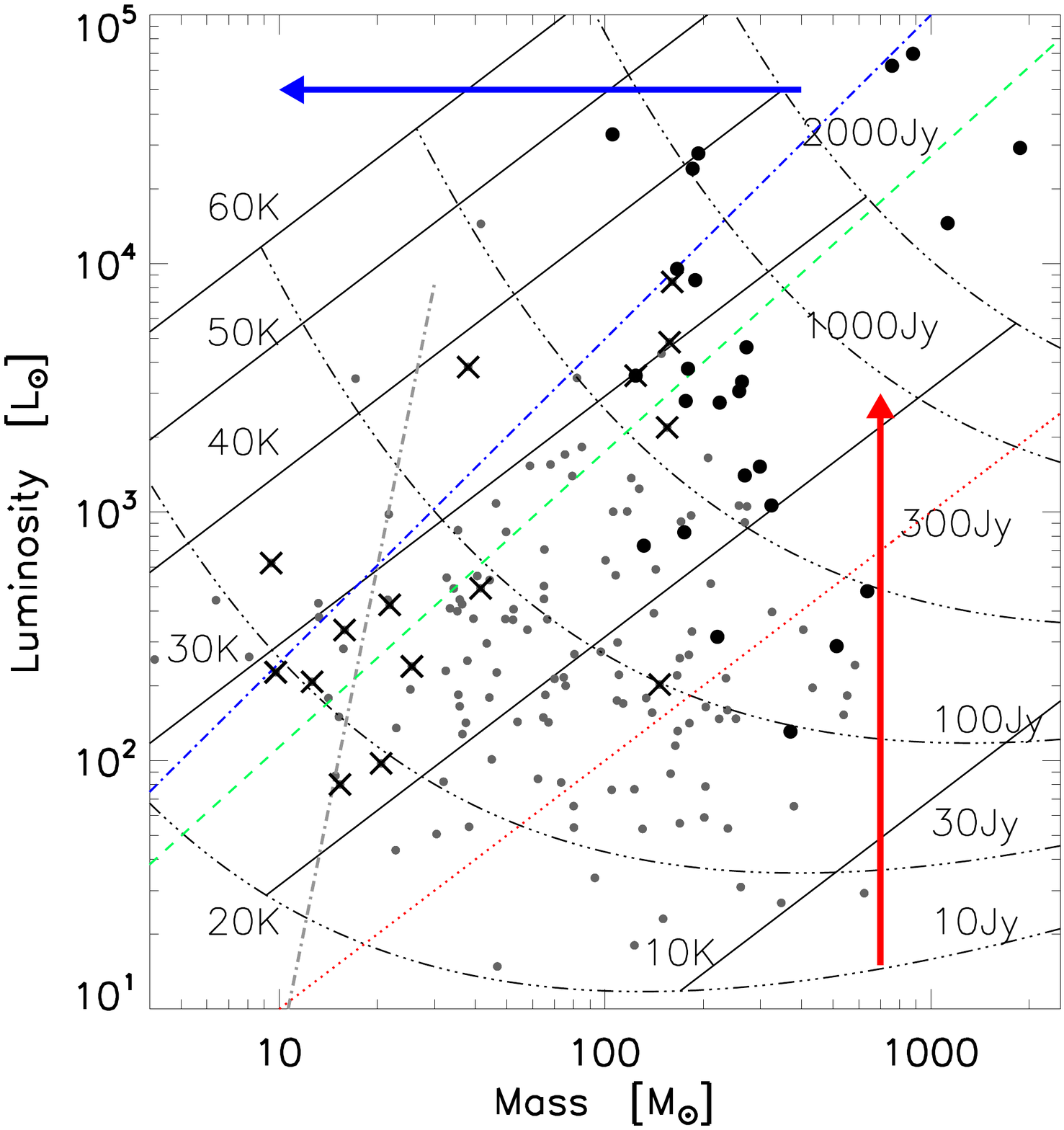} 
\caption{ Distribution of BLAST sources in the Cyg~OB2 complex in the
  \lm\ plane.
Those to the right of the steep dash-dot (gray) curve are above the
Bonner-Ebert mass.
Solid diagonal lines are loci of constant $T$ or $L/M$. The dotted line
(red) is $L=M$. Dot-dash curves ``orthogonal'' to these are for constant
250~\micron\ flux density.  
Dash (green) and dot-dash (blue) lines denote the location of 
sources powered by accretion and nuclear burning, respectively, as derived
empirically in Fig.~9 of \cite{mol2008}.
Cool sources discovered by BLAST with low $L/M$ still are externally
heated (stage E).  These appear to be gravitationally bound, but have no
significant internal power from star formation yet.
The vertical arrow indicates the direction of evolution in this diagram
as protostar formation takes hold within a clump.
The horizontal arrow indicates the direction of evolution as the
embedding material is dispersed by the formed stars/cluster, assuming
the surrounding dust still reprocesses most of the internally-generated
$L$.  Otherwise (as appears to be the case), the re-radiated $L$,
measured here, is less. 
Black filled circles are BLAST sources corresponding to clumps of
\cite{motte2007} (Table~\ref{tab:clump_cores}).
Crosses are sources with morphological evidence of mass-stripping from 
radiative interaction with Cyg~OB2; some at lower luminosity exhibit 
the effects of external ionization.
 }
\label{fig:lm}
\end{figure}


\begin{figure}
\centering
\includegraphics[width=\linewidth]{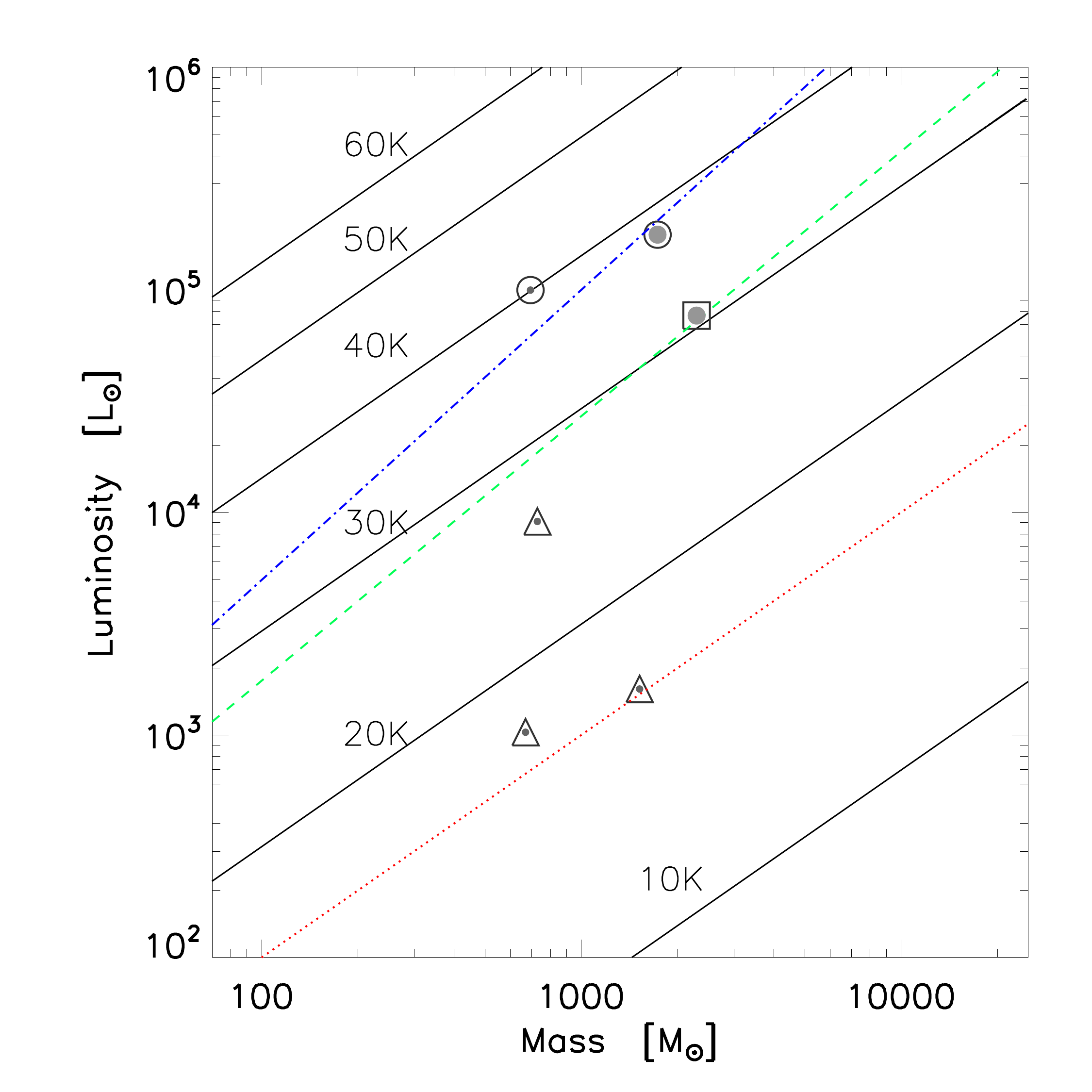}
\caption{Like Fig.~\ref{fig:lm}, but for sources more distant than
  Cyg~OB2 (\S~\ref{largedistance}). Triangles, circles, and square
  represent sources about 3.4, 6.8, and 8.5~kpc away, respectively.
  These include the most luminous and most massive sources in the
  survey.}
\label{fig:lmdistant}
\end{figure}


The $L - M$ diagram can be exploited to assess evolutionary stages
\citep{mol2008}.  Figure~\ref{fig:lm} shows our results in the $L - M$
plane for those sources at the distance of Cyg~OB2; the results for the
few distant sources (\S~\ref{largedistance}) are in
Figure~\ref{fig:lmdistant} (note the scales to larger masses and
luminosities). 

Lines of constant $T$ are diagonal lines (constant $L/M$) in this \lm\
diagram: a mass $M$ radiates a predictable $L$, depending on
$\kappa_0 r$ and $\beta$.  To be consistent with the analytical loci, we
plot $L$ from the SED fit rather than the only slightly larger $L_{\rm
  bol}$.  ``Orthogonal'' to these diagonal lines are loci of constant
250~\micron\ flux density.  Note that there are relatively fewer sources
between the 30 and 10~Jy loci because of the growing effect of cirrus
noise.

In this diagram the ``error ellipse'' determined by the Monte Carlo
technique is elongated along the locus for the source flux density and
the extent is well described by $\Delta T/T$.  We examined the histogram
of $\Delta T/T$ and rejected outliers $>0.19$; these were \baderrorelipse\
sources with poorly constrained SEDs, usually lacking in definitive data
near 100~\micron\ combined with their poor image quality at 500~\micron.  
The dashed and dot-dash thick lines, roughly lines of constant $T$, are
the loci obtained empirically by \cite{mol2008} for sources thought to
be in the accretion stage and the later nuclear burning stage (when
envelope dispersal begins), respectively.  For single low mass stars,
these would correspond to the spectroscopically defined ``class 0'' and
``class 1,'' respectively.  With BLAST, we detect sources of the size of
``clumps'' and furthermore even smaller angular-size high-mass ``cores''
are capable of forming multiple stars.  This warns against a simple
interpretation of this diagram based on single-star evolutionary tracks,
although it is possible that once high mass stars form, the most massive
will dominate the luminosity and ionization (this will depend on the
IMF, star formation efficiency, and also small-number statistics).

We think that the most illuminating way to think of pre-stellar
evolution in this diagram is in terms of the energy source for the
clump, which determines the appropriate equilibrium temperature $T$ for
the approximate SED.
In the very earliest stages being sought in submillimeter surveys, the
energy source for the clumps is predominantly external, namely the
impinging interstellar radiation field. In \Cyg\ the radiation field is
in principle higher than in the local interstellar medium because of all
the massive stars that have already formed, but this radiation is
attenuated by the dusty molecular material in which the pre-stellar
clumps are embedded.  A useful point of reference is the equilibrium
temperature corresponding to $L = M$, which for the adopted $\kappa_0 r$
and $\beta$ (\S~\ref{sec:sed}) is 16.2~K.

Massive sources located below this $L=M$ locus (see \S~\ref{sec:early})
still can have substantial $L$ by virtue of a lot of mass, but cannot
have any predominant internal source of energy (either accretion or
nuclear), and so could be called ``starless.''  This term is probably
best avoided for these clumps, within which there might actually be a
few low mass stars already forming and detectable by sensitive
telescopes like \textit{Spitzer} (\S~\ref{sec:lowlm}).  
The key consideration is what is the
dominant source of energy determining $T$.  As mentioned in the
introduction, these clumps could be said to be in stage E (``E'' for
``external'' or ``earliest''), which seems to us a better terminology
than ``starless,'' or ``class~$-1$'' for single-star pre-stellar cores.
Figure~\ref{fig:lm} shows that there are many clumps in this stage.  We
can calculate the Bonner-Ebert mass, above which (within the
assumptions) a clump is gravitationally unstable (e.g.,
\citealp{stahler2004}):
\begin{equation}
M_{ \rm BE} = 1\ [T_{\rm g}/(10\ {\rm K})]^{3/2} [n_{{\rm H}_2} /(10^4\ {\rm
    cm^{-3}})]^{-1/2}\ {\rm M}_{\odot}.
\label{Mbe}
\end{equation}
Thus, assuming that $T_{\rm g} \leq T$, the low-luminosity,
low-temperature BLAST clumps are unstable.  As discussed below, in the
simplest theory they would evolve into the higher-luminosity,
higher-temperature clumps seen higher in the diagram at the same $M$.

As gravitational collapse progresses, more and more mass accretes into
protostellar cores.  The actual luminosity in this accretion-powered
stage depends on the accretion rate and the potential well.  In
recognition of the underlying energetics, these clumps could be said to
be in stage A (``A'' for accretion-powered).  Consistent with this
interpretation, sources observed to be in this part of the \lm\ diagram
have a characteristic signature of active accretion
(\S~\ref{sec:accretion}).

Ultimately, nuclear fusion becomes the dominant source of power. Unlike
for low mass stars, massive pre-main sequence stars probably continue
accreting after first beginning nuclear burning, increasing their mass
further \citep{Zinn2007}.  When this accretion ends, an individual star
is at its final position on the ZAMS.  It will still be embedded, and so
optically obscured, but its vast power reprocessed will produce a
relatively warm far-infrared source.  For sufficiently massive stars,
significant ionization of the surroundings will ensue, producing a
hypercompact \HII\ region.  There could be many within a single clump,
for example \citep{rivera2010}.  These clumps would appear in the $L -
M$ diagram near the empirical nuclear burning locus (see
\S~\ref{sec:HII}).
 
As the massive stars clear their local environment through the expansion
of the \HII\ region and via radiation pressure on dust (successively
ultracompact then compact \HII\ regions), the individual objects will
become more extended far-infrared sources.  The strong short-wavelength
stellar radiation field will enhance both thermal emission and
non-equilibrium emission from PAHs in nearby photodissociation region
material, making the SED of the integrated re-emission of the clump
broader and more complex.  Evolution in the \lm\ diagram will
initially be at constant $L$ with decreasing $M$; ultimately, the
optical depth and/or covering factor would decrease so that the
reprocessed $L$ will fall.

A stellar cluster with an extended \HII\ region would emerge. Even
though such clusters are found in \Cyg, they are not detected as
compact BLAST sources and so do not appear in the \lm\ diagram; only
objects like their precursors would.

\subsection{Census}\label{sec:cen}

\begin{figure}
\centering
\includegraphics[width=\linewidth]{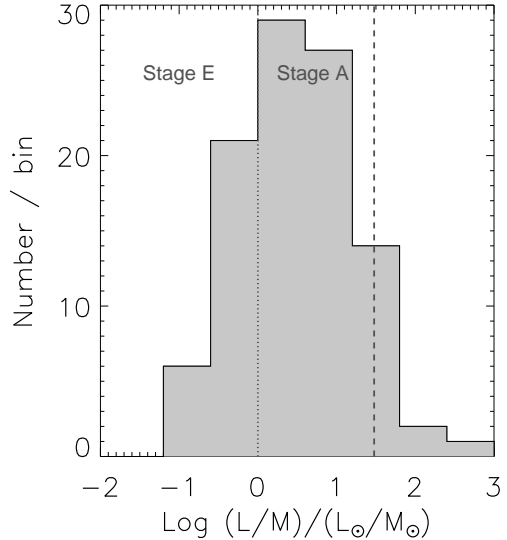}
\caption{ Histogram of $L/M$ ratio of the \Cyg\ compact sources in the
  mass range 50 -- 500~\msol. The dotted vertical line separates stage E
  sources below the locus $L = M$ in Fig.~\ref{fig:lm}.  Most sources
  are in the accretion-powered stage A, up to the dashed line 
  ($L/M \approx 30$~\lom), beyond which nuclear burning becomes predominant.}
\label{fig:loverm}
\end{figure}

It is interesting to ask how many objects are in each evolutionary
stage.  To address this we have concentrated on the mass range between
50 and 500~\msol\ where there is an abundance of sources (\stats\
omitting the sources more distant than Cyg~OB2) and the lower luminosity
end is minimally affected by completeness, at least down to $L = M$.
Figure~\ref{fig:loverm} presents a histogram showing the relative
populations in terms of $L/M$, which we have argued is at least
qualitatively related to the successive stages.  Note that this figure
looks quite similar to Figure~\ref{fig:temp}, there being a non-linear
mapping of $L/M$ into $T$.

Figure~\ref{fig:loverm} shows that there are many sources in stage E
($L/M < 1$~\lom), even though it is clear that there are selection effects
(basically the combination of low flux density and cirrus noise)
beginning to limit their detectability. Most of the sources are in the
accretion-dominated stage A ($L/M$ up to about 30 or $T = 30$~K).
The relatively fewer hotter sources, where nuclear burning is taking
over, presumably reflects the shorter lifetime of this stage, which
could be characterized by more rapid final collapse and then envelope
dispersal.


\begin{figure}
\centering
\includegraphics[width=\linewidth]{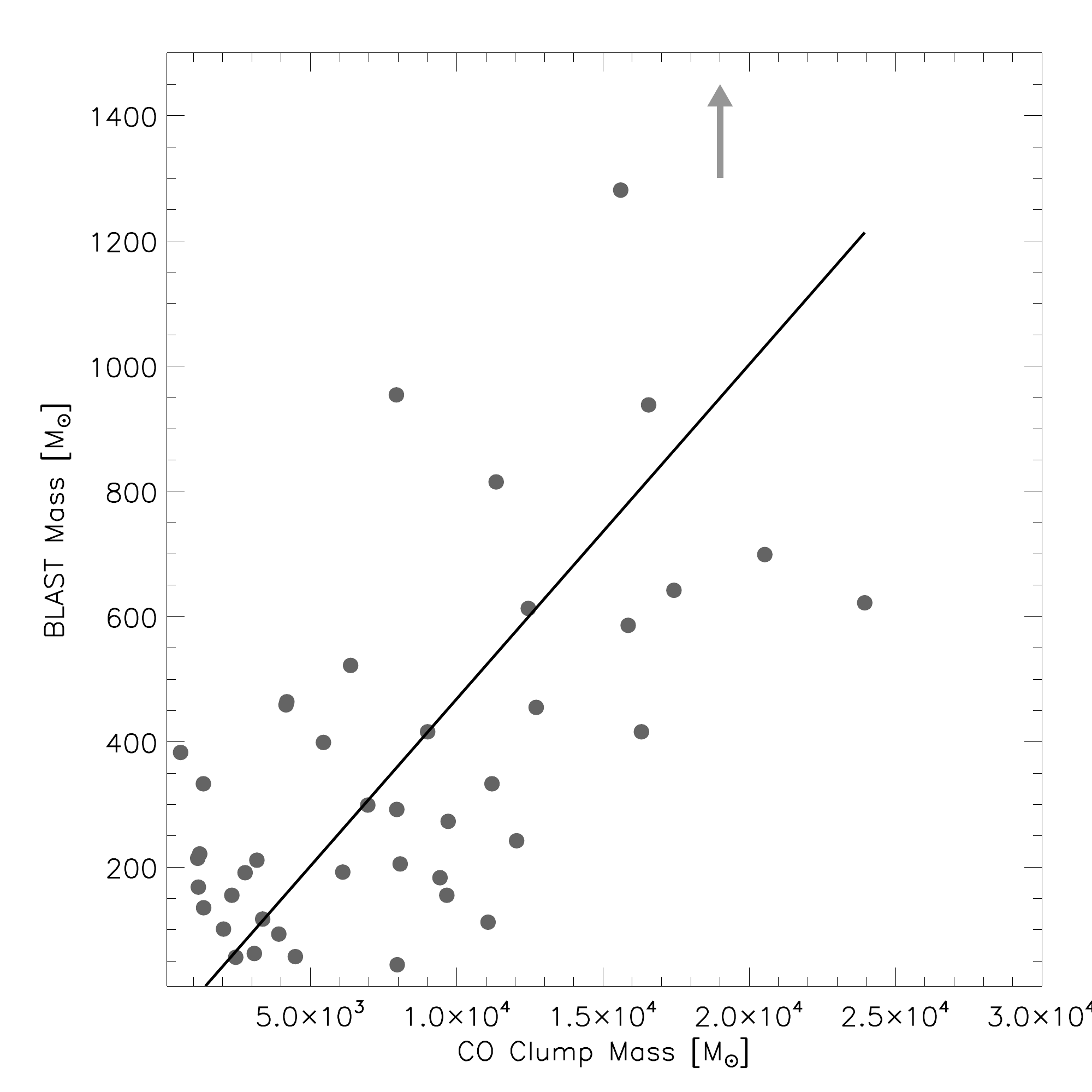}
\caption{Sum of the masses of the BLAST sources within CO clumps of
  \citet{sch2006} versus the CO-estimated clump mass.  Linear
  correlation has a slope of 0.05.  An outlier at 3000 \msol\ along the
  y direction, pointed to by the arrow, is not included in the fit.}
\label{fig:efficiency}
\end{figure}

\section{Evolutionary Stages}\label{evol}

Although our BLAST survey is ``blind and unbiased'', in surveying the
entire region rather than selecting sub-regions with, say, high
extinction \citep{sch2006,motte2007}, it is not unbiased in another
sense.  This region is obviously well known for its GMC and having
formed the Cyg OB2 association, which has dramatically influenced the
surrounding molecular material (\S~\ref{sec:ob2}).  In such a region one
expects there to be both triggered and/or sequential star formation,
supplementing spontaneous star formation.  The conditions could be quite
different now than what preceded the formation of Cyg OB2.  We find no
evidence for the precursor of another such massive compact association.
Nevertheless, within this reservoir several smaller, but still notable,
embedded clusters have been identified (\S~\ref{clusters}).  The OB
stars in the more evolved ones have produced extended \HII\ regions like
DR17 and DR22 \citep{mic1966}.

As discussed in \S~\ref{molecules}, in their
$^{13}$CO~J~=~2~$\rightarrow$ 1 data cubes, \citet{sch2006} have
identified many clumps (detailed in their Appendix~C).  Where the
coverage overlaps, each of the BLAST compact sources can be linked
through the morphology to one of these clumps (see examples in
\S~\ref{posters}).  In Figure~\ref{fig:efficiency} we compare the sum of
the masses of the BLAST sources (\S~\ref{sec:mass}) within each clump to
the tabulated CO-estimated clump mass.  These are fairly well
correlated, and we find typically that only a few percent of the
material over these extended regions is in the form of compact sources,
in some way related to the (potential) star formation efficiency.  This
plus the timescales for clump evolution indicate that there is still an
interesting future for star formation in this GMC.


From this perspective on the complex range of star formation going on in
\Cyg, our goal is to work toward submillimeter evidence for the earliest
stages of massive star formation.  The somewhat later stages have been
detected by a variety of earlier observations, and BLAST sees these too.
The many clusters in \Cyg\ are agents producing complex morphologies in
the medium, through expanding \HII\ regions and strong UV radiation
fields, and they might play a pivotal role in inducing further star
formation in the molecular clouds. We work in reverse chronological
order, finding what the outcome of a particular stage looks like in
multi-wavelength data, and then asking what this tells us to look for in
the stages that came before.

\subsection{The Influence of the Massive Cluster Cyg OB2}\label{sec:ob2}

Comparing the estimated mass of Cyg~OB2 (\S~\ref{clusters}), $4~
\rm{to}~ 10 \times 10^4$~\msol\ \citep{knodlseder2000} to the remaining
molecular reserve in \Cyg, $4 \times 10^6$~\msol\ (\S~\ref{molecules};
\citealp{sch2006}), indicates that formation of this OB2 association was
the major event for this GMC, perhaps not to be repeated.
Both of the central clusters, BBD1 and BBD2, are more massive than the
Trapezium.
Again, will there be more elsewhere in \Cyg?  There is no BLAST clump
that could be a precursor to such a prominent cluster and even adding
the mass of an embedding CO clump (the most massive of which is $2
\times 10^4$~\msol) would fall short.  Nevertheless, in unevolved
GMCs, massive and compact precursors to such clusters might be
detectable in the Planck Cold Core survey (\S~\ref{pcc};
\citealp{juv2010}).  It will of course depend on how long-lived the
precursor stage is.

Massive young stars, through ionization and radiation pressure,
dramatically influence their environment.  Cyg~OB2 has an age of about
$2.5 \times 10^6$~yr \citep{neg2008}, and in that time has created in
the ISM a lower density region of radius 1\degree\ (30 pc), ionized in
the interior (see Fig.~\ref{fig:omap}) and surrounded by distinctive
molecular cloud complexes.  The connectivity of these clouds in the CO
data cube and the signs of interaction through UV radiation (\MSX\ Band
A data) led \citet{sch2006} to conclude that they were all located at
the same distance.  While obviously not yet disrupted, the clouds have
been reorganized and so Cyg~OB2 might have induced some of the further
star formation that is on-going.

Above Cyg~OB2 in Figure~\ref{fig:omap} is a relative void (nevertheless,
it is projected on an emission and extinction plateau), with no
high-contrast structures of significant column density in any tracer
(molecules, dust, ionized gas).  The right hand edge is defined by CXR9
whereas the left, though well-defined and containing some BLAST emission
too, is unnamed.  This void could be a ``chimney'' blown out by Cyg~OB2,
channeling high pressure gas out of the Galactic disk.

\subsubsection{Pillars}\label{pillars}

\begin{figure}
\centering
\includegraphics[width=\linewidth]{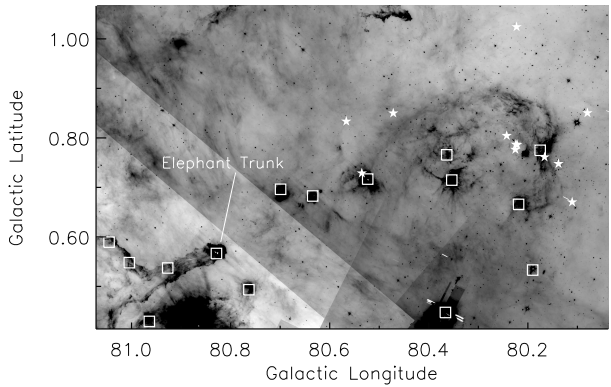}
\caption{\IRAC\ band 4 image showing a prominent pillar or ``elephant
  trunk'' structure pointing to Cyg~OB2. Squares are BLAST sources and
  stars show the position of the OB stars.}
\label{fig:trunk}
\end{figure}

Radiative forces sculpt pillars, where there are pre-existing dense
molecular structures.  The range of radiative influence of Cyg~OB2
extends to a radius of at least 2\degree\ (60~pc), as illustrated by the
prominent pillars near G79.4-1.1 and G81.4+2.2 in the \textit{Spitzer}
\IRAC\ and 24~\micron\ \MIPS\ images.
%
%
Figure~\ref{fig:trunk} shows an example of a long pillar or elephant
trunk pointing toward the Cyg~OB2 core.  At the end of the trunk is a
luminous BLAST clump C85.  In CO emission, it is called DR20NW by
\citet{sch2006}, with CO velocity $+12$~\kms\ (see also
Fig.~\ref{fig:dr20}).  The clump is the obstruction defining the trunk
structure.  Star formation is occurring in this clump, suggestive of
triggering by Cyg~OB2.  This geometry is seen elsewhere; for example,
C75 on the left edge of the chimney is also at the head of a pillar
facing Cyg~OB2 (see also the cometary tails below).

Such a peninsular structure is continually eroded, which would produce
eventually an isolated clump with a cometary tail, but this particular
trunk appears to be attached to the molecular cloud, curving round and
broadening out to include C91, C98, and C100 along the interface.  These
structures also evolve due to instabilities at the interface of the
\HII\ region and the neutral molecular cloud, and evidently on-going
star formation is commonly induced.


Note that DR20 (C88, C87, C90) is at a projected distance of
6\arcmin\ from this trunk, and yet does not display the same hallmark
interactions.  However, this complex is at a very different velocity
$-3$~\kms, more closely associated with DR21.  \citet{sch2006} note that
clump DR20W, at the same velocity, does have an elongated shape pointing
to Cyg~OB2 (see Fig.~\ref{fig:dr20}).  BLAST C80 is at the brighter head
of its CO emission and C83 in the dimmer tail.  Inspection of the
\textit{Spitzer} images shows signs of interaction with nearby stars in
Cyg~OB2, but at a different position angle than the CO clump which is
oriented more toward the core.


\citet{sch2006} also identify other molecular pillars in the CO channel
maps, the most prominent of which is the long ``DR17 pillar'' at a
velocity about $-13$~\kms\ (Fig.~\ref{fig:dr17}). From its direction of
elongation, it appears to be influenced by the cluster LK12 in DR17, but
there is not perfect alignment (possibly there is some influence of LK14
to the north).  This cluster does produce a bright mid-infrared
interface at BLAST C115, near the end of the pillar.  The molecular
pillar stretches from C115 through C114, C112, C113, and C108 to C109
and possibly C106 (Fig.~\ref{fig:dr17}).
The CO channel maps hint at a connection between this DR17 pillar and
the above-mentioned elephant trunk, but this pillar does not appear to
be influenced by Cyg~OB2.

\subsubsection{Cometary Tails}\label{tails}

Further striking evidence for interaction is provided by cometary tails
in more detached structures, pointing away from the core of Cyg~OB2.
These too are seen most clearly in \textit{Spitzer} \IRAC\ and
24~\micron\ \MIPS\ images.


DR18 is a large scale prototype, with a ionized edge facing the Cyg~OB2
core, the BLAST clump C73 in the neutral region next to this, and an
extended tail seen in both CO and dust emission (mid-infrared to submm).
It is actually more complicated than this in detail, because a loose
aggregate of early B stars has formed (triggered?) near the leading edge
adding additional ionizing power and sculpting of the immediate PDR
environment \citep{come1999}.


Another interesting example in the opposite direction is the cluster
ECX6-21 (DB12) containing BLAST C55 in the head.  C55 corresponds to OB2
globule 2 of \citet{sch2006} with CO velocity $-4.5$~\kms.  About
5\arcmin\ ``downstream'' is a miniature version containing C53.  These
are located on the inner edge of the ionized ridge CXR9.


Further afield from Cyg~OB2 is DR15, with a cluster and two BLAST
clumps, C32 and C28.  Here the putative tail (see Fig.~\ref{fig:dr15})
appears to be part of an extended structure prominent in presumed PAH
emission, which bends significantly toward lower latitude at a distance
of 17\arcmin, near C12.  There are several other BLAST clumps along this
structure, C21 before the bend, and C11, C9, C4, and C3 beyond.

\begin{figure}
\centering
\includegraphics[width=\linewidth]{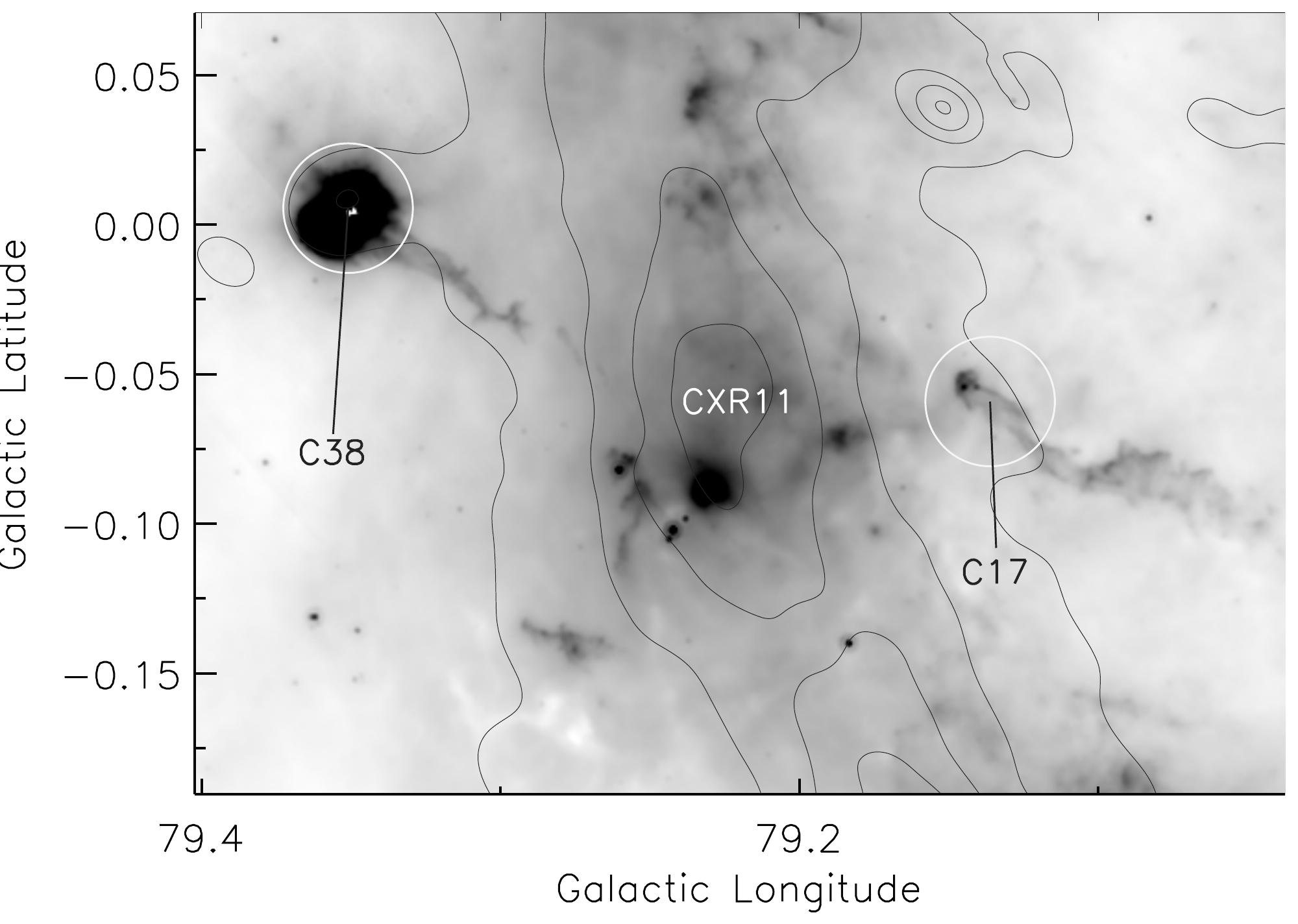}
\caption{24~\micron~\MIPS\ image showing two BLAST sources, C38 and C17,
  with cometary tails. The spatial extents of the tails are roughly 1.5
  and 3~pc, respectively. Contours of 21-cm radio emission from CGPS
  show ridge CXR11.}
\label{fig:cometary}
\end{figure}


Figure~\ref{fig:cometary} shows a field quite close by, spanning the
ionized ridge CXR11.  There are two BLAST clumps C38 (on the left) and
C17 with material ablated from the clump clearly being blown
away. However, in the BLAST image these cometary tails are not detected
because of low column density.  C38 is massive enough to have a
detectable ionized edge.  The C17 tail extends to an angular distance of
6\arcmin\ or about 3~pc in spatial scale.  If, for illustration, the
clump lifetime were $10^5$~yr the required material speed would be
$30~\rm{km s^{-1}}$ to reach such an extent.
  
There are many other sources like this with obvious interactions.  Their
positions in the \lm\ diagram are highlighted in Figure~\ref{fig:lm}.
For these exposed sources, it is clear that a sizable amount of mass is
lost due to this erosion, and so the more evolved sources occupy a
region towards the lower end of the mass axis in the \lm\ plot.


The ridge of submillimeter emission containing sources C59, C61, and C60
provides an interesting counter-example.  The superimposed CO ridge is
called ``OB2 globule 1" by \citet{sch2006} who suggest an interaction
due to its compactness and proximity to the core of Cyg~OB2, in
projection about 15\arcmin\ away.  However, inspection of the detailed
\textit{Spitzer} images now available reveals no signs of interaction.
Furthermore, the displaced ionized ridge parallel to the dust emission
is on the side away from Cyg~OB2 (see Fig.~\ref{fig:dr7}). Together with
the peculiar velocity, the evidence is that this object is more distant,
beyond the influence of Cyg~OB2.

\subsection{Clusters and Extended \HII\ Regions}\label{extended}

Interactions of clusters with the ISM produces large scale complex
structures and directly affect the next generation star formation
efficiency in the region.  The \Cyg\ region hosts numerous young open
clusters (\S~\ref{clusters}). In the BLAST region, there are five which
have well-resolved \HII\ regions with a fairly classical geometry.
Expansion of the ionized gas, stellar winds, and radiation pressure have
evacuated the core of the parent molecular cloud and produced a dense
shell which appears in an arc-shaped geometry surrounding the cavity and
cluster.  Partial shells are suggestive of a blow-out or blister
geometry.  The 21-cm radio emission traces the ionization front, and
slightly outside of this is the FIR and submillimeter continuum, where
dust in the shell absorbs the FUV radiation emitted by the star cluster.

\begin{figure}
\centering
\includegraphics[width=\linewidth]{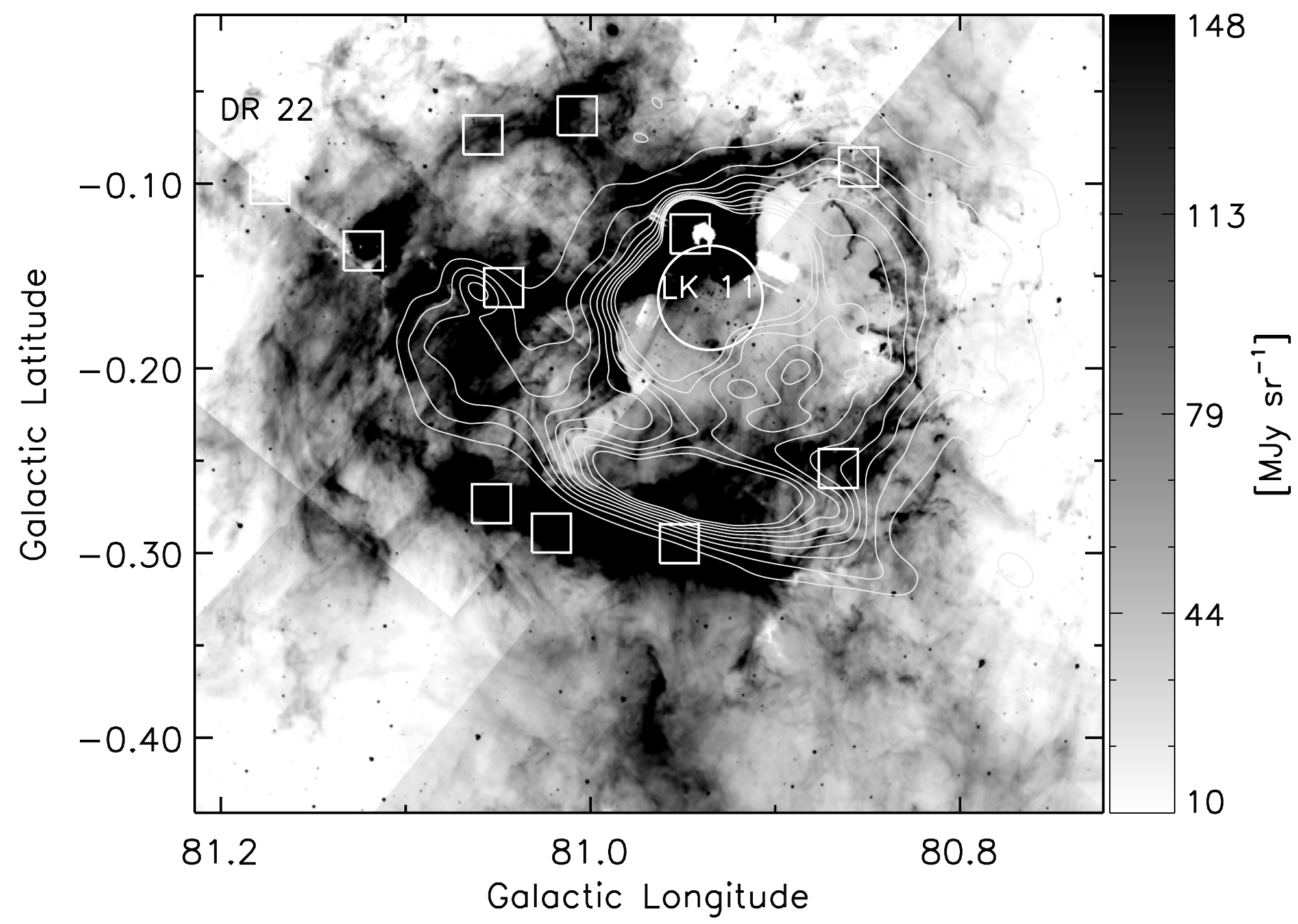}
\caption{ \IRAC\ band 4 (8.6~\micron) image of DR22. Circle marks the
  position and size ($R_{50}$) of the LK11 OB star cluster
  \citep{leduigou2002}. Overlaid contours are 21-cm radio continuum
  emission from the CGPS.  Squares indicate the BLAST sources, embedded
  in arcs of submillimeter emission outside the ionization front
  (Fig.~\ref{fig:dr22}).}
\label{fig:obstar}
\end{figure}

To illustrate this, Figure~\ref{fig:obstar} shows an \IRAC\ band 4 image
highlighting the most massive of these clusters, LK11, with about 80 OB
stars powering the DR22 region. Note that the PAH emission in this image
is from the PDR, outside the ionization front, whereas the
24~\micron\ emission in Figure~\ref{fig:dr22} shows greater correlation
with the ionized gas.  The BLAST sources occur along CO ridges to the
north and south-east, the latter clearly arc-shaped (making this the
``smiley nebula'').  The most luminous source C92, coincident with a
compact \HII\ region, must be internally powered (see below).  LK11 is
shaping this clump and both the geometry and proximity suggest that star
formation has been triggered there.

Similar geometrical statements can be made about 
LK09 and C30 (and C27, C33) in DR7 (Fig.~\ref{fig:dr7}),
LK12 and C115 in DR17 (Fig.~\ref{fig:dr17}), 
LK13 and C127 in G81.445+0.485 (the Diamond; Fig.~\ref{fig:dr21}),
and
DB22 and C182 (and C177, C178) in ECX6-33.

Although these \HII\ regions have evolved, the exciting clusters are
still quite young; their stellar density is typically five times higher
than for evolved open clusters \citep{leduigou2002}.  Thus their radii
$R_{50}$ (half population radius), about 1\arcmin\ (0.5~pc), are perhaps
indicative of the size of the precursor molecular clump.  The stellar
masses are estimated by \citet{leduigou2002} to be $300$ to
$2000$~\msol\ (a few times more for DR22) and so the precursor clump
would be more massive still.  BLAST clumps are of this angular size, but
not nearly as massive, not even up to these stellar masses let alone
allowing for inefficiency of star formation.  We conclude that there are
presently no BLAST clumps capable of forming such massive clusters.  The
embedding CO clumps are perhaps massive enough, but would have to
condense considerably to form a concentrated cluster.  The lifetime of
these condensations could be quite short, lowering the likelihood of
detecting this stage in a single GMC.

\subsection{Triggered Star Formation} \label{sec:trigger}

Spatially, a massive star cluster can promote further star formation in
nearby molecular material by driving winds and shocks, which both sweep
up material and overrun and compress pre-existing condensations.
Subsequent gravitational instability and/or radiatively-driven implosion
of the overrun condensations collectively lead to what is known as
``triggered'' star formation
\citep{Elmegreen1998,Zinn2007,koenig2008}. Thus a sequence of
generations of star formation can occur with spatial ordering
(recognized as ``sequential'' star formation).  For example,
\citet{koenig2008} studied these phenomena in W5 with \textit{Spitzer}.

As discussed, in the vicinity of the OB star clusters in \Cyg\ there is
evidence of further star formation.  The most persuasive, if
circumstantial, evidence for triggering is in the DR22, DR8, DR17,
and DR15 molecular clouds.

\subsection{Compact \HII\ Regions}\label{sec:HII} 
  
Earlier in their development, when massive protostars first become hot
enough to emit ionizing UV photons, they initially ionize only a dense
core, producing a hypercompact \HII\ region, potentially optically
thick at lower radio frequencies.  Subsequent expansion of the ionized
gas leads successively to ultracompact, compact, and then evolved
\HII\ regions \citep{churchwell}.  The initial stage, which occurs when
accretion is still strong, is accompanied by energetic bipolar outflows.

Circumstellar dust reprocesses the absorbed radiation from the
protostar, re-radiating the energy at infrared wavelengths.  Thus in the
Galactic Plane, embedded \HII\ regions are the most luminous objects
observed in the mid and far-infrared, first by \IRAS\ and then \MSX.

Hot cores, dense molecular clouds with temperature $\geq$ 100~K as
classified by \citet{kurtz2000}, are often thought to be precursors of
UC~\HII\ regions.  In our survey, we do not detect any source warmer
than $\sim 40$~K.
Nevertheless, some of the brightest sources detected by BLAST in
\Cyg\ are found among well-known \HII\ regions, with previous studies
revealing them to be compact or ultracompact; they are all detectable
in 21-cm radio continuum emission in the CGPS.  Five of these studied by
\citet{motte2007} are tablulated in Table~\ref{tab:HII} and discussed
in the next subsection.
\citet{motte2007} obtained masses of these sources assuming a dust
temperature of 40~K.  We are actually able to measure the temperature,
finding \hiitemprange. Still, it should be remembered that BLAST
measures dust emission including the outer envelope of these sources,
which is cooler than dust near the core; with higher angular resolution
\MAMBO\ should sample slightly warmer dust on average.

\begin{figure}
\centering
\includegraphics[width=\linewidth]{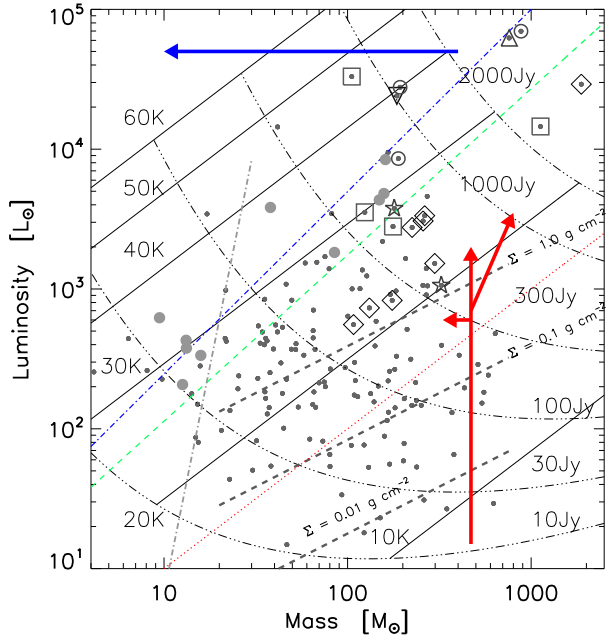} 
\caption{Same as Fig.~\ref{fig:lm}, but highlighting physical
properties deduced from ancillary data.
Many of the most luminous sources are ionizing and have associated
radio continuum emission.  Close to the nuclear burning locus, these
contain deeply embedded massive stars \citep{motte2007}:
$\triangle$~HC\HII; $\bigcirc$~UC\HII; $\triangledown$~C\HII.
Grey filled circles are other BLAST sources with compact 21-cm
continuum emission, outside the survey area of \MAMBO.
Protostars in stage A are at somewhat smaller 
$L/M$ than the compact \HII\ regions, near the accretion locus.
These are classified by \cite{motte2007} as $\Box$~HLIRPC;
$\diamond$~MIRQP; and $\bigstar$~IRQP.
At the lowest $L/M$, clumps are externally heated (stage E), while at
intermediate values the clumps are probably powered by low mass
protostars; the thick dashed lines are the model predictions of
\cite{krumholz2008} (eq.\ \ref{eq:lsig} in \S~\ref{sec:surfsigma}) for
surface densities $\Sigma~=$ 1.0, 0.1, and 0.01~g~cm$^{-2}$.
The vertical arrow signifies the result of evolution in $L$ as star
formation progresses in a clump.  This might be interpreted as the
result of the evolution of a single massive protostar \citep{mol2008}.
However, the initial rise seems more likely to be the result of power
from low-mass YSOs (\S~\ref{sec:lowlm}).  According to the model of
\cite{krumholz2008}, unless $\Sigma$ exceeds a critical value near
1.0~g~cm$^{-2}$, a massive star will not form and so the rise in $L$
will be arrested (\S~\ref{sec:vertical}); this possibility is indicated
schematically by the short horizontal arrow. Finally, the arrow segment
pointing to the upper right represents qualitatively the evolution in
the simulations by \citet{smith2009} and \citet{wang2010}, where mass is
fed into gravitational potential minima from larger scale structure.}
\label{fig:lm2}
\end{figure}

These five sources are marked in Figure~\ref{fig:lm2} with special
symbols.  They are clearly internally powered, beyond the transition
to predominant nuclear burning.  For reference, a 20~\msol\ zero-age
main sequence star of spectral type O9V has $L_{\rm bol} =
10^5$~\lsol\ \citep{schaerer1997}.  Even the most luminous of these
embedded stars at the distance of Cyg~OB2 is not that luminous.  Given
the mass of these clumps, $> 100$~\msol, they also seem likely to host
more than a single star.

The precursors to these sources should be cooler, at similar mass. In
the range of a few 100~\msol, there are many such BLAST sources in
Figure~\ref{fig:lm2}, the earlier stages to be discussed below.  However,
there do not seem to be the cooler equivalents of most massive sources
(W75N, DR21, DR21 OH), suggesting that the precursor stage to the most
massive (dense) clumps is relatively short-lived or that when cooler the
clump is too extended to have been classified as a BLAST source (\S~\ref{sec:vertical}).

Also recorded in Table~\ref{tab:HII} are another \hiibs\ instances
(including two distant ones) of BLAST sources with apparently associated
compact 21-cm radio emission, not targeted by \citet{motte2007}.  These
are slightly cooler, with median temperature 27~K.  These are marked in
Figures~\ref{fig:lmdistant} and \ref{fig:lm2} with grey filled circles.

\begin{figure}
\centering
   \includegraphics[width=\linewidth]{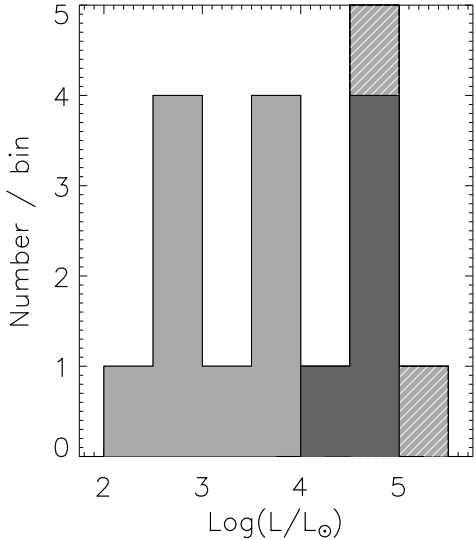} 
\caption{Histogram of $L_{\rm bol}$ for BLAST sources with compact
  \HII\ emission. Two of the most luminous (stripes) are at a larger
distance than Cyg~OB2.  The five noted by \citet{motte2007} are shaded black.
Other lower luminosity sources appear to have external ionization
(\S~\ref{sec:HII}).}
\label{fig:l_hii}
\end{figure}

A histogram of the bolometric luminosities of all of these sources,
including the luminosity corrections for the two distant ones, is given
in Figure~\ref{fig:l_hii}. The luminosities of the five marked sources
are at the high end, whereas most of the others have much lower
luminosities.  The luminosities can be used to calculate the
corresponding photoionizing flux $Q_0$ and check for consistent radio
emission (assuming no optical depth; see \S~\ref{ionization}).  The
lowest luminosity sources in Figure~\ref{fig:l_hii} would not account
for the associated \HII\ emission.  Inspection of the \IRAC\ image
reveals all but C76 to have cometary structure (those of higher mass and
luminosity were already mentioned above: C55, C73, C80; all but C14
appear to be influenced by Cyg~OB2) and our interpretation is that the
radio emission comes from external ionization.  C76 lies in close
projection to the O7~V star MT771.
%
\subsubsection{Specific Embedded \HII\ Regions in \Cyg}

Here we make brief comments on the five marked sources from
\citet{motte2007} in Table~\ref{tab:HII}, plus three others that appear
to be in the same advanced stage of evolution.


\textit{C1, AFGL 2591:} C1 coincides with AFGL~2591, an UC~\HII\ region
that is among the most luminous sources in \Cyg. \citet{motte2007}
identify a core S26 within a clump S5.  We find dust temperature $40$~K,
bolometric luminosity $2.7 \times 10^4$~\lsol, and mass $190$~\msol.
\citet{campbell1984} discovered that AFGL~2591 is comprised of a young
stellar group and deduced that the \HII\ region is being generated by B0
stars.  This region also has powerful outflows \citep{bally, poetzel}.
\citet{sch2006} argue that the cometary shape is due to the influence of
Cyg~OB2.


\textit{C88, DR20:} Compared to C1, C88 is slightly cooler (32~K), less
luminous ($8.6 \times 10^3$~\lsol), but equally massive ($190$~\msol).
It corresponds to clump N4 (core N10) of \citet{motte2007}.  The
embedding region (see Fig.~\ref{fig:dr20}) is bright in 21-cm radio
continuum, mid-infrared, and PAH emission.  C88 and C87 are two bright
sources along a ridge containing C90.  C88 is closest to the CGPS radio
peak, but offset to lower longitude, whereas the UC~\HII\ region
G80.86+0.4 observed by \cite{kurtz1994} is centered on C88.  C88 also
coincides with IRAS~20350+4126.  \citet{oden1990} suggest excitation by
an O5-6 ZAMS star, but this seems too early.

The source C87 (clump N6, core N14) is relatively cold, $27$ K, with no
free-free emission.  Its position on the \lm\ plot suggests most of the
luminosity is generated from accretion (see below).


\textit{C92, DR22:} C92, also known as IRAS~20375+4109, is a compact
\HII\ region in DR22 with a massive dense core, N58 \citep{motte2007}.
We find dust temperature $39$~K, bolometric luminosity $2.4\times
10^4$~\lsol, and mass $220$~\msol.  \citet{oden1986} studied this region
at both radio and far-infrared wavelengths and assuming it to be $45$ K
and $3$~kpc distant, they deduced that a single O6 ZAMS star was
accounting for a luminosity of $3\times10^5$~\lsol.  This would be
relaxed for the closer distance adopted here, and as mentioned above
there is likely more than a single star in a BLAST clump
(\S~\ref{sec:HII}).


\textit{C155, DR21:} C155 is the most luminous source in the BLAST
survey region with dust temperature $36$~K, luminosity $7.0 \times
10^4$~\lsol, and mass $880$~\msol.  DR21 has a dense core N46 with
signatures of outflows \citep{motte2007}, suggestive of an early stage
OB cluster.  Supportive of multiplicity, \citet{motte2007} detected
another two cores within a 1\arcmin\ radius in their clump N15; our mass
agrees with the total mass of the three cores.
This clump in the DR21 cloud complex (see Fig.~\ref{fig:dr21}) harbors
one of the most studied UC~\HII\ regions: microwave \citep{mic1966};
radio \citep{kurtz1994}; near-infrared \citep{davis2007}; and $^{13}$CO
\citet{sch2006}.  Studying hydrogen recombination and the ammonium line,
\citet{cyganowski} found cometary morphology associated with the
\HII\ region. It is also identified as MSX6C~G81.6802+0.5405, an
\MSX\ source, specifically bright at 8~\micron, and has saturated
\textit{Spitzer} sources.  There is an embedded near-infrared cluster,
W75S \citep{bica2003}.  The position of C155 in the \lm\ diagram
confirms that star formation is well established.


\textit{C169, W75N:} The W75N cloud complex shown in
Figure~\ref{fig:dr21} is a well-known massive star forming region.
After first being detected at low resolution in the radio by
\cite{westerhout,mic1966}, this complex was studied to decipher numerous
``protostellar physical conditions.'' A wide variety of sources was
found embedded in this cloud, e.g., a group of infrared sources
\citep{moor1991,persi,davis2007}, UC~\HII\ regions, and $\rm{H_{2}O}$
and OH masers \citep{hunter1994}.  \citet{haschick} identified three
ionized regions within W75N, namely W75N~(A); W75N~(B); and
W75N~(C). Subsequent higher resolution \cite{hunter1994} resolved
W75N~(B) into another three subregions.
\citet{motte2007} found three sources associated with the W75N cloud
(clump N13), of which W75N~(B) is the most massive object, with a dense
core N30. They classified this source as a HC~\HII\ region.  BLAST with
1\arcmin\ resolution observes only one source.  The derived properties
are dust temperature $36$~K, luminosity $6.2 \times 10^4$~\lsol, and
mass $760$~\msol, very close to C155.  The radio emission at 21-cm is
however much less, probably because of self absorption (there is
appreciable self-absorption in the spectrum of DR21 too,
\citealp{wendker1991}) but possibly because of a different IMF in the
embedded clusters.

There is another embedded cluster DB20 a few arc minutes to the north,
coincident with clump N14 of \citet{motte2007} and its core N28.


\textit{C32, DR15 cloud:} The DR15 complex was previously observed at
radio wavelengths by \citet{david}, and \citet{oden1990}, who predicted
that it hosts OB stars. Both free-free emission and PAH emission
observed by \MSX\ suggest on-going star formation.  The morphology of
this cloud is complex, with extended emission on a range of scales.  We
could resolve two sources, C32 and C28. The cluster LK08 and radio peak
are located between these (Fig.~\ref{fig:dr15}).  \citet{motte2007}
detected three embedded sources, of which two, MSX~79.2963+0.2835 and
MSX~79.3070+0.2768 are within C32. The \textit{Spitzer} sources are
saturated.  We find dust temperature 46~K, luminosity $3.3\times
10^4$~\lsol, and mass $110$~\msol. \cite{motte2007} designated core S41
in clump S11 (corresponding to C32) as a HLIRPC.  Our analysis based on
the location of C32 in the \lm\ plot suggests a slightly later stage
like the other embedded \HII\ regions being discussed.  However, it is a
complex region that will benefit from better resolution as provided by
\textit{Herschel}.  It is possible that the temperature of the dust is
influenced by the nearby cluster LK08, and fitting a single temperature
SED for this source would probably underestimate the clump mass.


\textit{C73, DR18:} DR18 was studied extensively in the near infrared by
\cite{come1999} who found a stellar group.  Most of the illumination is
due to a single B0.5~V star which can be observed in the visible as well
as the infrared.  The CGPS radio peak is at the head of the larger
cometary structure (see above) and probably is affected by external
ionization. Consistent with the cometary structure, C73 lies between the
ionized gas and the CO clump.  C73 has properties very similar to C88 in
DR20.  This interesting region is outside the coverage of
\citet{motte2007}.


\textit{C30, DR7:} Figure~\ref{fig:dr7} shows LK09 located in the centre
of the DR7 \HII\ region arc.  The CGPS \HII\ emission peak and ECX6-18
lie between BLAST sources C30 and C27, the former much more massive.
This region is not covered by \citet{motte2007}. The derived parameters
for C30 are very close to those for C32, assuming the same distance.
However, as remarked in \S~\ref{posters}, its velocity places it well
behind the Cyg~OB2 complex, at 3.6 to 7.5~kpc (we adopted 6.8~kpc in
\S~\ref{largedistance}).  Thus this source should be about 16 times more
massive and luminous, as listed in Table~\ref{tab:HII}, ranking it with
the most powerful sources in this survey.

\subsection{Stage A: Accretion Power, Protostars, and MYSOs} \label{sec:accretion}

As mentioned, BLAST cannot hope to see single stars, though the clump
might be dominated by the most massive member of the group.  During the
next earlier stage, precursor to the embedded \HII\ regions, matter
collapses under gravity at multiple sites inside the clump, which are
newly born stellar cores \citep{maclow2004}.  The luminosity of the
clump originates from the cumulative energy of accretion onto cores.
During this accretion stage, protostars build up luminosity with little
loss of envelope mass. In our \lm\ diagram (Fig.~\ref{fig:lm2}), the
evolution should be vertical, crossing the region between the dotted red
line ($L/M = 1$~\lom) and the dashed green line.  The former is coincidentally
close to diffuse cirrus emission in equilibrium with interstellar
radiation field at 17.5~K \citep{boulanger1996}. The latter, taken from
Figure~9 of \citet{mol2008}, is a best fit to class 0 sources.  Sources
plotting in this region would have cooler dust temperatures, in the
range $18~-~25$~K and $L/M$ $\geq$ 1~\lom\ would clearly indicate internally
powered sources.

\citet{motte2007} have identified several massive infrared quiet
protostellar cores (MIRQP), sources with mass greater than
$40$~\msol\ with at least some evidence of stellar activity and yet weak
emission in the mid infrared.  All of the massive infrared-quiet cores
are associated with high velocity SiO emission, providing convincing
evidence of powerful outflows and on-going accretion.  The SiO emission
of MIRQPs is typically stronger than from high luminosity infrared
protostellar cores (HLIRPCs), indicating that the former are in a more
vigorous accretion phase.

Of the 17 MIRQP, 8 are within our BLAST survey area and are readily
detected and characterized (Table~\ref{tab:HII}).  In the
\lm\ evolutionary diagram, they occupy the predicted region (see
diamonds in Fig.~\ref{fig:lm2}).  Judging from Figure~\ref{fig:lm2},
there are many other BLAST sources with similar properties.  The
sensitive \textit{Spitzer} images reveal the sub-structure, with many
accreting cores over a range of fluxes (see also \S~\ref{sec:lowlm}).

\textit{C157, DR 21(OH):} The most luminous MIRQP is in DR21(OH), which
is lacking in 21-cm radio free-free emission.  However, high-sensitivity
subarcsecond observations with the VLA at centimeter wavelengths by
\citet{araya2009} reveal a cluster of radio sources; the strongest
emission is toward the molecular core MM1, but is perhaps from
shock-ionized gas in a jet rather than from a nascent compact
\HII\ region. C157 is also cooler than the strongly ionizing sources
discussed above, as one might expect.  It has a dense core N14 along
with five neighboring fragments \citep{motte2007} in clump
N14. \citet{harvey} have observed 6.7 GHz methanol emission in all five
spots, but only two of them (DR21(OH) and DR21(OH)N) exhibited strong
peaks.  Our limited resolution integrates the dust emission from all
nearby fragments.  Though \cite{motte2007} has classified the core as
MIRQP, we note that the significant bolometric luminosity of C157, $2.9
\times 10^4$~\lsol, is probably indicative of the entire cluster,
detected in the near-infrared as DB19.

The next most luminous MIRQPs are in C7 and C115.  C7 is an isolated
source whereas C115 is at the tip of the DR17 molecular pillar.  C119 is
similarly at the tip of an adjacent pillar (Fig.~\ref{fig:dr17}). C36,
among the coolest, is just north of DR15, at the most evolved end of the
IRDC G79.34+0.33 discussed below(\S~\ref{irdcs}). There is a variety of
indications of star formation \citep{redman2003}, and \textit{Spitzer}
images indicate a rich stellar group.

There are two sources classified as infrared quiet protostar (IRQP, not
so massive as MIRQP) by \citet{motte2007} which appear to BLAST to be as
luminous as the MIRQP-containing sources.
The more luminous is C104 (N62 in clump N20 to the north-east of DR22)
in the molecular filament (velocity $-5$~\kms) extending from DR22
through MIRQP C107 to MIRQPs C141, C145 and cluster LK15 in DR23 (see
Fig.~\ref{fig:dr22}).  The most prominent molecular clump along this
filament contains C138 (N69, in clump N23); it defines the boundary of
brightest ionized part of DR23, powered by cluster DB17 to the east
(Fig.~\ref{fig:dr22}).
The other is C167 (N24 in clump N10 to the north-west of W75N). At CO
velocity $-3$~\kms, this is in the DR21 cloud complex
(Fig.~\ref{fig:dr21}).  The \textit{Spitzer} images indicate widespread
embedded star formation in the C167 cloud and adjacent BLAST sources.

The two HLIRPC (other than C32 discussed above) plot toward the high
luminosity (high temperature) side of this region in the \lm\ plane.
These correspond to the BLAST sources C85 and C87.
C85 contains source N6 in clump N3 of \citet{motte2007}.  It coincides
with DR20NW \citep{sch2006} at the head of the above-mentioned elephant
trunk.  There is an ionized rim on the face toward Cyg~OB2, but
apparently little internal ionization. Its position on the \lm\ plot
suggests most of the luminosity is generated from accretion.
Compared to C88 also in the DR20 molecular cloud, C87 (clump N6, core
N14) is relatively cold, $27$ K, with no free-free emission.

\subsection{Stage E: Externally-heated Cold Early Stage}\label{sec:early}

Prior to the stage of active accretion, there must pre-exist a cold
molecular clump with only external heating.  A low temperature and
corresponding low luminosity to mass ratio ($L/M \le 1$~\lom) would be
diagnostic of this stage.  As a forerunner of SPIRE on
\textit{Herschel}, BLAST was designed to search for the submillimeter
emission from such objects in the earliest stages of pre-stellar
evolution, and according to Figure~\ref{fig:lm} has been quite
successful, with \lmles\ sources down to the approximate
250~\micron\ completeness line of about 30~Jy.  These compact sources
are sufficiently cold and massive to be above the nominal Bonnor-Ebert
critical mass.
The higher mass objects can be considered as potential precursors of the
MIRQP accretion-powered stage.

\subsubsection{Cold Clouds with Nascent Star Formation}\label{coldcloudsf}

\begin{figure}
\centering
\includegraphics[width=\linewidth]{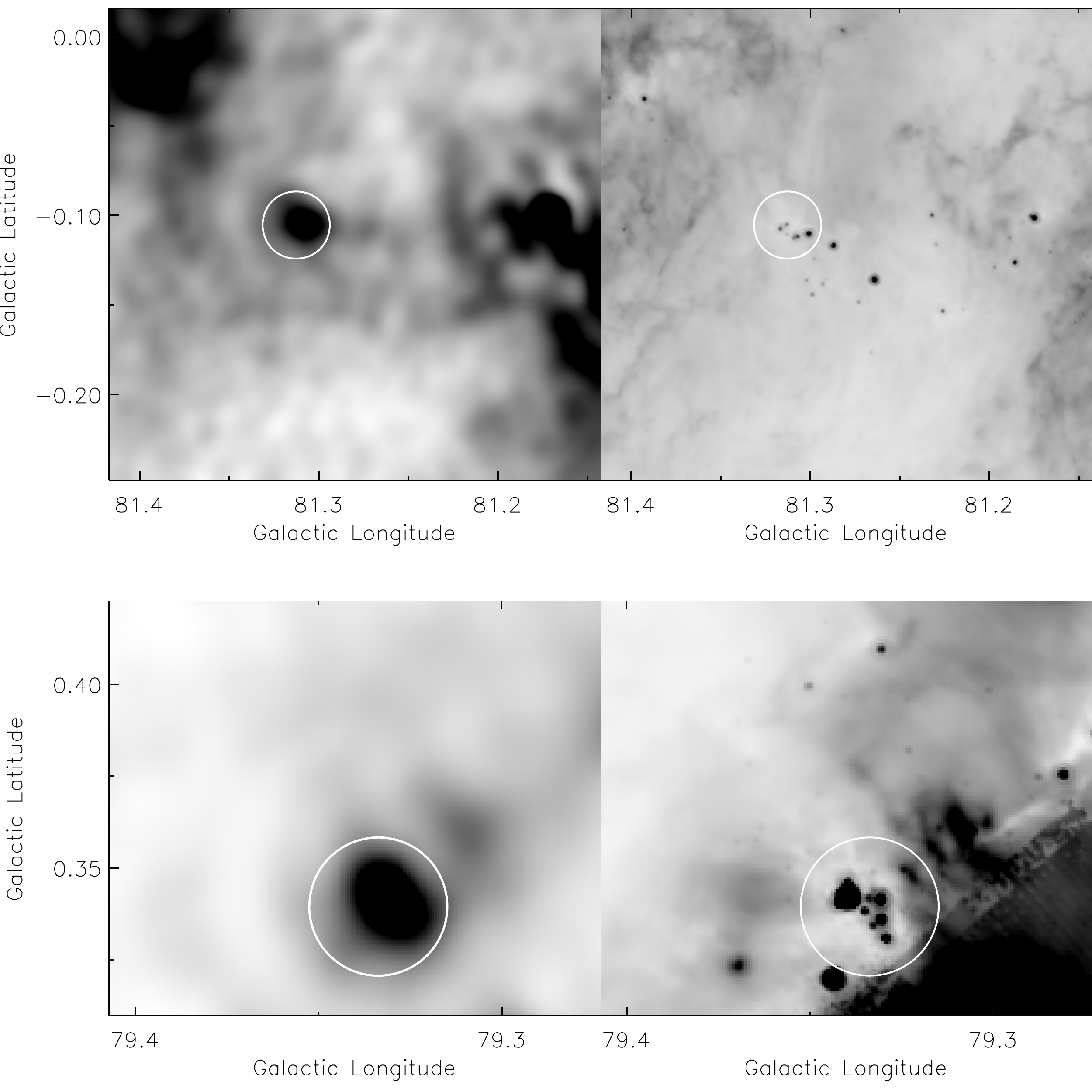} 
\caption{Left: cool BLAST sources at 250~\micron. Right: corresponding
  24~\micron\ \MIPS\ images, showing evidence for nascent star
  formation. Top: C116.  Bottom: C36 at the transition to MIRQP, in IRDC
  G79.34+0.33. Radius of circles 1.3\arcmin.}
\label{fig:cold}
\end{figure}

It is a general consensus that massive stars should begin to form within
stellar nurseries deeply embedded in a dense envelope of dust and
molecular gas.  Because \textit{Spitzer} is so sensitive and probes at
wavelengths where the dust can be penetrated, evidence for such nascent
star formation (not a lot of total power) is readily found.
Figure~\ref{fig:cold} shows two examples.  One is a cold BLAST source
C116 (along the DR22 to DR23 filament) corresponding to the
\citet{motte2007} core N70.  We find dust temperature 17.1$\pm$0.8~K,
bolometric luminosity $(2.2\pm0.4) \times 10^2$~\lsol, and mass
$(1.6\pm0.4)\times 10^2$~\msol. The other is C36, already containing a
low temperature MIRQP (\S~\ref{sec:accretion}), embedded in an IRDC
(\S~\ref{irdcs}).

\subsubsection{IRDCs}\label{irdcs}

Given the appropriate geometry, highly dense, massive molecular clouds
with significantly large extinction can be seen in silhouette as
infrared dark clouds (IRDC).  With the spatial resolution and
sensitivity of \MSX, many IRDCs have been found at Band A (8.6~\micron)
\citep{simon2006}.  The signature of an IRDC without a developed
protostar is lack of dust emission at mid-infrared \IRAS\ wavelengths
\citep{ega98}.  However, that does not mean there is no low-power
nascent star formation, as illustrated below.

\begin{figure}
\centering
\includegraphics[width=\linewidth]{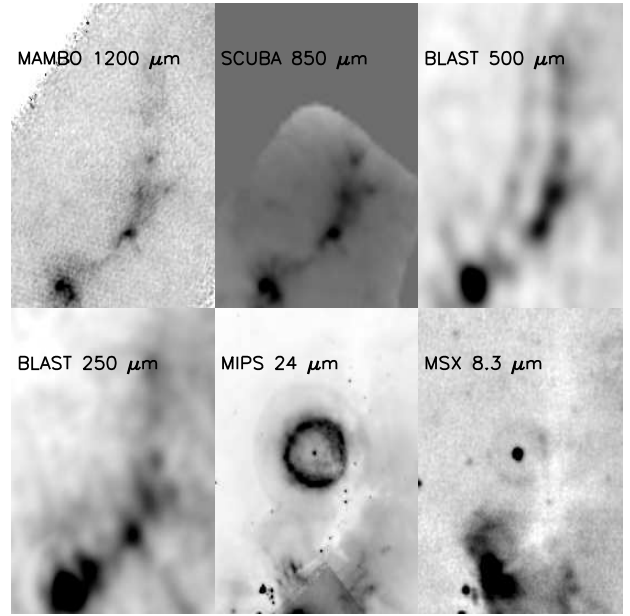} 
\caption{Multi-wavelength views of the IRDC ridge in \Cyg\ near $l$ $=$
  79\degree .23 and $b$ $=$ 0\degree .45. Each thumbnail has dimension
  12\arcmin $\times$ 16\arcmin.}
\label{fig:irdc}
\end{figure}

Because of the high column density, BLAST sees prominent IRDCs in
emission.  An example is the ridge (filament) shown in
Figure~\ref{fig:irdc} which contains the IRDCs G79.34+0.33 and
G79.27+0.38 \citep{ega98} with corresponding BLAST clumps C36 and C26,
respectively.  Although it was initially suspected that G79.34+0.33 was
associated with the DR15 cloud, later \citet{redman2003} found no such
evidence. This filamentary ridge is associated with distinct $^{13}$CO
emission at velocities near 1~\kms\ (Fig.~\ref{fig:dr15}).  G79.27+0.38
in the central part of the ridge was studied by \citet{wu}.

Finer details can be found with the \textit{Spitzer} images, both in
extinction and in the presence of point sources.  Both of these BLAST
sources show evidence of nascent star formation (Figs.~\ref{fig:cold}
and \ref{fig:irdc}).  In fact there are embedded mid-infrared objects
right along the filament, even the upper part of the ridge where the
less massive condensations are C25, C24, and C23.

Note that this ridge is seen in extinction against the shell of the
LBV star G79.29+0.46 \citep{wendker1991}, providing relative distance
information which locates the star behind the \Cyg\ region.  The
shell, even the faint outer halo seen at 24~\micron, shows no signs of
interaction with Cyg~OB2, indicating considerable separation.
\cite{kraemer2010lbv} have quantified this using the stellar
parameters, obtaining a distance of 3~kpc.

\begin{figure}
\centering
\includegraphics[width=\linewidth]{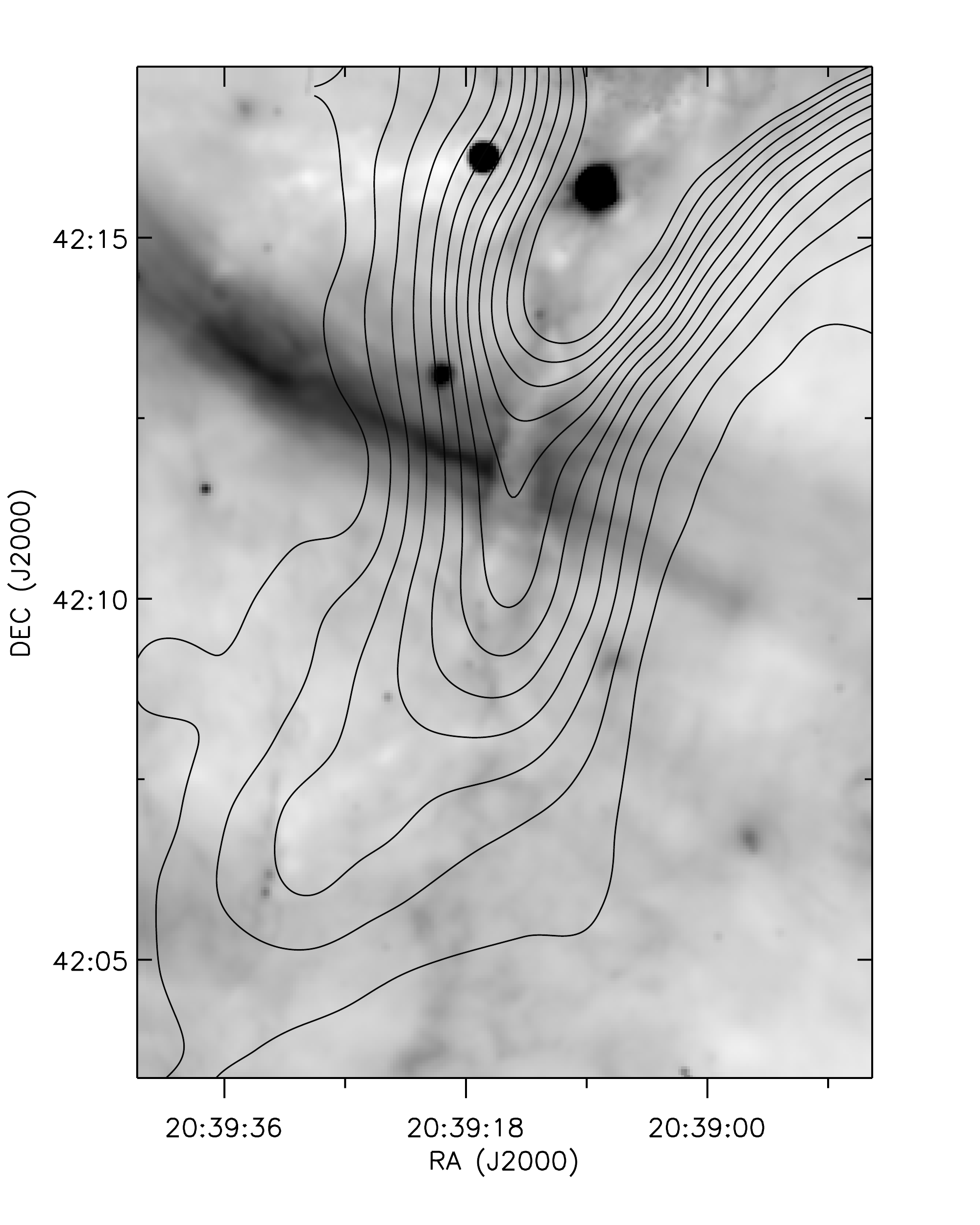} 
\caption{\MIPS\ 24~\micron\ image showing dark lane extending from DR21
  ridge. BLAST sees this lane in emission (Fig.~\ref{fig:dr21}). The
  contours are $^{13}$CO emission \citep{sch2006} integrated over $-7$
  to 1 \kms.}
\label{fig:dark}
\end{figure}

Figure~\ref{fig:dark} shows an interesting long filament that shows up
well in the 24~\micron\ image, in emission but changing to absorption in
front of a bright band of cirrus.  At \IRAC\ 8~\micron\ it is only in
emission, the bright cirrus band being absent.  This structure is
detected in BLAST emission as well as CO, extending the DR21 ridge to
the south (Fig.~\ref{fig:dr21}).  The characteristic CO velocity,
$-3$~\kms, confirms that this is related to the DR21 molecular cloud.
The extended mid-infrared source in the upper part is apparently warm
and has no BLAST counterpart.

\subsubsection{Starless Cold Clouds}\label{coldclouds}

Focusing on the earliest stage, are there any sources that are so far
starless?  The best candidates to search should be those in the region
with $L/M <1$~\lom.  In terms of internally generated power, they all
can be considered ``starless.''  But when examined in the
\MIPS\ 24~\micron\ image, most of these candidates do in fact have point
sources, indicating nascent star formation though probably only low mass
YSOs (\S~\ref{sec:lowlm}).

However, we found one instance that is apparently starless at
24~\micron, C81, an isolated feeble source with a weak CO signature at
$+6$~\kms.  While there are some point sources at \IRAC\ 8~\micron,
these do not have spectra rising to 24~\micron\ as might be expected of
YSOs; also there is no special concentration to the BLAST source and so
they are probably field star contamination.  The \IRAC\ image also shows
several small IRDCs.


Two other possibilities are C103 and C121.  (Curiously, C103 exhibits
diffuse emission at 24~\micron\ including being crossed by a partial
thin shell of radius 2.8\arcmin\ centered at G81.0425-0.1155.)
C108, clump N2 of \cite{motte2007} with no core, is crossed by IRDCs at
24 and 8~\micron, and might be another good example of a starless clump.

\section{Discussion}\label{sec:discussion}

The \lm\ diagram (Figs.~\ref{fig:lm} and \ref{fig:lm2}) can be adopted
as a diagnostic tool for describing the early evolutionary stages of
massive star formation \citep{mol2008}.  With our sample of compact
sources now covering a broad range of luminosity and mass,
particularly extending to low $L/M$, we are in a position to further assess
its utility.  We first review how independent empirical
evidence for evolutionary stages relates to the characteristic position
in the \lm\ diagram and then, to the extent possible, use theoretical
models to provide insight into the evolution.

\subsection{High $L/M$} \label{sec:lm_disc} 

In the \Cyg\ region, \hiis\ BLAST sources have detectable compact
21-cm continuum emission.  All have strong 100~\micron\ emission and
so are not background radio galaxies.  Our detections of these sources
are represented by filled light-grey circles in Figure~\ref{fig:lm2}.
Five of these were also noted and classified for \HII\ region
compactness by \cite{motte2007} and are marked as triangles, circles,
and inverted triangles.  All of these clumps appear in the higher $T$
($> 30$~K) and $L/M$ region, near the nuclear burning locus, above the
stage A sources and the empirically-obtained accretion line.  An
extensive discussion of individual embedded \HII\ regions in \Cyg\ has
been presented in \S~\ref{sec:HII}.  High mass protostars inside these
massive clumps account for the high ionization rate and luminosity.
We do not have the angular resolution to distinguish multiplicity, but
the clumps contain multiple sources when examined in the near and mid
infrared (\S~\ref{sec:lowlm}).

SiO (2~$\rightarrow$~1) emission is often used as a tracer of shocked
gas, associated with outflows which are an intrinsic consequence of
accretion process.  After the onset of nuclear burning stage, massive
stars build up an intense radiation pressure which halts further
accretion. In accordance with this expectation, the luminous sources
DR21 and W75N have only low intensity SiO (2~$\rightarrow$~1) emission
\citep{motte2007}.  In fact, all of their high luminosity IR
Protostellar cores (HLIRPC) correspond to BLAST sources with high
$L/M$ and also do not have a striking molecular emission accretion
signature.

Slightly earlier in the evolution, clumps are already quite luminous,
near the empirical accretion locus for massive stars; single-star
cores within them would be classified as class 0, at least for lower
mass stars.  On the basis of their position in this region in the \lm\
plot, we would expect these BLAST sources to show independent and
direct evidence of (bipolar) outflow and indeed this is the case
(\S~\ref{sec:accretion}).  \cite{motte2007} identified a class of
massive IR-quiet protostellar cores (MIRQP) with strong
SiO(2$\rightarrow$1) emission, indicative of on-going active
accretion.  The eight BLAST sources corresponding to their MIRQP are
marked by diamonds in Figure~\ref{fig:lm2} and all of them lie in the
stage~A accretion-dominated region of the diagram.

\subsection{Low $L/M$}\label{sec:lowlm}

In the \Cyg\ region we have discovered a significant fraction of the
clumps with low $L/M$ devoid of massive protostellar cores.  This low
$L/M$ regime is for us the most intriguing part of the \lm\ diagram,
potentially containing the story of the earliest stages of massive
star formation.  \cite{mol2008} show vertical tracks corresponding to
the evolution of a \emph{single and massive} protostar, and this is
suggested by the vertical arrow in our \lm\ diagram.  However, this is
probably a misleading interpretation for low $L/M$.  Even though these
clumps appear above the `$L=M$' line, and are arguably self-luminous,
there is no obvious signature of a high mass star being formed.  Given
that massive protostars probably evolve rapidly through this region,
it is statistically improbable to observe this stage.  

\begin{figure}
\centering
\includegraphics[width=\linewidth]{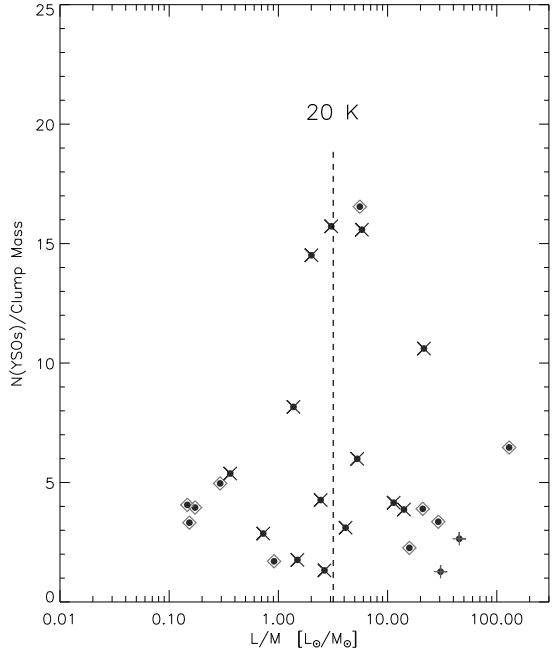}
\caption{Number of YSOs (classes 0, I, and II) per 100~$M_\odot$
  vs.\ $L/M$.  The cross, diamond, and plus symbols represent surface
  densities $\Sigma$ in the ranges 0.005 to 0.02, 0.02 to 0.08, and 0.08
  to 0.4 g~cm$^{-2}$, respectively. The vertical (dashed) line
  corresponds to 20~K, a typical value where $\Sigma$ in the model of
  \cite{krumholz2008} is about 1~g~cm$^{-2}$ for the masses of BLAST
  sources (see~\S~\ref{sec:surfsigma} and Fig.~\ref{fig:lm2}).}
\label{fig:ysosvslm}
\end{figure}

The more likely alternative energy source is the collective power of
many lower mass YSOs.  In \S~\ref{coldcloudsf} we have already previewed
some evidence for multiple YSOs in cold massive clumps.  Since the BLAST
observations were carried out, a census of YSOs has been obtained by
\cite{beerer2010} based on SEDs from \emph{Spitzer} imaging of the
northern part of the \Cyg\ region.  Using their map of \emph{Spitzer}
YSO positions (their Fig.~8) we find YSOs, often multiple YSOs,
associated with most of our clumps. For the brighter sources, we
probably underestimate the number of (crowded) YSOs when using their
Figure 8 rather than a catalog.  Figure~\ref{fig:ysosvslm} demonstrates
that the observed $L/M$ is correlated with the total number of YSOs per
unit clump mass, as expected energetically.  This also suggests
considerable fragmentation into low-mass stars, an inevitable result of
the initially low Jeans' masses in the cold gas \citep{krumholz2006}.

\begin{figure}
\centering
\includegraphics[width=\linewidth]{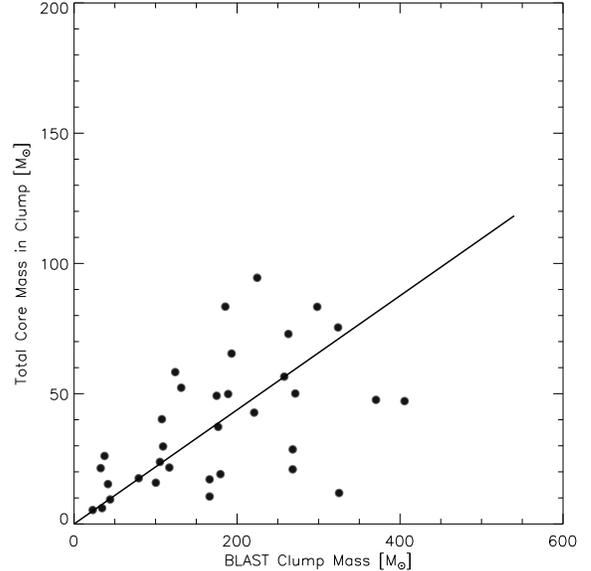}
\caption{Sum of the masses of cores within the BLAST clumps versus the
  clump mass.  The linear correlation shown has a slope 0.20.}
\label{fig:clumpvscoremass}
\end{figure}

There are (at least) two hypotheses for massive star formation, namely
the competitive accretion model \citep{bonnell1998} where massive stars
are formed due to dynamical coalescence of low-mass stars in a dense
clustered medium, and the turbulent core model \citep{MckeeTan2003} in
which massive cores with non-thermal support provided by the turbulence
evolve in a quasi-static way.  Both of these theories predict that high
mass star formation is preceded by an epoch of the low-mass star
formation, converting a larger proportion of the total mass in the
competitive accretion model.  Owing to low resolution, BLAST
observations cannot differentiate between these two hypotheses on this
basis.  Nor do we have masses for the \emph{Spitzer} YSOs.  However, the
intermediate resolution \MAMBO\ data resolves sub-structures (cores) of
size $\sim$ 0.1~pc within our clumps.  In
Figure~\ref{fig:clumpvscoremass} we compare the total mass of the cores
within each clump to the clump mass (assuming the cores are at the same
temperature as BLAST observes for the clump).  We find that (only) 20\%
of the clump mass is in cores, however, four bright sources (W75N, DR21,
DR21(OH) and C160) were excluded in the fit.

\subsection{Relationship to Surface Density} \label{sec:surfsigma}

According to the model of \cite{krumholz2008}, for the low $L/M$
regime the steady-state accretion luminosity generated by the low-mass
YSOs is:
\begin{equation}
L=390\left(\frac{\Sigma}{1~\rm g~cm^{-2}} \frac{M}{100~\rm
  M_{\odot}}\right)^{0.67}~\rm L_{\odot}.
\label{eq:lsig}
\end{equation}
In order to show the relative position of BLAST sources with respect
to this model prediction, we plot loci in the \lm\ diagram of 
Figure~\ref{fig:lm2} for $\Sigma~=$ 1.0, 0.1, and 0.01 g~cm$^{-2}$.

\begin{figure}
\centering
\includegraphics[width=\linewidth]{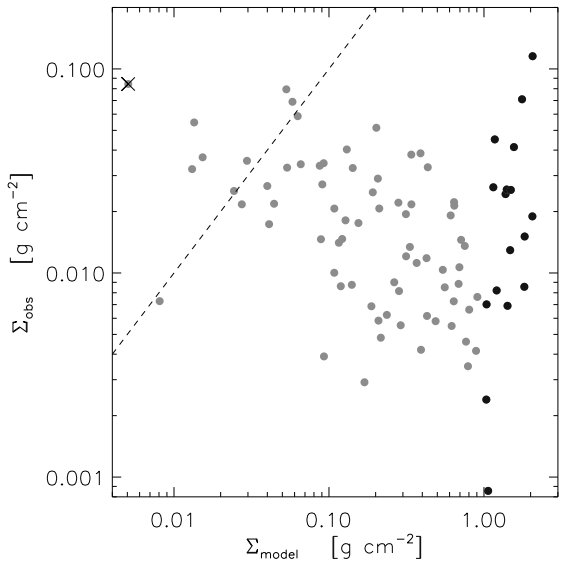} 
\caption{Observed surface density of BLAST clumps with $L/M~<$
  6~\lsol$/$\msol\ vs.\ the surface density required in the low-mass
  YSOs model of \cite{krumholz2008} to produce the observed $L$ given
  the observed $M$.  Sources above the one-to-one line (dashed) are
  under-luminous compared to the prediction from the model, and so
  possibly very young.  The source shown by a cross is near the map edge
  at 500~\micron. Sources with model surface density more than
  1~g~cm$^{-2}$ are highlighted.  }
\label{fig:sigmavssigma}
\end{figure}

If this theory is adopted then we can directly determine surface
density from equation~(\ref{eq:lsig}) for any source in the \lm\
diagram.  This is the surface density that would be required for a
clump of mass $M$ to produce a luminosity of $L$.  However, we can also
measure the surface density directly.  
In Figure~\ref{fig:sigmavssigma}, we plot surface densities of BLAST
sources versus the model surface density, for sources below $L/M~<$
6~\lsol$/$\msol\ where most of the luminosity arguably comes from
low-mass stars.
The measured surface density of the most of the sources is below the
one-to-one line. This might be because we have underestimated $\Sigma$
owing to poor angular resolution; qualitative support for this is the
evidence of sub-structure (\S~\ref{sec:lowlm}).
However, there are about 8 sources with surface density above the
one-to-one line.  These are very interesting, possibly a very early
stage where the presumed steady state of low mass YSO formation has not
been achieved.  We have confirmed that there is a lower number of YSOs
associated with these sources, except for the source shown by a cross
in Figure~\ref{fig:sigmavssigma} which was difficult to measure
near the edge of the image at 500~\micron.
    
\cite{krumholz2008} note that because cores with higher $\Sigma$ are
warmer due to the power of the low-mass YSOs, there is a critical
$\Sigma$ such that by the time the steady-state heating is achieved
fragmentation into further low-mass YSOs halts.  The threshold surface
density is about 1~g~cm$^{-2}$ and this is plotted in
Figure~\ref{fig:lm2}.  This characteristic surface density should be a
marker of (potential) massive star forming regions, and there is some
evidence consistent with this, at least for evolved objects.
In the BLAST survey we find that the few most luminous sources
containing compact \HII\ regions have such a high surface density.
The \MAMBO\ cores containing compact \HII\ regions and MIRQPs with
strong SiO emission also have a high measured surface density
(Table~\ref{tab:clump_cores}).
Similarly, \cite{garay2006a} found an average surface density of
0.8~g~cm$^{-2}$ for dense cores of ultracompact \HII\ regions.
We find that the typical surface density of massive embedded star
clusters observed by \cite{leduigou2002} in the \Cyg\ region is about
1~g~cm$^{-2}$ (allowing for the presence of some
additional gas -- now dispersed -- in the precursor).

An observed high $\Sigma$ associated with MYSOs might not be a signature
unique to the turbulent core model, or even a prerequisite.  In the
simulations by \citet{smith2009} of a competitive accretion model,
``massive stars were not formed from a single massive thermally
supported fragment, but instead from a smaller core which accreted
additional material channelled towards it by the potential of the
forming stellar cluster.''  Likewise, in the simulations by
\citet{wang2010} of a model of ``clump-fed massive star formation''
regulated by outflow feedback (ORCF), ``the most massive object is not
formed out of a pre-existing dense core'' but rather ``is controlled by
the global clump dynamics.''
Recently \cite{csengeri2010} observed a few of the massive young dense
cores (C107, C115, C119, DR21(OH), DR21) in \Cyg\ to study the kinematic
properties of the underlying environment. They found that non-thermal
support due to turbulence at the scale of the dense core is not enough
for a quasi-static evolution of the massive protostars as predicted by
the turbulent core model of \cite{MckeeTan2003}, and they favor
competitive accretion.

Therefore, a key question in investigating the mode of evolution is
whether there is evidence for cold high-$\Sigma$ clumps or cores before
high mass protostar(s), and even the preceding low-mass YSOs, are
established.  Even if the evolution timescale were favorably long, this
is difficult with BLAST, as can be seen directly from
equation~(\ref{eq:sigma}): at the spatial resolution of 0.5~pc about
4000~\msol\ is needed to produce a surface density of 1~g~cm$^{-2}$.  We
have noted that we do not see such massive cold objects; if such dense
cold objects do exist at lower mass, we cannot identify them by directly
measuring their true $\Sigma$.  Higher resolution is required for such
detections; for example, \MAMBO\ observations at the resolution of
0.1~pc in \Cyg\ need a mass of only 150~\msol\ to produce equivalent
contrast.  The predominantly low surface densities obtained by
\cite{elia2010} suggest that even the \emph{Herschel} resolution is not
sufficient for finding such dense sub-structures in most cold sources.

\subsection{Vertical Evolution in the \lm\ Diagram?} \label{sec:vertical}

Given the different theories, it is still unclear to us which of the low
$L/M$ BLAST clumps will give rise to massive star formation.  If there
is a universal IMF, then statistically a massive clump will be required
to give birth to massive protostars.  Therefore, the massive clumps
identified by BLAST are certainly of interest in this regard.  Naively,
the evolution in the \lm\ diagram would be vertical (constant $M$) as
the luminosity increases from accretion onto the massive protostar and
then nuclear burning.  Even this is complicated because the power
produced by low mass YSOs in most theories can mask the earliest low $L$
stage of the massive protostar.

However, it does not follow that massive clumps/cores always form
massive stars.  In fact, the model by \cite{krumholz2008} explicitly
predicts that this will not happen unless $\Sigma$ is above a critical
value of about 1~g~cm$^{-2}$.  Otherwise, their prediction is that the
clump is destined to continue to produce only low mass stars, at some
presumed steady-state rate.  Thus the vertical rise in $L$ is arrested
at a ceiling for a given $\Sigma$.  We have marked such a possibility in
Figure~\ref{fig:lm2} with a horizontal red arrow at a level which would
correspond to the sub-critical $\Sigma$ of the clump.  The end result is
a cluster or association of only low mass stars, i.e., one with a
truncated IMF.  Because the precursor is of low $\Sigma$ the resulting
assemblage of stars must be as well, making them more difficult to
detect, and there is no high mass star to draw attention to the region
by its high luminosity and ionization.

The empirical context, a snapshot of the stages of evolution, is that
there is a hierarchy of structures.  BLAST identifies compact sources of
size about 0.5 pc within the CO clumps of spatial extent about 10~pc
\citep{sch2006}.  Within most of the BLAST clumps, \MAMBO\ is able to
find smaller and denser sub-structures of scale 0.1~pc.  Presumably
these structures have been created and evolve by accumulating mass in
the dynamical environment of the larger reservoir and possibly also
simultaneously transfer mass to the immediately lower scale
sub-structures.  At any given time these structures will each have a
characteristic $\Sigma$.  At some point does this become frozen in time
so that we can discuss the fate of objects of fixed $\Sigma$?  Or
tracking a clump, could $\Sigma$ and $M$ evolve, allowing evolution
toward the upper right in the \lm\ diagram?  We have indicated this
latter possibility as well in Figure~\ref{fig:lm2} -- of course only
schematically.  Recent numerical simulations of the competitive
accretion model \citep{smith2009}, and the ORCF model \citep{wang2010}
which has features of both competitive accretion and the turbulent core
model, show a channeling of mass to bound potential energy minima of the
parsec-scale dense clumps studied, emphasizing the importance of
clump-level dynamics.
%
%
In these models, when the material is cold it is more distributed and so
would not be identified as a high mass, high $\Sigma$, cold clump.  By
the time that mass builds up in massive cores, accreting on massive
protostars, the compact source would have low mass YSOs as well and have
been warmed up.  Thus the cold precursors of dense clusters might be
more extended objects than clumps, a massive reservoir yet to be
channeled by gravity into a higher surface density state.  Some cold
IRDCs (\S~\ref{irdcs}, Fig.~\ref{fig:planck}), within which BLAST
identifies substructure as sources, might fit into this scenario, but
dynamical diagnostics would be needed to pursue the relevance of this
possibility.

\section{Conclusion and Future Work}\label{future}

An unbiased survey conducted by BLAST has enabled us to detect both
compact sources and the embedding diffuse emission as a basis for
studying the earliest stages of star formation.
We have quantified contrasting dust temperatures of diffuse extended
structures by correlating the fluctuating emission in \IRAS\ and BLAST
bands.
To reveal the global morphology, we have studied the relationship
between dust continuum emission and radio, mid-infrared, and $^{13}$CO
line emission. Comparing the total mass present in the BLAST compact
sources with the parent $^{13}$CO clump mass \citep{sch2006}, we have
obtained a linear correlation with slope of $0.05$, which might be
indicative of the star formation efficiency but also relates to the
relative lifetime of the compact clump phase.
With multi-wavelength photometry we have tightly constrained the dust
emission SEDs of the compact sources to measure temperature, luminosity
and mass.
We have diagnosed the earliest stages of star formation in \Cyg, and the
evolution, in the context of the luminosity -- mass diagram of the
compact sources.
We have studied eight DR regions which are known as massive star
formation sites in \Cyg. We have identified \hiis\ sources which have
compact \HII\ emission and found an average dust temperature of
33~K. Not all of these identified sources have sufficient luminosity to
be ionizing sources, pointing to the influence of Cyg~OB2 creating
external ionization.
The complex morphology along with the direction of cometary shapes of
some compact sources provide direct evidence of interaction with the
massive Cyg~OB2 association. Along the periphery of the evacuated
\HII\ region created by some of the more evolved stellar clusters, there
is a signature of triggered star formation.
We have observed a complex of infrared quiet dark clouds where star
formation has already started to take place. To BLAST, and also
\SCUBA\ and \MAMBO, this appears as an impressive ridge of cold dust
emission.  However, in sensitive high resolution 24 \micron\ \MIPS\ and
8 \micron\ \IRAC\ images, there is evidence of a stellar nursery,
indicating a potential site of massive star formation.
In our search for the precursors of clusters, however, we have not found
any cold compact clump which is massive enough to become a star cluster
like the Cyg~OB2 association or even the other lesser embedded clusters.
In the future, the unbiased Planck Cold Core survey might find such
objects in unevolved GMCs, possibly in a more extended lower surface
density state than in the final cluster.
Resolving multi-scale structures is a new challenge underlying
investigations of the earliest stages of star formation. BLAST with its
limited resolution could only resolve spatial structures of 0.5~pc at
1.7~kpc away.  On the other hand, \MAMBO\ with its 11\arcsec\ resolution
was successful in separating sub-structures on a scale of 0.1 pc;
further progress should be possible with \textit{Herschel}.  To
determine the role of surface density and more generally which cold high
mass clumps will evolve to produce high mass protostars even higher
resolution observations will be required, and so there appears to be a
rich future in uncovering the unseen story using SCUBA2
(8\arcsec\ resolution at 450~\micron, \citealp{holland2006}), CCAT
(3.5\arcsec\ at 350~\micron, \citealp{sebring2010}), and 
ALMA\footnote{http://www.almaobservatory.org/en/about-alma/essentials/numbers}
(down to sub-arcsecond resolution at submillimeter wavelengths

\acknowledgments
The BLAST collaboration acknowledges the support of NASA through grant
numbers NAG5-12785, NAG5-13301, and NNGO-6GI11G, the Canadian Space
Agency (CSA), the UK Particle Physics \& Astronomy Research Council
(PPARC), and Canada's Natural Sciences and Engineering Research
Council (NSERC). We would also like to thank the Columbia Scientific
Balloon Facility (CSBF) staff for their outstanding work.     

\clearpage
\begin{longtable*}{lcccllclccll}
\caption {}\label{tab:bsource}\\
\multicolumn{11}{c}{{MEASURED FLUX DENSITY}}\\\\
\hline \hline\\
\multicolumn{1}{l}{BLAST} &
\multicolumn{1}{c}{Source}&
\multicolumn{1}{c}{$l$} &
\multicolumn{1}{c}{$b$} &
\multicolumn{1}{l}{$S_{1200}$} &
\multicolumn{1}{l}{$S_{850}$} &
\multicolumn{1}{c}{$S_{500}$} &
\multicolumn{1}{l}{$S_{450}$} &
\multicolumn{1}{c}{$S_{350}$} &
\multicolumn{1}{c}{$S_{250}$} &
\multicolumn{1}{l}{$S_{100}$} &
\multicolumn{1}{l}{$S_{60}$} \\

\multicolumn{1}{l}{ID} &
\multicolumn{1}{c}{Name}&
\multicolumn{1}{c}{(Degree)} &
\multicolumn{1}{c}{(Degree)} &
\multicolumn{1}{l}{(Jy)} &
\multicolumn{1}{l}{(Jy)} &
\multicolumn{1}{c}{(Jy)} &
\multicolumn{1}{l}{(Jy)} &
\multicolumn{1}{c}{(Jy)} &
\multicolumn{1}{c}{(Jy)} &
\multicolumn{1}{l}{(Jy)} &
\multicolumn{1}{l}{(Jy)} \\ \hline
\\
\endfirsthead
\multicolumn{11}{c}{{\tablename} \thetable{} -- Continued} \\
\hline\hline\\
\multicolumn{1}{l}{BLAST} &
\multicolumn{1}{c}{Source}&
\multicolumn{1}{c}{$l$} &
\multicolumn{1}{c}{$b$} &
\multicolumn{1}{l}{$S_{1200}$} &
\multicolumn{1}{l}{$S_{850}$} &
\multicolumn{1}{c}{$S_{500}$} &
\multicolumn{1}{l}{$S_{450}$} &
\multicolumn{1}{c}{$S_{350}$} &
\multicolumn{1}{c}{$S_{250}$} &
\multicolumn{1}{l}{$S_{100}$} &
\multicolumn{1}{l}{$S_{60}$} \\

\multicolumn{1}{l}{ID} &
\multicolumn{1}{c}{Name}&
\multicolumn{1}{c}{(Degree)} &
\multicolumn{1}{c}{(Degree)} &
\multicolumn{1}{l}{(Jy)} &
\multicolumn{1}{l}{(Jy)} &
\multicolumn{1}{c}{(Jy)} &
\multicolumn{1}{l}{(Jy)} &
\multicolumn{1}{c}{(Jy)} &
\multicolumn{1}{c}{(Jy)} &
\multicolumn{1}{l}{(Jy)} &
\multicolumn{1}{l}{(Jy)} \\ \hline
\\
\endhead
\hline \\ \multicolumn{11}{l}{{Continued on Next Page\ldots}} \\
\endfoot
\endlastfoot
C0&  J203057+394248&$ 78.6760$&$      0.1902$& \nodata& \nodata& $   22.0\pm    3.9$& \nodata& $   85.8\pm    3.4$& $  110.7\pm    5.6$&     15.6\tablenotemark{u}& \nodata\\
C1&  J202925+401118&$ 78.8861$&$      0.7091$& $    8.1$& $   28.3$& $  151.5\pm    2.9$& $  226.9$& $  375.3\pm    8.0$& $ 1029.3\pm   35.2$&   5819.4&  5699.0\\
C2&  J202904+401506&$ 78.8980$&$      0.8006$& \nodata& \nodata& $   90.7\pm   32.4$& \nodata& $   79.1\pm   17.8$& $   59.9\pm    6.7$&     40.3&     1.3\\
C3&  J203329+394127&$ 78.9454$&$     -0.2135$& \nodata& \nodata& $   23.7\pm    1.3$& \nodata& $   50.9\pm    2.9$& $   85.4\pm    5.0$&     41.1\tablenotemark{u}&     7.7\tablenotemark{u}\\
C4&  J203323+394253&$ 78.9546$&$     -0.1860$& \nodata& \nodata& $   33.0\pm    1.3$& \nodata& $   68.8\pm    2.9$& $  130.7\pm    5.6$&    118.2\tablenotemark{u}&     6.9\tablenotemark{u}\\
C5&  J202653+403717&$ 78.9566$&$      1.3528$& \nodata& \nodata& $    5.1\pm    0.4$& \nodata& $    7.7\pm    0.5$& $   11.3\pm    0.9$&     11.2\tablenotemark{u}&     6.1\tablenotemark{u}\\
C6&  J203023+400926&$ 78.9702$&$      0.5401$& \nodata& \nodata& $   15.2\pm    0.9$& \nodata& $   50.3\pm    2.4$& $   74.4\pm    4.4$&  \nodata&     1.0\tablenotemark{u}\\
C7&  J203112+400310&$ 78.9778$&$      0.3528$& $    6.6$& $   17.6$& $  133.8\pm    3.0$& $  187.6$& $  287.8\pm    5.1$& $  512.5\pm   10.1$&    935.1&   688.1\\
\end{longtable*}

\begin{deluxetable*}{lrrr}
\tablecaption{Clusters and Stellar Groups}
\tablewidth{0pt}
\tablehead{
\colhead{Cluster\tablenotemark{a}} &
\colhead{$l$} &
\colhead{$b$} &
\colhead{Other}\\
\colhead{ID}&
\colhead{Degree}&
\colhead{Degree}&
\colhead{Names}}\\
\startdata
LK08            & 79.302  & 0.286  & DB10, ECX6-25  \\
LK09            &  79.301 &  1.291 &  DB11, ECX6-18 \\
DB12\tablenotemark{sg} &  79.879 &  1.178 &   ECX6-21  \\
ECX6-27         &  80.354 &   0.728 &   \nodata \\
DR18\tablenotemark{sg}  &  80.357 &   0.450 &    \nodata\\
LK11            &  80.935 &  -0.167 &  DB13, DB14\tablenotemark{sg}     \\
LK12            &  81.298 &  1.096  &    DB15   \\
LK13            &  81.445 &  0.483  &  DB16    \\
LK14            &  81.441 &  1.122  &  \nodata  \\
DB17            &  81.566 &  -0.719  &  \nodata  \\
LK15            &  81.610 &  0.142   &  \nodata  \\
DB18\tablenotemark{sg} &  81.661 &  -0.017  &  \nodata  \\
DR21            &  81.680 &  0.540   &   \nodata \\
DB19\tablenotemark{sg} &  81.711 &  0.582   &  \nodata  \\
W75N            &  81.870 &  0.779   &\nodata   \\
DB20            &  81.896 &  0.797   &  \nodata  \\
DB22            &  82.567 &  0.404   &  \nodata  \\

\enddata
\label{tab:clusters}
\tablenotetext{a}{LK, DB, and ECX6 are the cluster IDs by \cite{leduigou2002},
\cite{dutra2001}, and \cite{comeron2002}, respectively.}
\tablenotetext{sg}{ Stellar group.}
\end{deluxetable*}

\begin{deluxetable*}{lcccc}
\tablecaption{Structures observed at 1.2 mm associated with BLAST clumps}
\tablewidth{0pt}
\tablehead{
\colhead{BLAST} &
\colhead{Core\tablenotemark{a}} &
\colhead{Clump\tablenotemark{a}} &
\colhead{$\Sigma_{\rm core}$\tablenotemark{a}} &
\colhead{$\Sigma_{\rm clump}$\tablenotemark{a}} \\
\colhead{ID}&
\colhead{ID}&
\colhead{ID}&
\colhead{$\rm g~cm^{-2}$}&
\colhead{$\rm g~cm^{-2}$}}\\
\startdata
C1&S26&Cl-S5&0.58&0.33\\
C7&S30, S31&Cl-S7&0.21&0.18\\
C8&S29&Cl-S6&0.08&0.08\\
C26&S34&Cl-S9&0.08&0.08\\
C28&S39&\nodata&0.12&\nodata\\
C32&S40, S41&Cl-S11&0.13&0.11\\
C36&S36, S37, S38&Cl-S10&0.16&0.12\\
C85&N5, N6&Cl-N3&0.15&0.15\\
C87&N14&Cl-N6&0.48&0.14\\
C88&N10&Cl-N4&0.43&0.18\\
C90&N17&\nodata&0.15&\nodata\\
C92&N58, N59&Cl-N17&0.22&0.18\\
C104&N62&Cl-N20&0.08&0.09\\
C107&N63&Cl-N21&0.96&0.28\\
C115&N1, N2&Cl-N1&0.56&0.29\\
C116&N70&\nodata&0.07&\nodata\\
C119&N12, N13&Cl-N5&0.58&0.21\\
C126&N18&Cl-N9&0.07&0.07\\
C127&N29&Cl-N11&0.06&0.08\\
C128&N16&\nodata&0.13&\nodata\\
C132&N67, N69&Cl-N23&0.05&0.09\\
C134&N26&\nodata&0.06&\nodata\\
C135&N61&\nodata&0.07&\nodata\\
C139&N8&\nodata&0.12&\nodata\\
C141&N65, N64&Cl-N22&0.14&0.10\\
C145&N68&\nodata&0.29&\nodata\\
C149&N72&\nodata&0.05&\nodata\\
C155&N42, N46, N47&Cl-15&2.30&1.67\\
C157&N36, N38, N41, N44, N48&Cl-N14&1.92&1.24\\
C160&N37, N43, N51, N53, N54&Cl-N16&0.45&0.45\\
C163&N20&\nodata&0.18&\nodata\\
C166&N9&\nodata&0.05&\nodata\\
C167&N21, N22, N24&Cl-N10&0.08&0.10\\
C169&N30, N31, N32&Cl-N13&2.49&1.27\\
\enddata
\label{tab:clump_cores}
\tablenotetext{a}{Core, Clump ID, and surface density are from \cite{motte2007}}
\end{deluxetable*}

\clearpage
\begin{longtable*}{lccccc}
\caption{ {}\label{tab:sed}}\\
\multicolumn{5}{c}{{RESULTS FROM SED FITS}}\\\\
 \hline \hline\\
        \multicolumn{1}{l}{BLAST} &
        \multicolumn{1}{c}{$T$} &
        \multicolumn{1}{c}{$M$} &
        \multicolumn{1}{c}{$\Sigma$} &
        \multicolumn{1}{c}{$L$} &
        \multicolumn{1}{c}{$L_{\rm{bol}}$} \\

        \multicolumn{1}{l}{ID} &
        \multicolumn{1}{c}{(K)} &
        \multicolumn{1}{c}{(100~M$_\odot$)} &
        \multicolumn{1}{c}{(g cm$^{-2}$)} &
        \multicolumn{1}{c}{(100~L$_\odot$)} &
        \multicolumn{1}{c}{(100~L$_\odot$)} \\ \hline
\\
\endfirsthead

\multicolumn{5}{c}{{\tablename} \thetable{} -- Continued} \\
\hline \hline \\
        \multicolumn{1}{l}{BLAST} &
        \multicolumn{1}{c}{$T$} &
        \multicolumn{1}{c}{$M$} &
         \multicolumn{1}{c}{$\Sigma$} &
        \multicolumn{1}{c}{$L$} &
        \multicolumn{1}{c}{$L_{\rm{bol}}$} \\

        \multicolumn{1}{l}{ID} &
        \multicolumn{1}{c}{(K)} &
        \multicolumn{1}{c}{(100~M$_\odot$)} &
        \multicolumn{1}{c}{(g cm$^{-2}$)} &
        \multicolumn{1}{c}{(100~L$_\odot$)} &
        \multicolumn{1}{c}{(100~L$_\odot$)} \\\hline
\\
\endhead

 \hline \\ \multicolumn{5}{l}{{Continued on Next Page\ldots}}\\
\endfoot
\endlastfoot
                  C0 & $ 15.5  \pm  0.7 $ & $  1.8  \pm  0.4 $ & $  0.033 $ & $    1.4 \pm   0.2 $ & \nodata\\
                          C1 & $ 40.1  \pm  4.0 $ & $  1.9  \pm  0.4 $ & $  0.121 $ & $  277.2 \pm 100.7 $ & $  489.4 $\\
                  C3 & $ 16.6  \pm  1.1 $ & $  1.4  \pm  0.3 $ & $  0.021 $ & $    1.6 \pm   0.4 $ & \nodata\\
                  C4 & $ 17.5  \pm  0.7 $ & $  1.7  \pm  0.3 $ & $  0.022 $ & $    2.6 \pm   0.4 $ & \nodata\\
                  C5 & $ 13.2  \pm  1.3 $ & $  0.5  \pm  0.2 $ & $  0.022 $ & $    0.1 \pm   0.0 $ & \nodata\\
                          C6 & $ 19.6  \pm  0.6 $ & $  0.7  \pm  0.1 $ & $  0.009 $ & $    1.8 \pm   0.2 $ & $    2.0 $\\
                          C7 & $ 25.8  \pm  0.8 $ & $  2.6  \pm  0.3 $ & $  0.084 $ & $   33.4 \pm   3.3 $ & $   50.2 $\\
                  C8 & $ 17.3  \pm  1.6 $ & $  2.2  \pm  0.6 $ & $  0.038 $ & $    3.1 \pm   1.0 $ & \nodata\\
\end{longtable*}

\begin{deluxetable*}{lllllll}
\tablewidth{0pt}
\small
\tablecaption{Compact HII regions and protostars}
\tablewidth{0pt}
\tablehead{
\colhead{BLAST} &
\colhead{$T$}&
\colhead{$M$}&
\colhead{$L$} &
\colhead{Source} &
\colhead{\MAMBO}&
\colhead{Figure\tablenotemark{a}}\\
\colhead{ID}&
\colhead{(K)}&
\colhead{(100~M$_{\odot}$)}&
\colhead{(100~L$_{\odot}$)} &
\colhead{Name}&
\colhead{1.2 mm}&
\colhead{No}}
\startdata
  C1   &40 & 1.9 &489.0&        S26, AFGL2591 &         UC\HII & \ref{fig:dr15}\\
  C7   &26 & 2.6 & 50.2 &        S30, IRAS~20293+3952 &          MIRQP & \ref{fig:dr15}\\
 C14   &30 &  3.8 & 4.5 &          IRAS~20255+4032 &          \nodata &  \ref{fig:dr7}\\
 C30\tablenotemark{b}  &38 & 17.6 &1640.0 &      IRAS~20264+4042 &          \nodata &        \nodata\\
 C32  &46 & 1.1 &330.7 &  S41, IRAS~20306+4005 &        HLIRPC~ & \ref{fig:dr15}\\
 C36  &26& 2.6 & 30.6&  S37, IRAS~20305+4010 &         MIRQP~ & \ref{fig:dr15}\\
 C38  &38 &  0.4 & 48.7 &      IRAS~20319+3958 &       \nodata &        \nodata\\
 C49  &27 &  0.1 &  2.8 &            \nodata &       \nodata &        \nodata\\
\enddata
\tablenotetext{a}{Figure number of the images containing individual sources.}
\tablenotetext{b}{$M$ and $L$ corrected for distance (see \S~\ref{largedistance}).}
\label{tab:HII}
\end{deluxetable*}

\clearpage
\begin{deluxetable*}{lrrrrrr}
\tablewidth{0pt}

\tablecaption{\MSX\ counterparts}
\tablehead{
\colhead{BLAST} &
\colhead{$\Delta \alpha$\tablenotemark{a}} &
\colhead{$\Delta \delta$\tablenotemark{a}} &
\colhead{$S_8$} &
\colhead{$S_{12}$} &
\colhead{$S_{14}$} &
\colhead{$S_{25}$} \\
\colhead{ID} &
\colhead{(\arcsec)} &
\colhead{(\arcsec)} &
\colhead{(Jy)} &
\colhead{(Jy)} &
\colhead{(Jy)} &
\colhead{(Jy)}
}
\startdata
  C1&$  5.9$&$  3.0$&$ 313.8$&$ 568.0$&$ 798.5$&$1023.4$\\
  C6&$ -1.5$&$  6.0$&$   0.2$&$   0.8$&$   0.6$&$   1.6$\\
  C7&$ -7.8$&$ 13.7$&$  17.1$&$  23.5$&$  23.4$&$ 100.0$\\
 C14&$ 11.6$&$ -5.1$&$   0.4$&$   1.0$&$   1.2$&$   2.4$\\
 C15&$ 17.2$&$ -6.2$&$   4.6$&$   5.8$&$   3.7$&$  13.6$\\
 C19&$  4.4$&$  2.3$&$   1.8$&$   1.7$&$   0.9$&$   1.3$\\
 C22&$-27.6$&$ 15.4$&$   0.7$&$   1.3$&$   2.0$&$   2.5$\\
 C27&$ 15.4$&$ -6.3$&$   2.1$&$   4.7$&$   3.0$&$   8.7$\\
\enddata
\tablenotetext{a}{Offsets of \MSX\ sources with respect to BLAST counterparts.}
\label{tab:msx}
\end{deluxetable*}

\appendix
\section{Multi-wavelength photometry}\label{multiphot}

Flux densities at many supplementary wavelengths are presented in
Table~\ref{tab:bsource} and Table~\ref{tab:msx}.  They were obtained as
follows.


\textit{\MAMBO\ 1200~\micron}.
The \MAMBO\ camera mounted on the IRAM 30-m telescope imaged all regions
having column density greater than ($A_V\ge $ 15) at wavelength of 1.2
mm \citep{motte2007}.  With its 11\arcsec\ beam, \MAMBO\ has the ability
to resolve multiple cores (if present) within BLAST clumps.  Combining
our data with \MAMBO\ helps in multi-wavelength photometry, especially
when any cold source is undetected in the mid to far-infrared spectrum.
To obtain consistent photometry, we first convolved the \MAMBO\ map to
1\arcmin, about the resolution of the BLAST maps. This has two effects:
blending together any \MAMBO\ cores; and bringing in flux from the more
extended emission (halo or plateau) usually associated with these
multi-scale structures. Gaussian photometry was performed on the
convolved map.


\textit{\SCUBA\ 850 and 450~\micron}.
The archival data from the Submillimeter Common User Bolometer (\SCUBA)
on the 15-m James Clerk Maxwell Telescope (JCMT), \cite{difrancesco}
sample many high extinction regions of the Galactic Plane, including
\Cyg, with diffraction-limited beams of 14.5\arcsec\ and 7.5\arcsec, at
850 and 450~\micron, respectively.  Due to the higher spatial
resolution, \SCUBA\ can also identify numerous small-scale cores.  The
archival images consist of discontinuous smaller maps and so could not
be convolved to BLAST resolution prior to photometry.  Instead, we used
flux densities using the \SCUBA\ legacy catalogs of \citet{difrancesco}.
We obtained flux densities from the Extended Map Object Catalog (EMOC)
and added up all flux densities for sources within 1\arcmin\ of the
BLAST source.  Because of the small \SCUBA\ maps, some BLAST sources are
not completely scanned near the map edges and the resulting flux density
is low.


\textit{\IRAS\ 100 and 60 \micron}.
Bright dust emission is observed at 100 and 60~\micron\ throughout \Cyg,
and for most of the more luminous, hotter BLAST sources these \IRAS\ bands
cover the peak of the SED. The \IRAS\ point source catalog does not have
any entries for about half of our \Cyg\ region. This arose because of
the very bright source DR21 (C.~Beichman, private communication).
Nevertheless, the \IRAS\ images that have been made subsequently are not
seriously affected.  We extract flux densities at 100 and
60~\micron\ from the Infrared Galaxy Atlas images (IGA;
\citealp{cao1997}), made using HIRES \citep{aumann1990}, a resolution
enhancing algorithm.  There is some elongation of sources across the
dominant scan direction. The resolution at 100 and 60~\micron\ is about
2\arcmin\ and 1\arcmin, respectively, substantially better resolution
than in the original ISSA product.  Most of the isolated BLAST sources
are warm enough to have a counterpart in the IGA map. BLAST counterparts
at 100~\micron\ were measured with a circular aperture of 2.4\arcmin.
Similarly, at 60~\micron\ elliptical apertures of dimension
1.8\arcmin\ $\times$ 1.2\arcmin\ were placed with the major axis aligned
across the scan direction.  In the crowded regions, multiple Gaussian
photometry (elliptical in the case of 60~\micron) was carried out by
fixing centroids of nearby sources.

The counterparts to BLAST sources in the 60~\micron\ images are often
affected by extended cirrus-like structure and sometimes by
deconvolution artefacts. Such sources show inconsistent structure
between the 100 and 60~\micron\ images. We preferred not to use such
photometric data as detections to constrain the Wien part of the SED,
rather using them as upper limits through a penalty function
\citep{chapin2008}.  In crowded regions, the flux density measurements
of fainter sources are highly uncertain and for such cases we also used
the photometric data as an upper limit.  These upper limits are
indicated by a superscript `u' in Table~\ref{tab:bsource}.


\textit{\MIPS\ 70 \micron}.  \textit{Spitzer} \MIPS\ observations at
70~\micron\ have better resolution and sensitivity than \IRAS\ IGA
60~\micron\ data and so are particularly useful for detecting or placing
upper limits on colder and fainter BLAST sources.  The \MIPS\ images
were destriped, which is advantageous for probing faint sources, and
corrected at the pixel level for non-linearity \citep{dale2007}.
Measurements using a Gaussian model of fixed FWHM 40\arcsec\ (typical
extent of actual \MIPS\ sources) at the positions of 39 faint BLAST
sources where we have found no apparent counterparts in the \MIPS\ image
give an rms flux density of 5.7 Jy.  This provides a very useful upper
limit for these sources, constraining the SED on the Wien side of the
peak.
For flux densities of somewhat brighter sources, we first convolved the
70~\micron\ map to 1\arcmin, to be comparable to BLAST, and then made
measurements using multi-Gaussian photometry with a linear background
model.  This provided useful detections for 56 sources and upper limits
for the others.  For six BLAST sources, pixels at the peaks of the
bright \MIPS\ counterparts are blanked; however, the \IRAS\ data for these
are sufficient to constrain the SED.
The resolution of our convolved maps is also closer to the IGA
60~\micron\ resolution, facilitating a comparison with isolated sources
with a good \IRAS\ detection.  We find that the \MIPS\ flux densities are
slightly higher than at 60~\micron\ ($F_{60}/F_{70}$ $=$ 0.78), as expected for these SEDs.  At the
native resolution, without beam matching, the \MIPS\ flux densities are
systematically lower, as is discussed by \citet{mottram2010}.


\textit{\MSX\ 8, 12, 14, and 21~\micron}.
\MSX\ \citep{price2001} observed in A, C, D, and E bands at 8.3, 12.1,
14.7 and 21.3~\micron, respectively.  Among the four bands, A is the
most sensitive and is dominated by diffuse emission from PAH molecules.
It has angular resolution 18.3\arcsec.

\MSX\ sources have relatively small positional uncertainties (within
4\arcsec\ - 5\arcsec). \MSX\ sources were identified within a search
radius of 30\arcsec\ about the BLAST coordinate.  After visual
inspection, we rejected some \MSX\ ``matches'' that were either in a PDR
and/or excited by some nearby OB stars, and so had nothing to do with
the star formation history of the subject BLAST source.

We extracted flux densities from the map in its native resolution
directly by fitting multiple Gaussians, assuming a constant (not tilted)
background.  We find that our flux densities in all the bands are
systematically higher than reported in the \MSX\ point source catalog,
and the deviation is more noticeable for less bright sources.  This
trend was also observed by \citet{mol2008}.

Results of \MSX\ photometry are given in Table~\ref{tab:msx}.  In our
present analysis, \MSX\ flux densities in no way constrain temperature
and total mass.  We use \MSX\ flux densities in the SEDs only to
calculate a small correction to the bolometric luminosity by connecting
data points piecewise continuously \citep{chapin2008}.  See for example,
the SED for W75N in Figure~\ref{fig:sed1}; note the ``rising spectrum"
from 4 to 21~\micron, characteristic of a MYSO.  We find that this
correction does not change the results significantly as the
\Cyg\ sources detected by BLAST are quite cool.

\bibliographystyle{apj}


\end{document}